\documentclass[twocolumn]{aastex631}
\usepackage{epsfig,amsmath,footmisc,longtable,blindtext,hyperref}
\usepackage{gensymb}
\usepackage[normalem]{ulem}
\usepackage{soul}
\usepackage{xcolor}
\usepackage{natbib}

\newcommand{\xray}{\hbox{X-ray}}  

\newcommand{\hst}{{\it HST\/}}       

\newcommand{\chandra}{{\it Chandra\/}}

\newcommand{\xmm}{\hbox{\it XMM-Newton\/}}

\newcommand{\jwst}{{\it JWST\/}}
\newcommand{\spitzer}{{\it Spitzer\/ }} 
\newcommand{\galex}{{\it GALEX\/}}
\newcommand{\swift}{{\it Swift\/}}

\begin{document}

\title{X-ray to Mid-IR Spectral Energy Distributions: A Catalog of X-ray Selected AGNs in the XMM-COSMOS Survey}


\author[0000-0001-6924-9320]{\c{S}eyda \c{S}en}
\affiliation{Sabanc{\i} University, Faculty of Engineering and Natural Sciences, 34956, Istanbul, T\"{u}rkiye}

\author[0000-0002-6909-192X]{Eda Sonba\c{s}}
\affiliation{Department of Physics, The George Washington University, Washington, DC 20052, USA}
\affiliation{Sabanc{\i} University, Faculty of Engineering and Natural Sciences, 34956, Istanbul, T\"{u}rkiye}

\author[0000-0002-9119-2313]{Ece Kilerci}
\affiliation{Department of Astronomy and Space Sciences, Science Faculty, \.{I}stanbul University, Beyaz{\i}t 34119, \.Istanbul, T\"{u}rkiye}

\author[0000-0001-8694-1503]{Hasan Avdan}
\affiliation{T\"{u}rkiye National Observatories, DAG, 25050, Erzurum, T\"{u}rkiye}

\author[0000-0002-5422-4873]{Kalvir S. Dhuga}
\affiliation{Department of Physics, The George Washington University, Washington, DC 20052, USA}

\author[0000-0002-5274-6790]{Ersin G\"{o}\u{g}\"{u}\c{s}}
\affiliation{Sabanc{\i} University, Faculty of Engineering and Natural Sciences, 34956, Istanbul, T\"{u}rkiye}

\author[0000-0002-0167-2453]{W. N. Brandt}
\affiliation{Department of Astronomy and Astrophysics, 525 Davey Lab, The Pennsylvania State University, University Park, PA 16802, USA}
\affiliation{Institute for Gravitation and the Cosmos, The Pennsylvania State University, University Park, PA 16802, USA}
\affiliation{Department of Physics, 104 Davey Laboratory, The Pennsylvania State University, University Park, PA 16802, USA}

\begin{abstract}
\noindent We present a component-resolved analysis of the IR--X-ray connection in a sample of 104 spectroscopically confirmed Type~1 AGN. By combining \xmm, \swift/UVOT, and new high--angular resolution \jwst/NIRCam imaging for a subset of the sources, we constructed detailed SEDs in which the IR emission was explicitly decomposed into accretion-disk, torus, and polar-dust components. This approach enables us to isolate the intrinsic AGN-IR luminosities with minimal host-galaxy contamination, providing a clean assessment of the physical relationship between the accretion flow, circumnuclear dust, and the X-ray corona. By incorporating JWST/NIRCam observations into our SED decomposition, we significantly reduce host-galaxy contamination and isolate the nuclear infrared emission with greater fidelity, particularly at near-infrared (NIR) wavelengths. This improvement is most apparent at 1 $\mu\mathrm{m}$, where sources with JWST coverage exhibit a markedly steeper relation between $\nu$L$_{\nu}$ (2 keV) and the NIR luminosity than those without JWST data. This behavior is naturally explained by starlight dilution, whereby host-galaxy emission dominates lower-resolution measurements.
\end{abstract}


\keywords{AGN host galaxies (2017), Active galactic nuclei (16), Spectral energy distribution (2129) }

\section{Introduction} \label{sec:intro}

Active Galactic Nuclei (AGNs) are among the most energetic persistent sources in the Universe, powered by the accretion of material onto supermassive black holes (SMBH) residing at the centers of galaxies. This accretion produces emissions that span the entire electromagnetic spectrum, from radio to gamma rays. Connecting the emission across these regimes provides unique laboratories for understanding of the connection between black hole growth and galaxy evolution (e.g., \citealt{Fabian2012,Kormendy_Ho2013}).

The emission observed from AGNs originates from a variety of physical components, including the accretion disk, hot corona, obscuring torus, and relativistic jets.  The relative contributions of these components are reflected in their spectral energy distributions (SEDs), which ideally cover a broad range of wavelengths.

The $\xray$ band (2–10 keV) provides a particularly valuable window into AGN physics. Produced through Compton up-scattering of accretion disk photons in a hot corona (e.g., \citealt{Haardt_Maraschi1991, Reis_Miller2013}), these high-energy photons are significantly less affected by obscuration compared to UV or soft X-ray emission. While 2–10 keV emission remains relatively unaffected by column densities up to $N_{\mathrm{H}} \approx 10^{23}$ cm$^{-2}$, at higher column densities—particularly in the Compton-thick regime ($N_{\mathrm{H}} \gtrsim 10^{24}$ cm$^{-2}$)—photoelectric absorption and Compton scattering strongly suppress the observed flux \citep[e.g.,][]{Comastri2004, Ricci2017}. Optical and UV emission predominantly originates from the accretion disk \citep{Shakura_Sunyaev1973}, while infrared (IR) emission is attributed to reprocessing by surrounding circumnuclear dust structures \citep[e.g.,][]{Netzer2015, Mateos2016, RamosAlmeida2017}. These components, when observed together in a well-sampled SED, provide essential diagnostics for AGN physics, including accretion rates, obscuration levels, and black hole mass estimates \citep[e.g.,][]{Lusso2010, Padovani2017}.

The classical unified model of AGNs suggests that the observed variety in SEDs can be explained primarily through different viewing angles and varying levels of nuclear obscuration \citep{Antonucci1993, Urry1995}. However, more recent studies indicate that geometry alone is insufficient to explain the full diversity of AGN properties. Factors such as bolometric luminosity, accretion rate, and the complex, clumpy structure of the obscuring dust also play critical roles in shaping the observed emission \citep[e.g.,][]{Netzer2015, RamosAlmeida2017}. A key feature of orientation-based models is the presence of a circumnuclear, parsec-scale structure—commonly referred to as the dusty torus—that is both optically and geometrically thick \citep[e.g.,][]{Antonucci1993, Urry1995, Jaffe2004, Burtscher2013}. This structure obscures the inner regions of the AGN, including the accretion disk, X-ray emitting corona, and the broad-line region (BLR), depending on the viewing angle. According to these models, AGNs are classified as optically Type~1 when observed at low inclinations with respect to the polar axis of the obscuring torus, so that the line of sight does not intersect the obscuring material. In this case, the central region, including the black hole and the inner accretion flow, is directly visible. Conversely, when AGNs are viewed at higher inclinations, where the line of sight intersects the obscuring torus, the central regions are hidden from direct view, and the AGNs are classified as optically Type-2 \citep[see][]{Netzer2015, RamosAlmeida2017}. 

The presence of circumnuclear dust in AGN environments has been directly established through high-angular resolution observations, including infrared interferometry \citep[e.g.,][]{Jaffe2004, Tristram2007} and more recently with the Atacama Large Millimeter/submillimeter Array (ALMA) \citep[e.g.,][]{GarciaBurillo2016, AlonsoHerrero2018}. While the existence of these dusty structures is well known, their detailed geometry and physical nature remain an active area of investigation \citep[e.g.,][]{Mateos2016, RamosAlmeida2017}. X-ray observations (e.g., \citealt{Risaliti2002, Lamer2003, Markowitz2014}) indicate that the obscuring medium is neither homogeneous nor strictly axisymmetric. Consistent with these findings, modern theoretical and observational studies increasingly favor a multi-component circumnuclear environment over the classical smooth torus paradigm. In this framework, the dusty structure is described as a combination of a clumpy equatorial disk and a clumpy, radiatively driven polar wind (e.g., \citealt{Elitzur2006, Netzer2015, Stalevski2012}), where an ensemble of optically and geometrically thick clouds governs the infrared emission and obscuration properties across a range of spatial scales.

To better understand the interplay between the primary accretion flow, the high-energy corona, and the circumnuclear dust reprocessing, we focus on three specific regions of the SED, each probing distinct physical scales: (i) the near-infrared (NIR) region around $\sim$1 $\mu$m ($\sim$3 $\times$ 10$^{14}$ Hz), which captures the transition between the accretion disk and the hottest dust near the sublimation radius; (ii) the mid-infrared (MIR) region between $\sim$3–10 $\mu$m ($\sim$1 $\times$ 10$^{14}$ to $\sim$3 $\times$ 10$^{13}$ Hz), dominated by thermal emission from the dusty torus; and (iii) the X-ray region between 2–10 keV ($\sim$4.8 $\times$ 10$^{17}$ to $\sim$2.4 $\times$ 10$^{18}$ Hz), which provides a window into the high-energy corona. The primary objective of this work is to utilize the unprecedented sensitivity and angular resolution of \textit{JWST} to effectively isolate the intrinsic AGN infrared signature from the host-galaxy contribution. By combining these \textit{JWST} observations with archival X-ray and optical/UV data, we aim to provide robust constraints on the intrinsic SED shapes of X-ray selected Type~1 AGNs and investigate the physical connections between the central engine and the surrounding obscuring material.

The Cosmic Evolution Survey (COSMOS; \citealt{Scoville2007b}) provides an ideal combination of depth, area, and multi-wavelength coverage for studying AGNs. Covering a 2 deg$^{2}$ equatorial field, COSMOS has been observed by $\hst$ \citep{Scoville2007a, Koekemoer2007}, $\spitzer$ \citep{Sanders2007}, $\galex$ \citep{Zamojski2007}, $\xmm$ \citep{Hasinger2007, Cappelluti2009}, and $\chandra$ \citep{Civano2016}, alongside extensive coverage from ground-based facilities such as Subaru \citep{Taniguchi2007}, VLA \citep{Schinnerer2007}, CFHT \citep{McCracken2010}, and UKIRT \citep{McCracken2012}. This extensive dataset enables AGN identification through multiple techniques: X-ray \citep{Brusa2007, Brusa2010}, IR \citep{Donley2012}, radio \citep{Schinnerer2010}, and optical spectroscopy \citep[e.g.,][]{Lilly2007, Trump2007} or variability analyses \citep[e.g.,][]{DeCicco2015}.

In this paper, we present a new multi-wavelength SED catalog for a well-defined sample of 104 X-ray selected Type~1 AGNs from the XMM-COSMOS field. This catalog combines deep X-ray observations from XMM-Newton, UV/optical photometry from Swift/UVOT, and high-resolution infrared imaging from JWST/NIRCam and MIRI. Our study utilizes XMM-Newton data to maintain consistency with the primary XMM-COSMOS parent catalog \citep{Cappelluti2009, Brusa2010}. The XMM-Newton flux limits (reaching $\sim$$1.7 \times 10^{-15}$ erg cm$^{-2}$ s$^{-1}$ in the 0.5–2 keV band) are well-matched to the brightness of our sample, which exhibits a median 2–10 keV flux of $1.58 \times 10^{-14}$ erg cm$^{-2}$ s$^{-1}$, ensuring sufficient signal-to-noise for our multi-wavelength analysis. We aim to better constrain the intrinsic AGN SED by using \textit{JWST}’s high resolution to separate nuclear emission from host-galaxy stars and ISM dust more accurately than previous \textit{Spitzer} studies. Our analysis focuses on the MIR regime to study torus emission and revisit the X-ray–MIR luminosity relations. The SEDs are modeled using templates that account for the accretion disk, corona, and dusty torus, as well as host-galaxy stellar and ISM dust components.

The paper is organized as follows: Section 2 describes our sample selection and the reduction of $\xmm$, Swift/UVOT, and $\jwst$ data. Section 3 details our SED fitting methodology, including the treatment of absorption corrections and host galaxy subtraction. Results are presented in Section 4, where we compare our findings with previous studies and examine the implications for AGN unification models. We discuss the broader impacts of our work in Section 5, including implications for future $\jwst$ observations of AGNs. 

Throughout this work, we adopt a flat $\Lambda$CDM cosmology with $H_0 = 70$ km s$^{-1}$ Mpc$^{-1}$, $\Omega_M = 0.3$, and $\Omega_{\Lambda} = 0.7$.

\section{AGN Sample and Data Reduction}
\noindent \citet{Lusso2010} studied 545 X-ray-detected, radio-quiet Type~1 AGNs from the XMM-COSMOS survey. From this parent sample, we selected a representative subset of 104 AGNs based on the availability of reliable spectroscopic redshifts and comprehensive multi-wavelength coverage required for robust SED modeling. Specifically, our selection criteria require high-quality observations spanning the X-ray to MIR regime, including X-ray data from XMM-Newton, UV and optical observations compiled from Swift/UVOT and ground-based surveys incorporated into the COSMOS2020 catalog, near-infrared data from ground-based surveys and JWST, and mid-infrared observations from both Spitzer and JWST. All 104 sources in our subset are strictly radio-quiet AGNs, ensuring that the intrinsic X-ray luminosity ($L_X$; 2–10 keV) and SED measurements are not affected by jet-related emission. A small subset of the sample (five sources) consists of starburst galaxies. Although not a distinct AGN class, these objects were retained because their intense star formation can significantly influence the infrared SEDs. Their enhanced star formation rates produce strong HII-region emission, prominent nebular lines, and Polycyclic Aromatic Hydrocarbons (PAH) features that can contribute substantially to the observed infrared emission.

\begin{figure}[ht!]
\centering
\includegraphics[width=1.0\columnwidth]{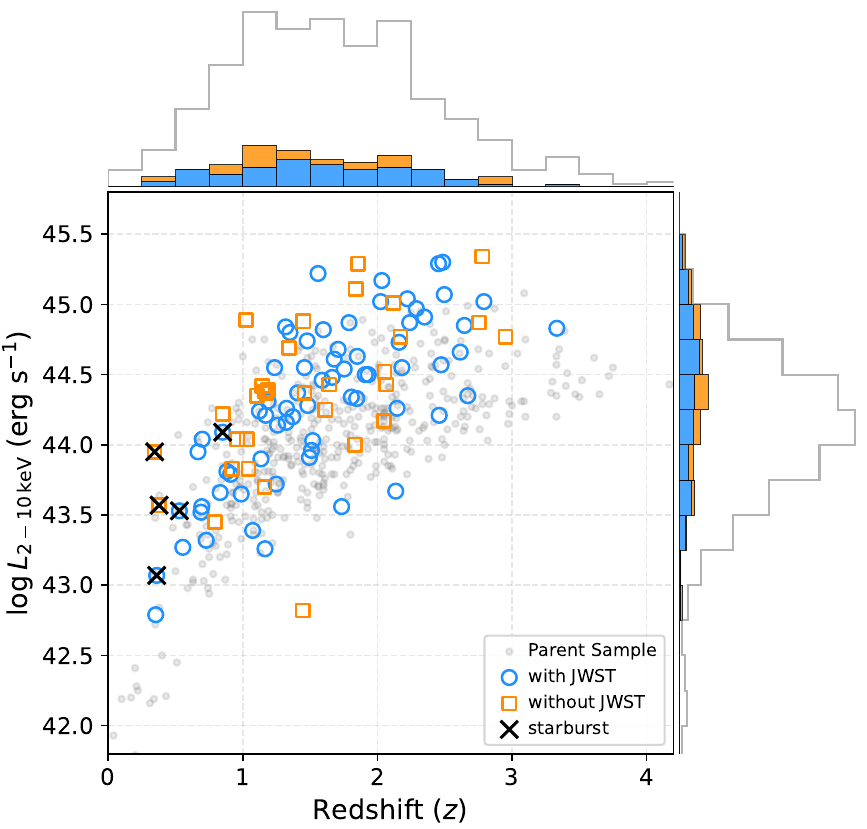}
\caption{Distribution of intrinsic 2--10 keV X-ray luminosity ($L_X$) versus redshift ($z$) for our sample. Open blue circles represent sources with \textit{JWST} coverage (71 sources), while open orange squares indicate the control sample without \textit{JWST} (33 sources). Black crosses highlight sources identified as starbursts. The light grey background points and the corresponding grey step histograms represent the parent Type~1 AGN population from the XMM-COSMOS survey \citep{Lusso2010}. The top and right marginal panels show the stacked redshift and $L_X$ distributions, respectively, providing a direct visual comparison between our subset and the parent population.}
\label{fig:Lxvsz1}
\end{figure}

We specifically identified 71 sources located within the archival JWST/NIRCam and MIRI footprints (see Section~\ref{sec:jwstdata} for details), while the remaining 33 sources, serve as a control sample to assess the impact of host-galaxy subtraction at lower angular resolutions. While we cross-matched the IDs with \citet{Brusa2009} for internal consistency, we refer to the sources throughout this work by their XID from \citet{Lusso2010} to maintain simplicity. The distribution of our selected sample in terms of redshift and $L_X$ is presented in Fig. \ref{fig:Lxvsz1}, where it is compared against the parent XMM-COSMOS Type~1 population. To statistically validate that this subset is a representative, unbiased representation of the X-ray selected Type~1 AGN population, we conducted a two-sample Kolmogorov-Smirnov (K-S) test. The resulting p-values for redshift ($p=0.32$) and X-ray luminosity ($p=0.065$) indicate no significant difference between the two distributions.

\begin{deluxetable*}{lcccrccrccccc}
\tablecolumns{13}
\tabletypesize{\scriptsize}
\setlength{\tabcolsep}{0.07in}
\tablewidth{0pt}
\rotate
\tablecaption{Unified Properties of the X-ray Detected COSMOS AGN Sample}
\label{tab:unified_catalog}
\tablehead{
\colhead{XID} & \colhead{RA} & \colhead{DEC} & \colhead{$z$} & \colhead{$i_{\mathrm{AGN}}$} & \colhead{$frac_{\mathrm{AGN}}$} & \colhead{$\log L_{\rm{AGN}}^{\rm{UV-IR}}$} & \colhead{$\log L_{\rm{AGN}}^{\rm{X-SED}}$} & \colhead{$\log \nu L_\nu^{\rm{1\mu m}}$} & \colhead{$\log \nu L_\nu^{\rm{6\mu m}}$} & \colhead{$\log \nu L_\nu^{\rm{12\mu m}}$} & \colhead{$\log L_{\rm{X}}^{\rm{2-10keV}}$} & \colhead{$\log \nu L_\nu^{\rm{2keV}}$} \\
\colhead{} & \colhead{(deg)} & \colhead{(deg)} & \colhead{} & \colhead{(deg)} & \colhead{} & \colhead{(erg/s)} & \colhead{(erg/s)} & \colhead{(erg/s)} & \colhead{(erg/s)} & \colhead{(erg/s)} & \colhead{(erg/s)} & \colhead{(erg/s)} \\
\colhead{(1)} & \colhead{(2)} & \colhead{(3)} & \colhead{(4)} & \colhead{(5)} & \colhead{(6)} & \colhead{(7)} & \colhead{(8)} & \colhead{(9)} & \colhead{(10)} & \colhead{(11)} & \colhead{(12)} & \colhead{(13)}
}
\startdata
1	&	150.105	&	1.981	&	0.373	&	20	&	0.15	$\pm$	0.00	&	44.65	$\pm$	0.02	&	44.66	$\pm$	0.02	&	43.61	$\pm$	0.02	&	43.04	$\pm$	0.02	&	42.83	$\pm$	0.02	&	43.96	$\pm$	0.03	&	43.75	$\pm$	0.03	\\
2	&	149.739	&	2.221	&	1.024	&	50	&	0.50	$\pm$	0.00	&	45.45	$\pm$	0.02	&	45.46	$\pm$	0.02	&	44.26	$\pm$	0.02	&	44.94	$\pm$	0.02	&	44.92	$\pm$	0.02	&	44.89	$\pm$	0.02	&	44.62	$\pm$	0.03	\\
3*	&	149.761	&	2.318	&	0.345	&	10	&	0.36	$\pm$	0.05	&	45.00	$\pm$	0.02	&	44.34	$\pm$	0.05	&	43.81	$\pm$	0.02	&	44.18	$\pm$	0.02	&	44.19	$\pm$	0.02	&	43.95	$\pm$	0.03	&	43.76	$\pm$	0.07	\\
6*	&	150.180	&	2.110	&	0.360	&	90	&	0.20	$\pm$	0.00	&	43.72	$\pm$	0.02	&	43.74	$\pm$	0.05	&	42.90	$\pm$	0.02	&	43.35	$\pm$	0.02	&	43.38	$\pm$	0.02	&	43.07	$\pm$	0.02	&	42.83	$\pm$	0.02	\\
8	&	150.054	&	2.590	&	0.699	&	0	&	0.20	$\pm$	0.00	&	44.37	$\pm$	0.02	&	44.53	$\pm$	0.03	&	43.38	$\pm$	0.02	&	42.81	$\pm$	0.02	&	42.62	$\pm$	0.02	&	44.04	$\pm$	0.08	&	43.97	$\pm$	0.11	\\
10	&	149.912	&	2.200	&	0.689	&	40	&	0.70	$\pm$	0.00	&	44.99	$\pm$	0.02	&	44.00	$\pm$	0.03	&	43.84	$\pm$	0.02	&	44.39	$\pm$	0.02	&	44.39	$\pm$	0.02	&	43.52	$\pm$	0.13	&	43.46	$\pm$	0.15	\\
13*	&	150.009	&	2.276	&	0.850	&	40	&	0.60	$\pm$	0.00	&	45.08	$\pm$	0.02	&	44.58	$\pm$	0.06	&	43.79	$\pm$	0.02	&	44.58	$\pm$	0.02	&	44.56	$\pm$	0.02	&	44.09	$\pm$	0.05	&	43.87	$\pm$	0.08	\\
15	&	150.499	&	2.445	&	2.033	&	50	&	0.70	$\pm$	0.00	&	46.57	$\pm$	0.02	&	45.69	$\pm$	0.02	&	45.46	$\pm$	0.02	&	45.16	$\pm$	0.02	&	45.06	$\pm$	0.02	&	45.17	$\pm$	0.04	&	44.92	$\pm$	0.06	\\
16	&	150.472	&	2.410	&	0.667	&	10	&	0.05	$\pm$	0.00	&	44.94	$\pm$	0.02	&	44.46	$\pm$	0.02	&	43.76	$\pm$	0.02	&	43.87	$\pm$	0.02	&	43.86	$\pm$	0.02	&	43.95	$\pm$	0.08	&	43.64	$\pm$	0.14	\\
17	&	149.852	&	1.998	&	1.236	&	0	&	0.23	$\pm$	0.07	&	46.41	$\pm$	0.02	&	45.23	$\pm$	0.02	&	45.31	$\pm$	0.02	&	44.87	$\pm$	0.02	&	44.70	$\pm$	0.02	&	44.55	$\pm$	0.01	&	44.44	$\pm$	0.04	\\
18	&	150.133	&	2.303	&	1.598	&	30	&	0.40	$\pm$	0.00	&	45.49	$\pm$	0.02	&	45.46	$\pm$	0.02	&	44.40	$\pm$	0.02	&	44.96	$\pm$	0.02	&	44.97	$\pm$	0.02	&	44.82	$\pm$	0.10	&	44.62	$\pm$	0.13	\\
20	&	150.245	&	2.432	&	0.695	&	0	&	0.20	$\pm$	0.01	&	44.89	$\pm$	0.02	&	44.11	$\pm$	0.02	&	43.68	$\pm$	0.02	&	43.91	$\pm$	0.02	&	43.92	$\pm$	0.02	&	43.56	$\pm$	0.07	&	43.42	$\pm$	0.08	\\
22	&	150.195	&	2.068	&	0.554	&	10	&	0.41	$\pm$	0.05	&	44.98	$\pm$	0.02	&	43.77	$\pm$	0.06	&	43.76	$\pm$	0.02	&	44.07	$\pm$	0.02	&	44.09	$\pm$	0.02	&	43.27	$\pm$	0.02	&	43.14	$\pm$	0.14	\\
23*	&	149.790	&	2.321	&	0.378	&	40	&	0.20	$\pm$	0.00	&	44.33	$\pm$	0.02	&	44.20	$\pm$	0.07	&	43.26	$\pm$	0.02	&	43.79	$\pm$	0.02	&	43.78	$\pm$	0.02	&	43.57	$\pm$	0.03	&	43.32	$\pm$	0.05	\\
24	&	150.103	&	2.530	&	1.318	&	70	&	0.15	$\pm$	0.00	&	45.60	$\pm$	0.02	&	45.47	$\pm$	0.05	&	44.50	$\pm$	0.02	&	44.06	$\pm$	0.02	&	43.90	$\pm$	0.02	&	44.84	$\pm$	0.03	&	44.61	$\pm$	0.07	\\
25	&	150.102	&	2.105	&	2.289	&	30	&	0.64	$\pm$	0.12	&	46.51	$\pm$	0.02	&	45.59	$\pm$	0.04	&	45.33	$\pm$	0.03	&	45.50	$\pm$	0.03	&	45.49	$\pm$	0.03	&	44.97	$\pm$	0.02	&	44.71	$\pm$	0.11	\\
28	&	150.306	&	2.602	&	1.342	&	40	&	0.40	$\pm$	0.00	&	45.08	$\pm$	0.02	&	45.39	$\pm$	0.06	&	44.01	$\pm$	0.02	&	44.55	$\pm$	0.02	&	44.54	$\pm$	0.02	&	44.69	$\pm$	0.03	&	44.21	$\pm$	0.12	\\
30	&	149.956	&	2.028	&	1.753	&	30	&	0.35	$\pm$	0.10	&	46.42	$\pm$	0.02	&	45.22	$\pm$	0.05	&	45.26	$\pm$	0.03	&	45.28	$\pm$	0.03	&	45.26	$\pm$	0.03	&	44.54	$\pm$	0.05	&	44.31	$\pm$	0.15	\\
31	&	149.946	&	2.369	&	0.909	&	20	&	0.20	$\pm$	0.00	&	45.26	$\pm$	0.02	&	44.52	$\pm$	0.06	&	44.07	$\pm$	0.02	&	44.20	$\pm$	0.02	&	44.20	$\pm$	0.02	&	43.79	$\pm$	0.01	&	43.65	$\pm$	0.08	\\
34	&	149.994	&	2.301	&	1.789	&	30	&	0.10	$\pm$	0.00	&	45.64	$\pm$	0.02	&	45.44	$\pm$	0.06	&	44.45	$\pm$	0.02	&	44.62	$\pm$	0.02	&	44.61	$\pm$	0.02	&	44.87	$\pm$	0.09	&	44.48	$\pm$	0.14	\\
37	&	149.917	&	2.385	&	1.123	&	10	&	0.06	$\pm$	0.02	&	45.75	$\pm$	0.02	&	44.79	$\pm$	0.02	&	44.58	$\pm$	0.02	&	44.69	$\pm$	0.02	&	44.68	$\pm$	0.02	&	44.24	$\pm$	0.05	&	43.98	$\pm$	0.07	\\
38	&	150.245	&	1.900	&	1.559	&	20	&	0.30	$\pm$	0.00	&	45.93	$\pm$	0.02	&	45.97	$\pm$	0.08	&	44.74	$\pm$	0.02	&	44.93	$\pm$	0.02	&	44.92	$\pm$	0.02	&	45.22	$\pm$	0.02	&	44.64	$\pm$	0.12	\\
39	&	150.498	&	2.660	&	0.851	&	50	&	0.38	$\pm$	0.17	&	44.93	$\pm$	0.04	&	44.85	$\pm$	0.10	&	43.68	$\pm$	0.04	&	44.36	$\pm$	0.04	&	44.33	$\pm$	0.04	&	44.22	$\pm$	0.05	&	43.91	$\pm$	0.11	\\
42	&	150.512	&	2.410	&	0.988	&	30	&	0.03	$\pm$	0.02	&	45.19	$\pm$	0.02	&	44.22	$\pm$	0.10	&	44.02	$\pm$	0.03	&	44.19	$\pm$	0.03	&	44.19	$\pm$	0.03	&	43.65	$\pm$	0.03	&	43.64	$\pm$	0.14	\\
44	&	150.215	&	2.204	&	1.850	&	20	&	0.60	$\pm$	0.01	&	45.70	$\pm$	0.03	&	45.21	$\pm$	0.08	&	44.53	$\pm$	0.03	&	44.65	$\pm$	0.03	&	44.65	$\pm$	0.03	&	44.63	$\pm$	0.03	&	44.41	$\pm$	0.10	\\
47	&	150.318	&	2.602	&	0.959	&	20	&	0.46	$\pm$	0.14	&	45.26	$\pm$	0.03	&	44.48	$\pm$	0.07	&	44.07	$\pm$	0.08	&	44.32	$\pm$	0.08	&	44.33	$\pm$	0.08	&	44.04	$\pm$	0.16	&	43.73	$\pm$	0.21	\\
48	&	149.771	&	2.258	&	2.222	&	10	&	0.27	$\pm$	0.07	&	46.16	$\pm$	0.02	&	45.62	$\pm$	0.02	&	45.08	$\pm$	0.02	&	45.50	$\pm$	0.02	&	45.53	$\pm$	0.02	&	45.04	$\pm$	0.02	&	44.74	$\pm$	0.12	\\
49	&	150.073	&	2.004	&	0.353	&	40	&	0.44	$\pm$	0.08	&	43.42	$\pm$	0.08	&	43.41	$\pm$	0.09	&	42.29	$\pm$	0.07	&	42.83	$\pm$	0.07	&	42.82	$\pm$	0.07	&	42.79	$\pm$	0.05	&	42.56	$\pm$	0.07	\\
51	&	150.059	&	2.015	&	2.497	&	30	&	0.16	$\pm$	0.05	&	46.47	$\pm$	0.03	&	45.58	$\pm$	0.06	&	45.31	$\pm$	0.10	&	45.13	$\pm$	0.10	&	45.07	$\pm$	0.10	&	45.07	$\pm$	0.09	&	44.67	$\pm$	0.12	\\
52	&	150.068	&	1.851	&	1.135	&	30	&	0.01	$\pm$	0.02	&	45.17	$\pm$	0.02	&	44.52	$\pm$	0.06	&	44.04	$\pm$	0.10	&	43.86	$\pm$	0.10	&	43.80	$\pm$	0.10	&	43.90	$\pm$	0.03	&	43.81	$\pm$	0.08	\\
53	&	150.314	&	2.804	&	1.459	&	40	&	0.56	$\pm$	0.12	&	45.25	$\pm$	0.03	&	45.00	$\pm$	0.07	&	44.18	$\pm$	0.06	&	44.71	$\pm$	0.06	&	44.70	$\pm$	0.06	&	44.37	$\pm$	0.02	&	44.18	$\pm$	0.07	\\
56	&	150.004	&	2.237	&	1.407	&	10	&	0.60	$\pm$	0.00	&	45.29	$\pm$	0.02	&	45.01	$\pm$	0.07	&	44.21	$\pm$	0.02	&	44.64	$\pm$	0.02	&	44.67	$\pm$	0.02	&	44.37	$\pm$	0.01	&	44.15	$\pm$	0.15	\\
57	&	150.491	&	2.775	&	1.449	&	0	&	0.10	$\pm$	0.04	&	45.83	$\pm$	0.03	&	45.69	$\pm$	0.08	&	44.74	$\pm$	0.07	&	44.85	$\pm$	0.07	&	44.85	$\pm$	0.07	&	44.88	$\pm$	0.02	&	44.55	$\pm$	0.14	\\
60	&	150.216	&	1.989	&	2.240	&	30	&	0.01	$\pm$	0.02	&	45.64	$\pm$	0.03	&	45.49	$\pm$	0.05	&	44.48	$\pm$	0.10	&	44.50	$\pm$	0.10	&	44.48	$\pm$	0.10	&	44.87	$\pm$	0.02	&	44.59	$\pm$	0.05	\\
61	&	150.124	&	2.358	&	0.728	&	40	&	0.40	$\pm$	0.01	&	44.47	$\pm$	0.02	&	44.07	$\pm$	0.04	&	43.40	$\pm$	0.02	&	43.94	$\pm$	0.02	&	43.93	$\pm$	0.02	&	43.32	$\pm$	0.05	&	43.28	$\pm$	0.15	\\
66	&	149.868	&	2.331	&	1.478	&	20	&	0.10	$\pm$	0.01	&	45.84	$\pm$	0.02	&	45.31	$\pm$	0.11	&	44.62	$\pm$	0.04	&	44.88	$\pm$	0.04	&	44.89	$\pm$	0.04	&	44.74	$\pm$	0.02	&	44.28	$\pm$	0.09	\\
69	&	149.895	&	2.174	&	1.323	&	40	&	0.39	$\pm$	0.02	&	44.93	$\pm$	0.02	&	44.58	$\pm$	0.08	&	43.86	$\pm$	0.03	&	44.40	$\pm$	0.03	&	44.39	$\pm$	0.03	&	44.16	$\pm$	0.14	&	43.93	$\pm$	0.31	\\
71	&	149.956	&	2.502	&	1.458	&	20	&	0.10	$\pm$	0.00	&	44.90	$\pm$	0.02	&	45.28	$\pm$	0.09	&	43.84	$\pm$	0.02	&	44.28	$\pm$	0.02	&	44.29	$\pm$	0.02	&	44.55	$\pm$	0.02	&	44.23	$\pm$	0.09	
\enddata
\end{deluxetable*}

\addtocounter{table}{-1}
\begin{deluxetable*}{lcccrccrccccc}
\tablecolumns{13}
\tabletypesize{\scriptsize}
\setlength{\tabcolsep}{0.07in}
\tablewidth{0pt}
\rotate
\tablecaption{\textit{(Continued)}}
\tablehead{
\colhead{XID} & \colhead{RA} & \colhead{DEC} & \colhead{$z$} & \colhead{$i_{\mathrm{AGN}}$} & \colhead{$frac_{\mathrm{AGN}}$} & \colhead{$\log L_{\rm{AGN}}^{\rm{UV-IR}}$} & \colhead{$\log L_{\rm{AGN}}^{\rm{X-SED}}$} & \colhead{$\log \nu L_\nu^{\rm{1\mu m}}$} & \colhead{$\log \nu L_\nu^{\rm{6\mu m}}$} & \colhead{$\log \nu L_\nu^{\rm{12\mu m}}$} & \colhead{$\log L_{\rm{X}}^{\rm{2-10keV}}$} & \colhead{$\log \nu L_\nu^{\rm{2keV}}$} \\
\colhead{} & \colhead{(deg)} & \colhead{(deg)} & \colhead{} & \colhead{(deg)} & \colhead{} & \colhead{(erg/s)} & \colhead{(erg/s)} & \colhead{(erg/s)} & \colhead{(erg/s)} & \colhead{(erg/s)} & \colhead{(erg/s)} & \colhead{(erg/s)} \\
\colhead{(1)} & \colhead{(2)} & \colhead{(3)} & \colhead{(4)} & \colhead{(5)} & \colhead{(6)} & \colhead{(7)} & \colhead{(8)} & \colhead{(9)} & \colhead{(10)} & \colhead{(11)} & \colhead{(12)} & \colhead{(13)}
}
\startdata
74	&	150.450	&	2.246	&	0.882	&	30	&	0.10	$\pm$	0.00	&	45.33	$\pm$	0.02	&	44.37	$\pm$	0.05	&	44.14	$\pm$	0.02	&	44.31	$\pm$	0.02	&	44.31	$\pm$	0.02	&	43.81	$\pm$	0.03	&	43.66	$\pm$	0.14	\\
75	&	150.132	&	1.799	&	1.679	&	40	&	0.38	$\pm$	0.09	&	45.50	$\pm$	0.04	&	45.30	$\pm$	0.08	&	44.30	$\pm$	0.07	&	44.37	$\pm$	0.07	&	44.35	$\pm$	0.07	&	44.61	$\pm$	0.02	&	44.46	$\pm$	0.12	\\
79	&	150.354	&	2.342	&	1.708	&	0	&	0.05	$\pm$	0.00	&	45.70	$\pm$	0.06	&	45.18	$\pm$	0.07	&	44.49	$\pm$	0.02	&	44.05	$\pm$	0.02	&	43.88	$\pm$	0.02	&	44.68	$\pm$	0.08	&	44.40	$\pm$	0.12	\\
83	&	150.351	&	2.678	&	2.754	&	50	&	0.33	$\pm$	0.16	&	46.05	$\pm$	0.10	&	45.59	$\pm$	0.08	&	44.62	$\pm$	0.12	&	45.30	$\pm$	0.12	&	45.28	$\pm$	0.12	&	44.87	$\pm$	0.01	&	44.65	$\pm$	0.19	\\
84	&	150.300	&	2.507	&	1.495	&	0	&	0.05	$\pm$	0.00	&	45.51	$\pm$	0.02	&	44.25	$\pm$	0.08	&	44.41	$\pm$	0.02	&	43.97	$\pm$	0.02	&	43.80	$\pm$	0.02	&	43.91	$\pm$	0.08	&	43.80	$\pm$	0.36	\\
86	&	150.509	&	2.699	&	0.794	&	30	&	0.10	$\pm$	0.03	&	44.76	$\pm$	0.02	&	44.29	$\pm$	0.11	&	43.57	$\pm$	0.05	&	43.75	$\pm$	0.05	&	43.74	$\pm$	0.05	&	43.45	$\pm$	0.05	&	43.44	$\pm$	0.09	\\
87	&	150.102	&	1.848	&	1.664	&	10	&	0.20	$\pm$	0.00	&	45.54	$\pm$	0.02	&	45.35	$\pm$	0.09	&	44.45	$\pm$	0.02	&	44.88	$\pm$	0.02	&	44.91	$\pm$	0.02	&	44.48	$\pm$	0.03	&	44.25	$\pm$	0.08	\\
88	&	150.518	&	2.522	&	2.779	&	20	&	0.31	$\pm$	0.16	&	44.43	$\pm$	0.06	&	45.12	$\pm$	0.21	&	44.94	$\pm$	0.16	&	44.59	$\pm$	0.16	&	44.60	$\pm$	0.16	&	45.34	$\pm$	0.06	&	45.06	$\pm$	0.11	\\
91	&	150.496	&	2.413	&	1.371	&	30	&	0.24	$\pm$	0.11	&	45.34	$\pm$	0.02	&	44.89	$\pm$	0.05	&	44.12	$\pm$	0.02	&	44.30	$\pm$	0.02	&	44.29	$\pm$	0.02	&	44.20	$\pm$	0.01	&	43.99	$\pm$	0.08	\\
93	&	150.292	&	2.545	&	2.646	&	10	&	0.05	$\pm$	0.00	&	45.96	$\pm$	0.02	&	45.60	$\pm$	0.10	&	44.78	$\pm$	0.02	&	44.89	$\pm$	0.02	&	44.89	$\pm$	0.02	&	44.85	$\pm$	0.01	&	44.48	$\pm$	0.21	\\
94	&	150.200	&	2.191	&	1.510	&	20	&	0.01	$\pm$	0.00	&	45.61	$\pm$	0.02	&	44.57	$\pm$	0.12	&	44.43	$\pm$	0.02	&	44.57	$\pm$	0.02	&	44.56	$\pm$	0.02	&	43.96	$\pm$	0.02	&	43.85	$\pm$	0.07	\\
95	&	150.029	&	2.210	&	1.258	&	50	&	0.70	$\pm$	0.00	&	45.27	$\pm$	0.02	&	44.48	$\pm$	0.02	&	44.08	$\pm$	0.02	&	44.76	$\pm$	0.02	&	44.74	$\pm$	0.02	&	44.14	$\pm$	0.22	&	43.88	$\pm$	0.25	\\
96	&	150.402	&	2.884	&	2.117	&	10	&	0.48	$\pm$	0.15	&	46.60	$\pm$	0.02	&	45.64	$\pm$	0.06	&	44.48	$\pm$	0.02	&	45.25	$\pm$	0.02	&	45.18	$\pm$	0.02	&	45.01	$\pm$	0.02	&	44.78	$\pm$	0.08	\\
98	&	150.573	&	2.500	&	1.106	&	30	&	0.45	$\pm$	0.16	&	45.46	$\pm$	0.04	&	44.75	$\pm$	0.10	&	44.27	$\pm$	0.05	&	44.45	$\pm$	0.05	&	44.44	$\pm$	0.05	&	44.35	$\pm$	0.12	&	44.02	$\pm$	0.17	\\
103	&	150.117	&	1.930	&	1.519	&	50	&	0.60	$\pm$	0.03	&	45.21	$\pm$	0.08	&	44.75	$\pm$	0.09	&	42.99	$\pm$	0.11	&	44.84	$\pm$	0.11	&	44.83	$\pm$	0.11	&	44.03	$\pm$	0.01	&	43.84	$\pm$	0.12	\\
109	&	150.457	&	2.648	&	2.050	&	20	&	0.31	$\pm$	0.11	&	45.98	$\pm$	0.09	&	45.21	$\pm$	0.10	&	44.79	$\pm$	0.10	&	45.05	$\pm$	0.10	&	45.06	$\pm$	0.10	&	44.17	$\pm$	0.09	&	44.05	$\pm$	0.16	\\
110	&	150.092	&	2.399	&	2.473	&	60	&	0.50	$\pm$	0.13	&	45.64	$\pm$	0.08	&	45.11	$\pm$	0.09	&	44.31	$\pm$	0.06	&	45.08	$\pm$	0.06	&	45.03	$\pm$	0.06	&	44.57	$\pm$	0.04	&	44.46	$\pm$	0.35	\\
111	&	150.334	&	2.561	&	1.834	&	20	&	0.44	$\pm$	0.16	&	46.23	$\pm$	0.02	&	45.01	$\pm$	0.08	&	44.50	$\pm$	0.07	&	44.61	$\pm$	0.07	&	44.49	$\pm$	0.07	&	44.00	$\pm$	0.10	&	43.82	$\pm$	0.13	\\
113	&	149.898	&	2.094	&	1.910	&	20	&	0.07	$\pm$	0.03	&	45.44	$\pm$	0.03	&	45.08	$\pm$	0.08	&	45.04	$\pm$	0.03	&	45.17	$\pm$	0.03	&	45.17	$\pm$	0.03	&	44.50	$\pm$	0.03	&	44.18	$\pm$	0.11	\\
114	&	150.312	&	1.978	&	2.350	&	20	&	0.19	$\pm$	0.11	&	45.52	$\pm$	0.03	&	45.20	$\pm$	0.16	&	44.24	$\pm$	0.05	&	44.38	$\pm$	0.05	&	44.37	$\pm$	0.05	&	44.91	$\pm$	0.15	&	44.73	$\pm$	0.19	\\
115	&	150.545	&	2.507	&	1.161	&	20	&	0.35	$\pm$	0.10	&	45.85	$\pm$	0.05	&	45.22	$\pm$	0.10	&	44.32	$\pm$	0.12	&	44.58	$\pm$	0.12	&	44.58	$\pm$	0.12	&	44.38	$\pm$	0.13	&	44.22	$\pm$	0.16	\\
118	&	149.895	&	2.239	&	1.734	&	30	&	0.54	$\pm$	0.05	&	45.84	$\pm$	0.05	&	44.39	$\pm$	0.05	&	44.68	$\pm$	0.09	&	44.82	$\pm$	0.09	&	44.81	$\pm$	0.09	&	43.56	$\pm$	0.02	&	43.65	$\pm$	0.18	\\
123	&	150.383	&	2.560	&	2.065	&	20	&	0.11	$\pm$	0.06	&	45.29	$\pm$	0.02	&	44.95	$\pm$	0.17	&	44.67	$\pm$	0.05	&	44.84	$\pm$	0.05	&	44.84	$\pm$	0.05	&	44.43	$\pm$	0.02	&	44.14	$\pm$	0.23	\\
127	&	150.004	&	2.389	&	1.846	&	10	&	0.01	$\pm$	0.00	&	44.89	$\pm$	0.02	&	45.32	$\pm$	0.09	&	44.11	$\pm$	0.10	&	44.36	$\pm$	0.10	&	44.37	$\pm$	0.10	&	44.33	$\pm$	0.03	&	44.25	$\pm$	0.16	\\
128	&	150.199	&	2.133	&	2.161	&	40	&	0.32	$\pm$	0.20	&	46.55	$\pm$	0.02	&	45.28	$\pm$	0.06	&	43.68	$\pm$	0.04	&	43.92	$\pm$	0.04	&	43.92	$\pm$	0.04	&	44.73	$\pm$	0.04	&	44.65	$\pm$	0.19	\\
129	&	150.253	&	1.997	&	1.170	&	50	&	0.02	$\pm$	0.03	&	44.93	$\pm$	0.02	&	44.43	$\pm$	0.16	&	45.36	$\pm$	0.08	&	45.24	$\pm$	0.08	&	45.19	$\pm$	0.08	&	44.21	$\pm$	0.17	&	43.96	$\pm$	0.20	\\
132	&	149.992	&	2.132	&	2.136	&	20	&	0.30	$\pm$	0.01	&	45.37	$\pm$	0.02	&	44.57	$\pm$	0.28	&	43.78	$\pm$	0.11	&	43.73	$\pm$	0.11	&	43.69	$\pm$	0.11	&	43.67	$\pm$	0.21	&	43.57	$\pm$	0.27	\\
133	&	150.272	&	2.230	&	2.615	&	10	&	0.09	$\pm$	0.05	&	46.26	$\pm$	0.02	&	45.59	$\pm$	0.13	&	44.31	$\pm$	0.02	&	44.74	$\pm$	0.02	&	44.76	$\pm$	0.02	&	44.66	$\pm$	0.39	&	44.57	$\pm$	0.44	\\
136	&	149.667	&	2.286	&	1.029	&	30	&	0.09	$\pm$	0.03	&	44.53	$\pm$	0.07	&	44.59	$\pm$	0.10	&	45.06	$\pm$	0.04	&	45.29	$\pm$	0.04	&	45.30	$\pm$	0.04	&	44.04	$\pm$	0.05	&	43.66	$\pm$	0.23	\\
137	&	149.958	&	2.003	&	1.806	&	50	&	0.80	$\pm$	0.02	&	45.49	$\pm$	0.04	&	44.91	$\pm$	0.08	&	43.38	$\pm$	0.14	&	43.55	$\pm$	0.14	&	43.55	$\pm$	0.14	&	44.34	$\pm$	0.02	&	44.07	$\pm$	0.11	\\
141	&	150.139	&	1.877	&	0.832	&	30	&	0.11	$\pm$	0.06	&	45.24	$\pm$	0.04	&	44.36	$\pm$	0.13	&	44.31	$\pm$	0.04	&	44.98	$\pm$	0.04	&	44.96	$\pm$	0.04	&	43.66	$\pm$	0.05	&	43.24	$\pm$	0.18	\\
146	&	149.910	&	2.081	&	2.791	&	40	&	0.30	$\pm$	0.02	&	45.74	$\pm$	0.04	&	45.48	$\pm$	0.05	&	44.05	$\pm$	0.04	&	44.22	$\pm$	0.04	&	44.22	$\pm$	0.04	&	45.02	$\pm$	0.03	&	44.67	$\pm$	0.13	\\
153	&	150.285	&	2.395	&	1.932	&	20	&	0.08	$\pm$	0.03	&	45.58	$\pm$	0.04	&	45.40	$\pm$	0.06	&	43.49	$\pm$	0.04	&	45.33	$\pm$	0.04	&	45.34	$\pm$	0.04	&	44.50	$\pm$	0.20	&	44.33	$\pm$	0.24	\\
161	&	150.286	&	2.015	&	2.671	&	20	&	0.15	$\pm$	0.06	&	46.16	$\pm$	0.02	&	45.03	$\pm$	0.11	&	44.39	$\pm$	0.11	&	44.64	$\pm$	0.11	&	44.65	$\pm$	0.11	&	44.35	$\pm$	0.06	&	44.30	$\pm$	0.20	\\
162	&	150.402	&	2.791	&	0.920	&	40	&	0.07	$\pm$	0.04	&	44.56	$\pm$	0.04	&	44.40	$\pm$	0.06	&	44.95	$\pm$	0.04	&	45.20	$\pm$	0.04	&	45.21	$\pm$	0.04	&	43.83	$\pm$	0.09	&	43.50	$\pm$	0.20	\\
163	&	150.310	&	2.467	&	1.165	&	20	&	0.06	$\pm$	0.15	&	44.34	$\pm$	0.02	&	43.77	$\pm$	0.12	&	43.35	$\pm$	0.14	&	43.23	$\pm$	0.14	&	43.18	$\pm$	0.14	&	43.26	$\pm$	0.03	&	43.11	$\pm$	0.13	\\
164*	&	150.327	&	1.929	&	0.529	&	30	&	0.39	$\pm$	0.17	&	44.75	$\pm$	0.02	&	44.09	$\pm$	0.14	&	43.14	$\pm$	0.24	&	43.39	$\pm$	0.24	&	43.40	$\pm$	0.24	&	43.53	$\pm$	0.11	&	43.32	$\pm$	0.16	\\
165	&	150.180	&	2.231	&	2.146	&	40	&	0.56	$\pm$	0.05	&	45.14	$\pm$	0.06	&	45.03	$\pm$	0.11	&	43.56	$\pm$	0.04	&	43.73	$\pm$	0.04	&	43.73	$\pm$	0.04	&	44.26	$\pm$	0.06	&	44.05	$\pm$	0.27	
\enddata
\end{deluxetable*}

\addtocounter{table}{-1}
\begin{deluxetable*}{lcccrccrccccc}
\tablecolumns{13}
\tabletypesize{\scriptsize}
\setlength{\tabcolsep}{0.07in}
\tablewidth{0pt}
\rotate
\tablecaption{\textit{(Continued)}}
\tablehead{
\colhead{XID} & \colhead{RA} & \colhead{DEC} & \colhead{$z$} & \colhead{$i_{\mathrm{AGN}}$} & \colhead{$frac_{\mathrm{AGN}}$} & \colhead{$\log L_{\rm{AGN}}^{\rm{UV-IR}}$} & \colhead{$\log L_{\rm{AGN}}^{\rm{X-SED}}$} & \colhead{$\log \nu L_\nu^{\rm{1\mu m}}$} & \colhead{$\log \nu L_\nu^{\rm{6\mu m}}$} & \colhead{$\log \nu L_\nu^{\rm{12\mu m}}$} & \colhead{$\log L_{\rm{X}}^{\rm{2-10keV}}$} & \colhead{$\log \nu L_\nu^{\rm{2keV}}$} \\
\colhead{} & \colhead{(deg)} & \colhead{(deg)} & \colhead{} & \colhead{(deg)} & \colhead{} & \colhead{(erg/s)} & \colhead{(erg/s)} & \colhead{(erg/s)} & \colhead{(erg/s)} & \colhead{(erg/s)} & \colhead{(erg/s)} & \colhead{(erg/s)} \\
\colhead{(1)} & \colhead{(2)} & \colhead{(3)} & \colhead{(4)} & \colhead{(5)} & \colhead{(6)} & \colhead{(7)} & \colhead{(8)} & \colhead{(9)} & \colhead{(10)} & \colhead{(11)} & \colhead{(12)} & \colhead{(13)}
}
\startdata
166	&	150.167	&	2.798	&	1.042	&	20	&	0.24	$\pm$	0.11	&	45.38	$\pm$	0.02	&	44.61	$\pm$	0.08	&	44.12	$\pm$	0.06	&	44.65	$\pm$	0.06	&	44.64	$\pm$	0.06	&	43.83	$\pm$	0.06	&	43.78	$\pm$	0.14	\\
167	&	149.887	&	2.117	&	2.048	&	40	&	0.48	$\pm$	0.04	&	44.92	$\pm$	0.05	&	45.22	$\pm$	0.07	&	44.20	$\pm$	0.04	&	44.34	$\pm$	0.04	&	44.33	$\pm$	0.04	&	44.57	$\pm$	0.05	&	44.29	$\pm$	0.19	\\
168	&	150.104	&	2.666	&	2.951	&	60	&	0.49	$\pm$	0.18	&	45.90	$\pm$	0.33	&	45.23	$\pm$	0.09	&	43.87	$\pm$	0.05	&	44.41	$\pm$	0.05	&	44.40	$\pm$	0.05	&	44.77	$\pm$	0.05	&	44.62	$\pm$	0.24	\\
171	&	150.367	&	2.305	&	1.187	&	60	&	0.20	$\pm$	0.19	&	44.65	$\pm$	0.06	&	44.78	$\pm$	0.10	&	43.99	$\pm$	0.05	&	44.76	$\pm$	0.05	&	44.71	$\pm$	0.05	&	44.31	$\pm$	0.14	&	43.79	$\pm$	0.22	\\
176	&	150.454	&	2.806	&	1.613	&	30	&	0.36	$\pm$	0.15	&	45.62	$\pm$	0.05	&	44.86	$\pm$	0.09	&	43.35	$\pm$	0.06	&	44.11	$\pm$	0.06	&	44.07	$\pm$	0.06	&	44.25	$\pm$	0.04	&	44.05	$\pm$	0.14	\\
180	&	150.209	&	2.482	&	3.333	&	60	&	0.47	$\pm$	0.13	&	45.71	$\pm$	0.16	&	45.76	$\pm$	0.19	&	44.45	$\pm$	0.05	&	44.62	$\pm$	0.05	&	44.61	$\pm$	0.05	&	44.83	$\pm$	0.06	&	44.64	$\pm$	0.24	\\
186	&	149.895	&	2.047	&	2.182	&	10	&	0.12	$\pm$	0.05	&	45.64	$\pm$	0.02	&	44.14	$\pm$	0.02	&	42.22	$\pm$	0.16	&	45.47	$\pm$	0.16	&	45.48	$\pm$	0.16	&	44.55	$\pm$	0.45	&	44.36	$\pm$	0.47	\\
187	&	150.241	&	2.659	&	3.356	&	50	&	0.20	$\pm$	0.00	&	44.48	$\pm$	0.02	&	45.53	$\pm$	0.24	&	44.45	$\pm$	0.05	&	44.68	$\pm$	0.05	&	44.69	$\pm$	0.05	&	44.34	$\pm$	0.07	&	44.05	$\pm$	0.12	\\
188	&	149.624	&	2.181	&	1.188	&	60	&	0.52	$\pm$	0.07	&	44.32	$\pm$	0.10	&	44.89	$\pm$	0.06	&	43.28	$\pm$	0.02	&	43.96	$\pm$	0.02	&	43.93	$\pm$	0.02	&	44.39	$\pm$	0.14	&	44.13	$\pm$	0.16	\\
189	&	150.127	&	2.627	&	1.839	&	60	&	0.10	$\pm$	0.03	&	44.91	$\pm$	0.02	&	45.68	$\pm$	0.04	&	44.95	$\pm$	0.10	&	45.03	$\pm$	0.10	&	44.80	$\pm$	0.10	&	45.11	$\pm$	0.01	&	44.59	$\pm$	0.18	\\
192	&	150.251	&	2.737	&	2.172	&	10	&	0.24	$\pm$	0.10	&	46.01	$\pm$	0.03	&	45.29	$\pm$	0.18	&	43.65	$\pm$	0.02	&	44.41	$\pm$	0.02	&	44.37	$\pm$	0.02	&	44.77	$\pm$	0.10	&	44.28	$\pm$	0.23	\\
196	&	149.837	&	2.009	&	1.483	&	20	&	0.42	$\pm$	0.04	&	45.56	$\pm$	0.03	&	44.74	$\pm$	0.08	&	44.81	$\pm$	0.11	&	44.92	$\pm$	0.11	&	44.92	$\pm$	0.11	&	44.28	$\pm$	0.11	&	44.00	$\pm$	0.15	\\
197	&	150.164	&	2.598	&	1.589	&	50	&	0.60	$\pm$	0.00	&	45.10	$\pm$	0.02	&	45.33	$\pm$	0.42	&	44.34	$\pm$	0.03	&	44.60	$\pm$	0.03	&	44.61	$\pm$	0.03	&	44.46	$\pm$	0.24	&	44.43	$\pm$	0.43	\\
199	&	149.744	&	2.028	&	2.454	&	20	&	0.40	$\pm$	0.00	&	46.41	$\pm$	0.05	&	45.93	$\pm$	0.05	&	43.90	$\pm$	0.02	&	44.58	$\pm$	0.02	&	44.55	$\pm$	0.02	&	45.29	$\pm$	0.00	&	45.00	$\pm$	0.07	\\
216	&	150.243	&	1.869	&	2.024	&	20	&	0.32	$\pm$	0.04	&	45.75	$\pm$	0.05	&	45.50	$\pm$	0.20	&	45.22	$\pm$	0.05	&	45.48	$\pm$	0.05	&	45.49	$\pm$	0.05	&	45.02	$\pm$	0.09	&	44.94	$\pm$	0.18	\\
219	&	149.866	&	2.003	&	1.248	&	40	&	0.01	$\pm$	0.00	&	44.93	$\pm$	0.02	&	44.49	$\pm$	0.07	&	43.82	$\pm$	0.02	&	43.46	$\pm$	0.02	&	43.34	$\pm$	0.02	&	43.72	$\pm$	0.08	&	43.57	$\pm$	0.20	\\
221	&	150.555	&	2.641	&	1.144	&	40	&	0.54	$\pm$	0.07	&	45.05	$\pm$	0.07	&	45.00	$\pm$	0.08	&	43.91	$\pm$	0.07	&	44.45	$\pm$	0.07	&	44.44	$\pm$	0.07	&	44.42	$\pm$	0.06	&	44.03	$\pm$	0.20	\\
228	&	150.252	&	2.486	&	1.073	&	0	&	0.25	$\pm$	0.05	&	45.40	$\pm$	0.03	&	44.03	$\pm$	0.08	&	44.17	$\pm$	0.05	&	44.52	$\pm$	0.05	&	44.55	$\pm$	0.05	&	43.39	$\pm$	0.09	&	43.33	$\pm$	0.23	\\
236	&	150.416	&	2.526	&	1.445	&	20	&	0.18	$\pm$	0.08	&	44.77	$\pm$	0.06	&	43.63	$\pm$	0.08	&	43.61	$\pm$	0.13	&	43.86	$\pm$	0.13	&	43.87	$\pm$	0.13	&	42.82	$\pm$	0.03	&	42.84	$\pm$	0.19	\\
237	&	150.183	&	2.247	&	2.485	&	30	&	0.10	$\pm$	0.00	&	45.98	$\pm$	0.02	&	45.83	$\pm$	0.14	&	44.88	$\pm$	0.02	&	45.44	$\pm$	0.02	&	45.45	$\pm$	0.02	&	45.30	$\pm$	0.07	&	45.11	$\pm$	0.12	\\
239	&	150.397	&	2.902	&	1.642	&	30	&	0.04	$\pm$	0.12	&	44.47	$\pm$	0.32	&	44.53	$\pm$	0.23	&	43.31	$\pm$	0.41	&	43.87	$\pm$	0.41	&	43.87	$\pm$	0.41	&	44.43	$\pm$	0.07	&	44.19	$\pm$	0.16	\\
245	&	150.254	&	2.331	&	2.459	&	10	&	0.10	$\pm$	0.02	&	45.39	$\pm$	0.02	&	44.73	$\pm$	0.13	&	44.22	$\pm$	0.08	&	44.33	$\pm$	0.08	&	44.32	$\pm$	0.08	&	44.21	$\pm$	0.10	&	44.03	$\pm$	0.21	\\
249	&	150.096	&	2.145	&	1.325	&	30	&	0.10	$\pm$	0.00	&	44.69	$\pm$	0.02	&	44.98	$\pm$	0.07	&	43.61	$\pm$	0.03	&	44.17	$\pm$	0.03	&	44.17	$\pm$	0.03	&	44.26	$\pm$	0.02	&	43.94	$\pm$	0.10	\\
250	&	150.065	&	2.329	&	1.350	&	30	&	0.10	$\pm$	0.00	&	44.90	$\pm$	0.02	&	45.47	$\pm$	0.21	&	42.47	$\pm$	0.02	&	44.49	$\pm$	0.02	&	44.53	$\pm$	0.02	&	44.80	$\pm$	0.05	&	44.48	$\pm$	0.34	\\
265	&	150.176	&	1.760	&	1.161	&	40	&	0.10	$\pm$	0.08	&	44.72	$\pm$	0.33	&	44.29	$\pm$	0.13	&	43.68	$\pm$	0.33	&	43.22	$\pm$	0.33	&	43.21	$\pm$	0.33	&	43.70	$\pm$	0.06	&	43.54	$\pm$	0.11	\\
273	&	150.383	&	2.723	&	2.052	&	10	&	0.25	$\pm$	0.14	&	45.38	$\pm$	0.03	&	44.61	$\pm$	0.12	&	44.18	$\pm$	0.07	&	44.41	$\pm$	0.07	&	44.42	$\pm$	0.07	&	44.52	$\pm$	0.27	&	44.24	$\pm$	0.26	\\
275	&	150.159	&	2.825	&	1.856	&	20	&	0.51	$\pm$	0.04	&	45.77	$\pm$	0.02	&	46.10	$\pm$	0.06	&	44.71	$\pm$	0.02	&	45.14	$\pm$	0.02	&	45.16	$\pm$	0.02	&	45.29	$\pm$	0.03	&	44.87	$\pm$	0.10	\\
281	&	150.147	&	2.718	&	1.177	&	40	&	0.01	$\pm$	0.00	&	44.83	$\pm$	0.02	&	44.97	$\pm$	0.13	&	43.66	$\pm$	0.02	&	43.74	$\pm$	0.02	&	43.72	$\pm$	0.02	&	44.37	$\pm$	0.10	&	43.99	$\pm$	0.21
\enddata
\tablecomments{Galaxies marked with an asterisk ($^{\ast}$) are classified as starburst broad-line AGNs. 
The columns are defined as follows: 
Columns (1)--(4): Galaxy identification number (XID), Right Ascension, Declination (both in degrees), and spectroscopic redshift ($z$), fundamentally anchored to the master catalog of \citet{Lusso2010}.
Columns (5)--(6): Torus inclination angle ($i_{\mathrm{AGN}}$) and AGN fractional contribution ($frac_{\mathrm{AGN}}$) derived from SED fitting.
Columns (7)--(8): Total AGN luminosities from SED fitting ($\log L_{\rm{AGN}}^{\rm{UV-IR}}$ and $\log L_{\rm{AGN}}^{\rm{X-SED}}$).
Columns (9)--(11): Monochromatic infrared emissions at 1, 6, and 12 $\mu$m derived from SED fitting to map out the dust emission profile.
Columns (12)--(13): Intrinsic X-ray luminosities ($\log L_{\rm{X}}^{\rm{2-10keV}}$ and $\log \nu L_\nu^{\rm{2keV}}$) computed through empirical XSPEC modeling, completely decoupled from SED-inferred statistical values.}
\end{deluxetable*}

\subsection{X-ray Observations} 
\noindent In this study, we carried out X-ray spectral analysis for a sample of 104 AGNs utilizing data from the three detectors of the European Photon Imaging Camera (EPIC; \citealt{Jansen2001}) onboard \textit{XMM-Newton}: the two MOS \citep{Turner2001} and the pn detectors \citep{Struder2001}. Observation data for each source were obtained from the \textit{XMM-Newton} Science Archive. Data reduction was performed using the \textit{XMM-Newton} Science Analysis Software (SAS), version 19.1.0. Raw event files were generated with the SAS tasks \texttt{emproc} (for MOS) and \texttt{epproc} (for pn). 

To select the optimal dataset for sources with multiple archive observations, background light curves were first generated for all candidate pointings to screen for soft-proton background flaring. High-background flare intervals were excluded by defining Good Time Intervals (GTIs). The single observation yielding the longest clean net exposure time post-filtering was then selected for analysis. This approach provides a homogeneous dataset with high spectral quality while avoiding strong flare contamination, large off-axis PSF distortions, and spectral/flux variability across different epochs.

Events were filtered to retain valid X-ray photon patterns (patterns 0--12 for MOS and 0--4 for pn). Source spectra were extracted using circular regions with radii of $30^{\prime\prime}$--$35^{\prime\prime}$, optimized to balance the source encircled energy fraction against background contamination, brightness, and local environment. Background spectra were extracted from nearby source-free regions on the same CCD chip. Response matrix files (RMFs) and ancillary response files (ARFs) were generated individually for each target using \texttt{rmfgen} and \texttt{arfgen}.

Spectral fitting was performed using XSPEC version 12.10.0 \citep{Arnaud1996}. To explicitly account for redshifted intrinsic obscuration, we adopted an absorbed power-law model framework of the form \texttt{phabs $\times$ zTBabs $\times$ powerlaw}. Foreground Galactic absorption was modeled using \texttt{phabs}, with hydrogen column densities ($N_{\mathrm{H}}$) fixed to the coordinate-specific value for each target obtained from the HI4PI survey \citep{HI4PI2016}. Intrinsic line-of-sight absorption at the source redshift ($z$) was modeled using \texttt{zTBabs}, with the intrinsic column density ($N_{\mathrm{H, int}}$) treated as a free parameter along with the photon index ($\Gamma$) and the power-law normalization.

To ensure unbiased parameter estimation across low- and high-count spectra without relying on standard Gaussian assumptions, spectral fitting was carried out using Poisson-native Cash statistics (\texttt{c-stat}; \citealt{Cash1979}). Spectra were lightly binned using \texttt{grppha} to a minimum of 5 counts per spectral bin to ensure valid execution within XSPEC. All spectra were fitted across the full 0.2--10~keV energy range to properly anchor the primary continuum, with MOS and pn data modeled simultaneously. Instrument relative normalizations were allowed to vary, while all physical model parameters were tied across detectors. Finally, since the CIGALE SED fitting code requires absorption-corrected X-ray inputs \citep{Boquien2019}, we utilized the \texttt{cflux} component in XSPEC to compute unabsorbed intrinsic fluxes strictly within the rest-frame 2--10~keV energy band based on the best-fit spectral model for each source.

The derived intrinsic column densities are predominantly below $10^{22}\ \mathrm{cm}^{-2}$ (typically $\log N_{\mathrm{H, int}} \sim 20\text{--}21.5\ \mathrm{cm}^{-2}$). These low absorption levels confirm that our sample consists of unobscured or mildly obscured systems, fully consistent with expectations for optically classified Type~1 AGNs \citep[e.g.,][]{Mateos2005, Tozzi2006, Mainieri2007, Merloni2014}. 

The X-ray spectral properties derived in this work are summarized in Table~\ref{tab:unified_catalog}, including the source ID, the intrinsic 2--10 keV luminosity, and the rest-frame 2 keV monochromatic luminosity obtained from the spectral fitting analysis.

\subsection{UV Observations} \label{sec:uvdata}

\noindent UV data for the selected sources were retrieved from the {\it Swift} Ultraviolet/Optical Telescope (UVOT) archive via the High Energy Astrophysics Science Archive Research Center (HEASARC) interface. Circular source extraction regions were centered on the coordinates listed in Table 1. A 3 arcsec extraction radius was adopted to balance maximizing the enclosed source flux against minimizing the contamination from the stellar contribution of nearby sources. The flux density values (in units of erg s$^{-1}$ cm$^{-2}$ \AA$^{-1}$) were measured using the {\scshape uvotsource} task within the {\scshape heasoft} software package (High Energy Astrophysics Software, version 6.27.1). The measured flux densities were converted to magnitudes using the corresponding zeropoint\footnote{\url{https://swift.gsfc.nasa.gov/analysis/uvot_digest/zeropts.html}} and conversion factor values\footnote{\url{https://swift.gsfc.nasa.gov/caldb/docs/uvot/uvot_caldb_counttofluxratio_10wa.pdf}} provided in the UVOT calibration database. To correct for Galactic extinction, extinction values were obtained using the NASA/IPAC Extragalactic Database (NED) extinction calculator. Due to the similarity in their central wavelengths, the extinction values provided for the HST/WFC3 F275W, F225W, and F218W filters were adopted for the UVOT filters UVW1, UVM2, and UVW2, respectively. The extinction-corrected magnitudes were then converted into spectral flux densities in Jansky units for use in the SEDs.

A total of 76 sources without definitive flux measurement in at least one of the {\it Swift}/UVOT UV filters (UVW1, UVM2, or UVW2), resulting in upper limits, as well as those lacking UVOT observations were further examined using the avaliable {\it XMM-Newton} Optical Monitor (OM) data. Among these, 40 sources had available OM measurements that provided reliable detections. The data reductions were carried out using the {\scshape omichain} pipeline task within the Science Analysis Software ({\scshape sas}, version 18.0.0), which provided source detection and photometry in the UVW1, UVM2, and UVW2 bands.  OM photometric analysis was performed following the standard procedure (using a standard aperture size of 5.7 arcsec) to ensure consistent flux conversion and calibration. OM source detections were cross-checked against the coordinates listed in Table~\ref{tab:unified_catalog}, and matches with positional offsets smaller than 1 arcsec were included in the analysis. Similar to the UVOT procedure, extinction corrections were applied using values from the NED extinction calculator, and the resulting magnitudes were converted into spectral flux densities in Jansky units.\footnote{\url{https://www.cosmos.esa.int/web/xmm-newton/sas-watchout-uvflux}}

\subsection{JWST Observations} \label{sec:jwstdata}

\noindent We analyzed imaging data from the \textit{James Webb Space Telescope} (JWST) for a sample of AGN primarily using five filters: NIRCam/F115W, F150W, F277W, F444W, and MIRI/F770W. The observations were obtained from the Mikulski Archive for Space Telescopes (MAST) archive and consist of calibrated Stage 2 mosaics (\texttt{*.i2d.fits} files). Data processing and photometric analysis were carried out using the \texttt{Astropy} \citep{astropy:2022}, \texttt{Photutils} \citep{photutils}, and custom Python routines.

For each filter, we extracted the science extension of the FITS image and normalized the pixel values by the \texttt{PHOTMJSR} keyword to convert surface brightness units (MJy~sr$^{-1}$) to Jy~pixel$^{-1}$. These images were used for both visualization and photometric measurement.

We estimated the 2D background using the \texttt{Background2D} algorithm, with a size of $25 \times 25$ pixels and a $3 \times 3$ pixel median filter. Sigma clipping with a $3\sigma$ threshold was applied to exclude bright sources. A coverage mask was used to avoid zero-value pixels. The background model and RMS noise were calculated using the \texttt{MMMBackground} estimator.

Aperture photometry was performed at target positions using a circular aperture of radius 5 pixels (corresponding to an angular radius of $0.1560$ arcsec for NIRCam and $0.555$ arcsec for MIRI), with a sky annulus from 8 to 11 pixels. Background-subtracted aperture sums were computed using \texttt{aperture\_photometry}, and errors were estimated from the background RMS. We applied aperture corrections to recover total fluxes using empirically derived correction factors for each individual filters.

Instrumental magnitudes were converted to Vega magnitudes using filter-specific zero points computed as:
\begin{equation}
\mathrm{ZP}_\mathrm{Vega} = -2.5 \log_{10} \left( \frac{\mathrm{PHOTMJSR} \times \mathrm{PIXAR\_SR}}{ZP_{\nu} \times 10^6} \right),
\end{equation} 
where $ZP_{\nu}$ is the Vega zero point in Jy. The final corrected magnitudes were used to derive fluxes in both cgs units (erg~cm$^{-2}$~s$^{-1}$) and Jy:
\begin{equation}
f = ZP \times 10^{-0.4 \times m_\mathrm{corr}}, \quad \log f = \log ZP - 0.4 m_\mathrm{corr}.
\end{equation}
\\
Uncertainties were propagated using standard error propagation, combining background RMS and Poisson noise. The final catalogs include corrected magnitudes, flux densities in erg~cm$^{-2}$~s$^{-1}$ and Jy, and their respective uncertainties.

The angular resolution of JWST substantially reduces host-galaxy contamination compared to previous infrared facilities. The physical scale corresponding to an angular size $\theta$ is given by

\begin{equation}
s({\rm pc}) \simeq 4.848 \times D({\rm Mpc}) \times \theta({\rm arcsec})
\end{equation}

\noindent where $D$ is the distance to the source. For the redshift range covered by our sample ($z \approx 0.3$-4), the NIRCam and MIRI circular aperture radius used for the photometric extraction of 0.156 arcsec and 0.555 arcsec correspond to physical scales of approximately 0.7-1.3 kpc and 2.5-4.7 kpc, respectively. The NIRCam observations probe the central kiloparsec-scale regions of the host galaxies, while the MIRI observations sample a larger circumnuclear environment.

For consistency, the same aperture radius of 5 pixels was adopted for the photometric extraction of each source in both the NIRCam and MIRI images. Because the two instruments have different pixel scales, this corresponds to different angular aperture radii, $0.1560$ arcsec for NIRCam and $0.555$ arcsec for MIRI. Consequently, the effective physical regions sampled by the two instruments are not identical. The higher angular resolution of NIRCam provides a cleaner measurement of the nuclear emission, whereas the MIRI photometry is expected to include a larger contribution from surrounding stellar and dust emission. Thus, although the same aperture size was used in pixel units, this does not imply identical levels of host-galaxy contamination in the two instruments. Nevertheless, both NIRCam and MIRI provide a substantial improvement over the lower spatial resolution of previous facilities such as Spitzer, significantly reducing host-galaxy dilution and enabling a more reliable characterization of the nuclear infrared emission.

\subsection{Multi-wavelength data from literature} \label{sec:litdata}
\noindent The COSMOS2020 catalog \citep{Weaver2022} provides homogeneous $2''$ aperture photometry measured using \texttt{SExtractor} \citep{Bertin1996}. We use the Classic version of COSMOS2020 and select sources with continuous coverage from the UV to the mid-IR.

For overlapping photometric bands, we adopt the $u^{\star}$ band from the CLAUDS survey \citep{Sawicki2019}, which provides full coverage of the COSMOS field, and the $g$, $r$, $i$, and $z$ bands from the Hyper Suprime-Cam (HSC), which are significantly deeper than the Subaru Suprime-Cam data.

Out of our total sample, 83 sources have direct counterparts in the COSMOS2020 catalog within a $0.6''$ cross-matching radius. For the remaining 21 sources lacking COSMOS2020 counterparts, we supplement the multi-wavelength photometry using the legacy catalog of \citet{Capak2007}, based on their original COSMOS identifiers. This angular separation refers to the positional matching tolerance between COSMOS2020 source coordinates and the legacy optical/NIR counterpart positions. The \citet{Capak2007} catalog provides deep optical and near-infrared photometry in 15 bands covering $0.3$--$2.4\,\mu$m, obtained from observations with Subaru, CFHT, KPNO, CTIO, and \textit{HST}. The connection between the legacy catalog and X-ray sources is established via the XMM-COSMOS identifiers (XID) from \citet{Lusso2010}, cross-matched through the intermediate catalog of \citet{Brusa2010}, which links COSMOS IDs to updated coordinates.

Galactic extinction corrections are applied using the $E(B-V)$ values reported in Table~11 of \citet{Capak2007}, and \textit{Spitzer}/MIPS $24\,\mu$m measurements are included when available.

\section{SED Analysis} \label{sec:SED}

\noindent {Based on the availability of JWST imaging, we divide our AGNs into two sub-samples: the first sample (S1) consists of 73 AGN located within the JWST/NIRCam footprints, while the second sample (S2) includes those lacking JWST coverage. For S1 objects with available MIRI observations, these data are included in the SED modeling to better constrain the mid-infrared emission and the torus component.

\subsection{CIGALE SED Analysis} \label{sec:cigaleSED}
\noindent The Code Investigating GALaxy Emission \citep[\texttt{CIGALE},][]{Boquien2019} is used to fit the SEDs of our sample. \texttt{CIGALE} performs SED fitting based on the energy balance principle, where the intrinsic UV/optical radiation emitted by the stellar populations and the AGN accretion disk is absorbed by dust and subsequently re-emitted in the mid- to far-infrared wavelengths. \texttt{CIGALE} uses multi-wavelength-band photometric data for user-selected modules \citep[e.g.,][]{Boquien2019}.

We utilise \texttt{CIGALE} version 2022.1 to fit the SEDs with the selected modules listed in Table \ref{tab:cigale} that include star formation history (SFH), dust attenuation law, dust emission, and single stellar population (SSP). 
We use the delayed SFH with optional exponential burst model (sfhdelayed) of \citet{Boquien2019}. We utilize the stellar templates (bc03) from \citet{Bruzual2003} with the initial mass function of \citet{Salpeter1955}. 

Since our galaxy sample spans a wide redshift range, we adopt a broad grid of stellar population ages—ranging from 100 to 6000~Myr—to account for both young and evolved stellar populations. The main star formation history is modeled with e-folding times ($\tau$) spanning 200 to 10000~Myr. For the late starburst component, we set the burst ages between 1 and 5~Myr, with $\tau$ values also ranging from 200 to 10000~Myr.
We note that most of the listed parameter values in Table~ \ref{tab:cigale} are chosen as a result of many trials of the CIGALE on our sample.
We include the nebular emission (continuum and line nebular emission). 
The dust attenuation model of \citet{Charlot2000} (dustatt$\_$modified$\_$CF00) is used. This dust attenuation law model includes the birth cloud and the interstellar medium (ISM) contributions as two power-laws \citep{LoFaro2017,Buat2018} and \citep{Buat2018}. In the fitting framework, the slopes of these components were allowed to vary within the predefined model boundaries. For dust emission the model (dl2014) given by dust templates of \citep{Draine2014} is used. 
Since our sample includes a few starburst galaxies, to be able to account for different  PAHs mass fraction we included most of the possible values between 0.47 and 7.32, with a large range of  minimum radiation field (U$_{min}$) values within the possible range between 0.1 and 50.0. The dust power-law slope was also allowed to be between 1.0 and 3.0. For the fraction of illuminated, we considered a few values within the lowest and highest possible values.
We used the AGN emission model of  \citet{Stalevski2012, Stalevski2016} (skirtor2016). 
While $i_{AGN} = 0^{\circ}$ represents Type~1 AGN, and $i_{AGN} = 90^{\circ}$ represents Type~2 AGN, we still included a large range of  $i_{AGN}$ to perform an analysis independent of optical spectral observation data. For the AGN contribution fraction (${\rm frac}_{\rm AGN}$), since optical spectral classification may not necessarily result in a higher  ${\rm frac}_{\rm AGN}$, we allowed a large set of values between low  and high limits of 0.01 -- 0.9.
We included the X-ray emission module and used fixed redshifts during the fitting. 
To perform an SED analysis independent of X-ray analysis, we did not fix the AGN Photon index and allowed it to be in 1.4 -- 2.1 range. 

Representative examples of the obtained SEDs are shown in Figure \ref{fig:figB3} and \ref{fig:figB4}. The black line is the best-fitted model line that is produced by adding all of the emission components.

The multiwavelength SEDs are assembled from archival data spanning approximately 12 years, primarily between the X-ray observations in 2012 and the JWST observations in 2024. Although not simultaneous, the impact of this time gap is mitigated by the nature of AGN emission; while Type~1 AGNs exhibit variability—typically $\sim$0.05--0.2 dex in the optical \citep[e.g.,][]{VandenBerk2004, MacLeod2010} and up to $\sim$0.3 dex in X-rays \citep{Liu2017,Timlin2020}—mid-infrared variability is generally much smaller ($\lesssim$0.05 dex) due to the large spatial scale of the dust-emitting region \citep[e.g.,][]{Kozlowski2010, Assef2013, Koshida2014}. 
Crucially, our torus parameters are constrained by high-sensitivity JWST mid-infrared photometry, which enables an effective decomposition of host-galaxy and AGN emission, thereby minimizing systematic degeneracies in the SED fitting. 
 
To rigorously evaluate the reliability and scientific validity of the derived best-fit SED parameters, we performed a comprehensive CIGALE mock analysis on a representative control subsample of 50 galaxies spanning the full parameter space of our catalog ($z \approx 0.35 - 3.33$, $\log L_{\mathrm{X}} \approx 42.7 - 45.3$, and $f_{\mathrm{AGN}} \approx 0.01 - 0.90$). The mock workflow generates a synthetic catalog based on the primary best-fit grid, injects Gaussian noise mirroring our empirical photometric uncertainties, and re-fits the artificial data under identical module configurations (\texttt{sfhdelayed}, \texttt{bc03}, \texttt{dustatt\_modified\_CF00}, \texttt{dl2014}, \texttt{skirtor2016}, and \texttt{x\_ray}). We present detailed plots in Appendix~\ref{app:mock_check}, and present the input (exact) and recovered (estimated) parameters with good convergence across all critical parameters. In particular, we find tight 1:1 correlations for the AGN luminosity ($r^2 = 0.99$) and total dust luminosity ($r^2 = 1.00$), ensuring that the overall energy balance is robustly preserved. Furthermore, the AGN fraction ($f_{\mathrm{AGN}}$) and the X-ray spectral index ($\alpha_{\mathrm{ox}}$) are successfully decoupled with recovery rates of $r^2 = 0.93$ and $r^2 = 0.87$, respectively. While structural and geometry-dependent parameters like the inclination angle ($i$, $r^2 = 0.89$) show a marginally higher scatter due to intrinsic SED degeneracies, they remain well-bounded within their physical regimes, confirming that our final catalog parameters are robust and free from systematic modeling artifacts (see Appendix~\ref{app:mock_check} for the complete figure).

For a subset of host-dominated systems, the inferred AGN fractions are exceptionally low ($f_{\rm AGN} \approx 0$) and exhibit formally small or near-zero uncertainties, alongside tightly constrained AGN MIR luminosities. Within the Bayesian framework of CIGALE, this behavior does not signify a high-signal detection of a weak central engine. Instead, it reflects a scenario where the broadband photometry strongly rules out substantial torus emission, forcing the posterior probability distribution functions to tightly concentrate against the lower boundary of the model grid. This sharp truncation inherently produces very narrow credible intervals, indicating a high statistical confidence that the AGN contribution is securely bounded at a minimal level rather than providing a well-measured nuclear component \citep[see e.g.,][]{Yang2022, Yang2023}.

Physical parameters of our AGN sample measured from the SED analysis are listed in Table~\ref{tab:unified_catalog}; these include the AGN inclination ($i_{\mathrm{AGN}}$), $\log L_{\rm{AGN}}^{\rm{UV-IR}}$ which represents the integrated AGN core luminosity covering the spectral range from $0.1\,\mu\rm{m}$ to $1000\,\mu\rm{m}$, the intrinsic AGN X-ray luminosity ($L_{\rm{X, AGN}}$) in the rest-frame $2-10\,\rm{keV}$ band computed through empirical XSPEC spectral modeling to ensure independence from SED-inferred statistical scaling laws, and the monochromatic infrared luminosities $\log \nu L_\nu^{\rm{1\mu m}}$, $\log \nu L_\nu^{\rm{6\mu m}}$, and $\log \nu L_\nu^{\rm{12\mu m}}$.

\begin{table*}
        \centering
        \caption{List of modules and parameter settings for \texttt{CIGALE} fitting.}
        \label{tab:cigale}
        \tabletypesize{\scriptsize}
        \tablewidth{0pt}
        \begin{tabular}{cc}
             \hline
             \textbf{Parameters} & \textbf{Value} \\
             \hline
             \multicolumn{2}{c}{\textbf{Star Formation History Module}: \texttt{sfhdelayed}} \\
             \hline
             e-folding time of the main stellar & 200, 300, 400, 500, 800, 900, 1000, 1100, 1200, 1400, 4500,  \\
             population [$10^6$ yr] &  1500, 1600, 1800, 2000, 2500, 3000, 4000,\\ 
             & 5000, 5500, 6000, 6500, 7000, 8000, 9000, 10000\\
             Age of the galaxy's main stellar & 100, 200, 300, 400, 500, 600, 700, 800, 900, 1000,  \\
             population [$10^6$ yr] &2000, 3000, 4000, 5000, 6000 \\
             e-folding time of the late starburst population [$10^6$ yr] & 200, 500, 2000, 10000\\
             Age of the late burst [$10^6$ yr] & 1.0, 2.0, 5.0\\
             Mass fraction of the late burst population & 0.00, 0.01, 0.05, 0.1 \\
             \hline
             \multicolumn{2}{c}{\textbf{Stellar Population Module}: \texttt{bc03}} \\
             \hline
             Initial mass function & \cite{Salpeter1955} \\
             Metallicity & 0.02 \\
             Age of separation between the young and  & 10.0 \\
             old star populations &  \\
             \hline
             \multicolumn{2}{c}{\textbf{Nebular Emission Module}: \texttt{nebular}} \\
             \hline
             Ionisation parameter &  $-2.0$ \\
             Gas metallicity & 0.02 \\
             \hline
             \multicolumn{2}{c}{\textbf{Dust Attenuation Module}: \texttt{dustatt\_modified\_CF00}} \\
             \hline
             Logarithm of the V-band attenuation in the ISM &  0.1, 0.17, 0.28, 0.46, 0.77, 1.29, 2.15, 3.59, 5.99, 10.0\\
             Ratio of V-band attenuation from old and young stars & 0.44 \\
             Power-law slope of the attenuation in the ISM & $-10.0$, $-5.0$, $-1.3$, $-1.0$, $-0.9$, $-0.7$, $-0.5$, $-0.1$ \\
             Power-law slope of the attenuation in the birth cloud & $-10.0$, $-5.0$, $-3.0$, $-1.3$, $-1.0$, $-0.7$\\
             \hline
             \multicolumn{2}{c}{\textbf{Dust Emission Module}: \texttt{dl2014}} \\
             \hline
             Mass fraction of PAH & 0.47, 1.12, 1.77, 2.50, 3.19, 3.90, 5.26, 5.95, 6.63, 7.32 \\
             Minimum radiation field  & 0.1, 0.2, 0.7, 1.0, 1.2, 1.5, 1.7, 2.0, 3.0, 5.0, 6.0, 7.0, 10.0, 12.0, 17.0, \\
             (${\rm U}_{\rm min}$) &20.0, 30.0, 35.0, 40.0, 50.0 \\
             Power-law slope $\alpha$ ($\frac{{\rm dU}}{{\rm dM}} \propto U^\alpha$) & 1.0, 1.4, 1.5, 1.9, 2.0, 2.3, 2.5, 2.8, 3.0 \\
             Fraction illuminated from ${\rm U}_{\rm min}$ to ${\rm U}_{\rm max}$ & 0.01, 0.1, 0.5, 0.9 \\
             \hline
             \multicolumn{2}{c}{\textbf{AGN Emission Module}: \texttt{skirtor2016}} \\
             \hline            
             Optical depth at 9.7 $\mu$m & 0.1, 0.3, 6.0\\
            $i_\mathrm{AGN}$ & 0.0, 10.0, 20.0, 30.0, 40.0, 50.0, 60.0, 70.0, 80.0, 90.0\\
            AGN contribution fraction (${\rm frac}_{\rm AGN}$)  & 0.01, 0.1--0.9 (step 0.1) \\
             \hline
                          \multicolumn{2}{c}{\textbf{X-ray Emission Module}: \texttt{xray}} \\
             \hline
            $\alpha_{ox}$ & $-1.7$, $-1.6$, $-1.5$, $-1.4$, $-1.3$, $-1.2$, $-1.1$, $-1.0$ \\
            Photon index ($\Gamma$) & 1.4, 1.5, 1.6, 1.7, 1.8, 1.9, 2.0, 2.1 \\
             \hline     
        \end{tabular}
\tablecomments{
References for the respective modules: \texttt{sfhdelayed} \citep{Boquien2019}; \texttt{bc03} \citep{Bruzual2003}; \texttt{dustatt\_modified\_CF00} \citep{Charlot2000}; \texttt{dl2014} \citep{Draine2014}; \texttt{skirtor2016} \citep{Stalevski2012, Stalevski2016}.
\\
For the \texttt{skirtor2016} module, the AGN geometry and dust parameters are fixed to $pl=1.0$, $q=1.0$, $R=20$, $M_{\rm cl}=0.97$, $\delta=-0.36$, and $\texttt{disk\_type}=1$. The torus opening angle varies between $10^\circ$ and $80^\circ$. The polar dust is modeled with an SMC extinction law, $E(B-V)=0.03$, temperature $T_{\rm polar}=1000$~K, and emissivity index $\epsilon=1.6$. \\
For the \texttt{X-ray} module, we adopt $E_{\rm cut}=300$~keV, $\texttt{max\_dev\_alpha\_ox}=0.2$, $\texttt{angle\_coef}=(0.5,0.0)$, and $\texttt{det\_lmxb}=\texttt{det\_hmxb}=0$.
}
\end{table*}

\section{Results} \label{sec:results}
\subsection{Comparison with Literature} \label{sec:compare}
\noindent Before discussing our main results and comparing this study with the available literature, we briefly mention a number of extracted parameters from the SED fits; in particular, the AGN-fraction (defined as the AGN contribution to the total infrared luminosity integrated between $1\,\mu\mathrm{m}$ and $1000\,\mu\mathrm{m}$) and the inclination angle (see the entries in Table~\ref{tab:unified_catalog}). We find that the AGN-fraction turns out to be small ($\sim$$5\%$) for quite a few ($\sim$$20\%$) of the AGN,  which would seem (at first pass) at variance with these sources being Type~1 AGNs. Equally somewhat surprising is that the fits lead to quite a few of the sources exhibiting an inclination angle greater than $\sim$$50^\circ$. This again appears at odds with a strict Type~1 classification. A closer examination of these parameters (the results of which are presented in the Appendix) and the procedures deployed in the SED fits leads us to believe that the extracted parameters are not necessarily out of line with expectations based on the fact that the observational properties defining a Type~1 AGN can span a wide range \citep{Urry_Padovani1995} depending on the degree of dust obscuration and the line-of-sight inclination angle.

Accurate measurements of AGN X-ray luminosities are important for characterizing the radiative output of accreting supermassive black holes and for enabling comparisons across different AGN populations. Several previous studies provide useful benchmarks for these quantities. For example, \citet{Lusso2010} derived 2-10 keV X-ray luminosities for a large sample of Type~1 AGNs using observed (unabsorbed) fluxes from $\xmm$ and $\chandra$ observations. Their analysis did not apply corrections for intrinsic absorption, instead relying on observed luminosities to compute bolometric corrections. 

Figure~\ref{fig:LxvsLlusso} compares the $2$--$10$ keV intrinsic X-ray luminosities derived in this work with those reported by \citet{Lusso2010} for our sample, differentiated by JWST/NIRCam coverage and starburst classification. Overall, we find strong agreement with the 1:1 identity line across more than two orders of magnitude in luminosity ($\log L_{\mathrm{X}} \sim 43.0$--$45.3\ \text{erg s}^{-1}$), confirming the reliability of our X-ray spectral extraction. Both the JWST-covered (open circles) and non-JWST (open squares) subsamples follow the same tight identity trend, with starburst sources (red symbols) also anchoring well along the reference relation at low-to-intermediate luminosities. Minor deviations observed in individual sources (such as XID 236, where $\Delta \log L_{\mathrm{X}} \approx 0.22\text{ dex}$) are well within expected systematic uncertainties. These slight differences arise primarily from our detailed spectral fitting procedure in XSPEC—utilizing redshift-corrected intrinsic absorption (\texttt{ztbabs}) and Poisson statistics (\texttt{cstat}) while allowing the photon index ($\Gamma$) and column density ($N_{\mathrm{H}}$) to vary simultaneously—whereas large-scale literature catalogs often adopt fixed or averaged parameters.

\begin{figure}[ht!]
\centering
\includegraphics[width=0.48\textwidth]{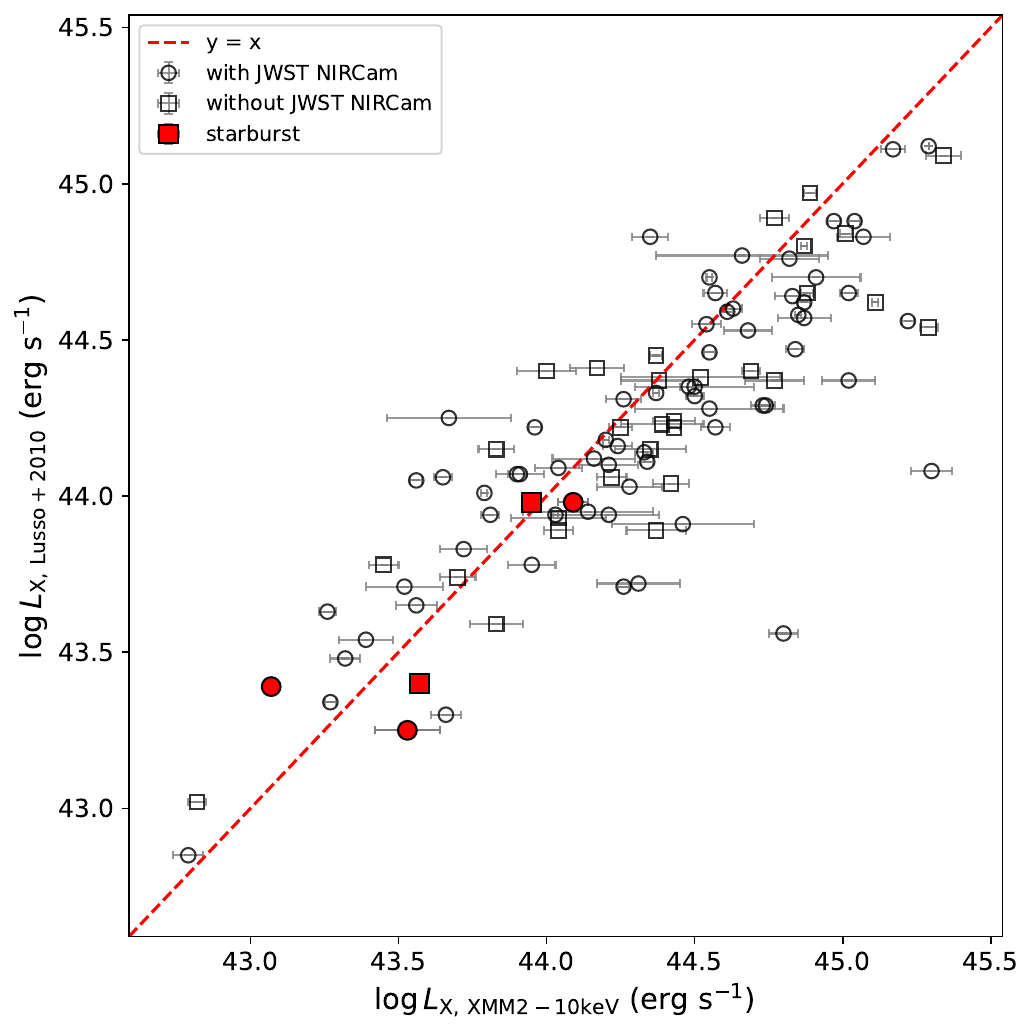}
\caption{Comparison between the 2–10 keV X-ray luminosities derived in this work and those reported by \citet{Lusso2010}. Open circles denote sources with JWST/NIRCam coverage, open squares denote sources without JWST/NIRCam coverage, and red filled symbols mark starburst broad-line AGNs. The dashed red line indicates the one-to-one relation.}
\label{fig:LxvsLlusso}
\end{figure}

\subsection{The X-ray to Mid-IR Relations} \label{sec:Xray-midIR}
\noindent Previous studies have published a close connection between an AGN's mid-infrared continuum and its hard X-ray emission, as both are believed to trace the intrinsic accretion power of the central regime. The 6-$\mu$m emission originates primarily from dust heated by the AGN, while the 2--10~keV X-ray luminosity probes the coronal emission from the inner accretion flow. This link, despite arising from physically distinct regions, serves as a robust diagnostic of AGN bolometric output and obscuration.

Early work by \citet{Lutz2004} studied the low-resolution \textit{ISO} spectra of local Seyfert galaxies. It demonstrated a strong correlation between the nuclear mid-IR and absorption-corrected X-ray luminosities, though the luminosity range was limited to sources with intrinsic X-ray luminosities $\log L_{\mathrm{X, AGN}} \gtrsim 42.7\text{ erg s}^{-1}$ to ensure a reliable overlap with the literature sample. Later, \citet{Gandhi2009} used sub-arcsecond \textit{VLT/VISIR} imaging to isolate the nuclear component from the host galaxy, confirming the relation with reduced scatter. \citet{Asmus2015} extended this approach to a larger, high-resolution sample, reinforcing the need to minimize host contamination.

At higher luminosities and redshifts, wide-field X-ray surveys combined with \textit{Spitzer} and \textit{WISE} photometry enabled broader statistical analyses. \citet{Fiore2009} and \citet{Lanzuisi2009} explored the relation for X-ray- and mid-IR-selected AGNs, noting that host contamination and selection biases can shift the normalization. \citet{mateos2015} presented one of the most comprehensive calibrations for Type~1 AGNs, finding a nearly linear slope ($a\approx0.94$) over $42\lesssim \log L_X \lesssim 46$. \citet{Stern2015} extended this relation to quasars up to $\log L_X\sim46$, providing a polynomial fit to account for mild luminosity-dependent curvature. Overall, these studies agree that unobscured AGNs follow a near-linear $L_{\mathrm{6\,\mu m}}$–$L_{\mathrm{X}}$ relation. While obscured and Compton-thick sources can exhibit lower observed X-ray luminosities at fixed mid-IR power because of line-of-sight absorption, studies using absorption-corrected X-ray luminosities show that obscured and unobscured AGNs broadly follow the same global \(L_{\mathrm{6\,\mu m}}\)–\(L_{\mathrm{X}}\) relation, although modest differences associated with MIR anisotropy and heavy obscuration may remain.

Figure~\ref{fig:LxvsL6} (a plot of the $2-10\,\mathrm{keV}$ X-ray luminosity ($L_{X}$) and the $6\,\mu\mathrm{m}$ luminosity ($L_{6\,\mu\mathrm{m}}$)) compares our Type~1 AGN sample with the relations found in the literature. Open circles denote sources with JWST/NIRCam detections, open squares indicate those without, and red-filled symbols mark starburst broad-line AGNs. Literature fits are shown as blue \citep{mateos2015}, green \citep{Lutz2004}, and magenta \citep{Gandhi2009} lines. Our best-fit relations were derived using Orthogonal Distance Regression (ODR), implemented via the \texttt{scipy.odr} module from the \texttt{SciPy} Python package. This method accounts for measurement uncertainties in both axes; the resulting relation, based on absorption-corrected X-ray luminosities, is shown in black:}
\begin{equation}
\label{eq3}
\log L_{6\,\mu{\rm m}} = (1.02\pm0.08)\,(\log L_X - 44) + 44.16,\quad R=0.68.
\end{equation}

The measured slope ($1.02 \pm 0.08$) is consistent with unity ($a \approx 1.0$) and agrees well within error bars with the values reported in literature studies \citep[e.g.,][]{Lutz2004, Gandhi2009, mateos2015}. The correlation is robust i.e., $R=0.68$ (see Equation~\ref{eq3}). This tight connection strongly suggests that the mid-IR emission is primarily driven by the central engine's bolometric power via the thermal reprocessing of X-ray/UV photons by the circumnuclear dusty torus. Moreover, the nearly linear slope indicates that the $L_{6,\mu{\rm m}}/L_{\rm X}$ ratio remains largely constant across nearly three orders of magnitude in luminosity, suggesting little luminosity-dependent variation in the relative strength of the torus emission within our Type-1 AGN population. Interestingly, the subsample without JWST coverage yields a slightly shallower slope ($a=0.93\pm0.11$) than the JWST-detected subsample ($a=1.05\pm0.10$), although the two slopes are statistically consistent. The slope closer to unity for the JWST subsample is consistent with reduced host-galaxy contamination enabled by NIRCam's higher spatial resolution, supporting an approximately linear intrinsic $L_{\rm X}$--$L_{6,\mu{\rm m}}$ relation.

\begin{figure*}[ht!]
\centering
\includegraphics[width=0.98\textwidth]{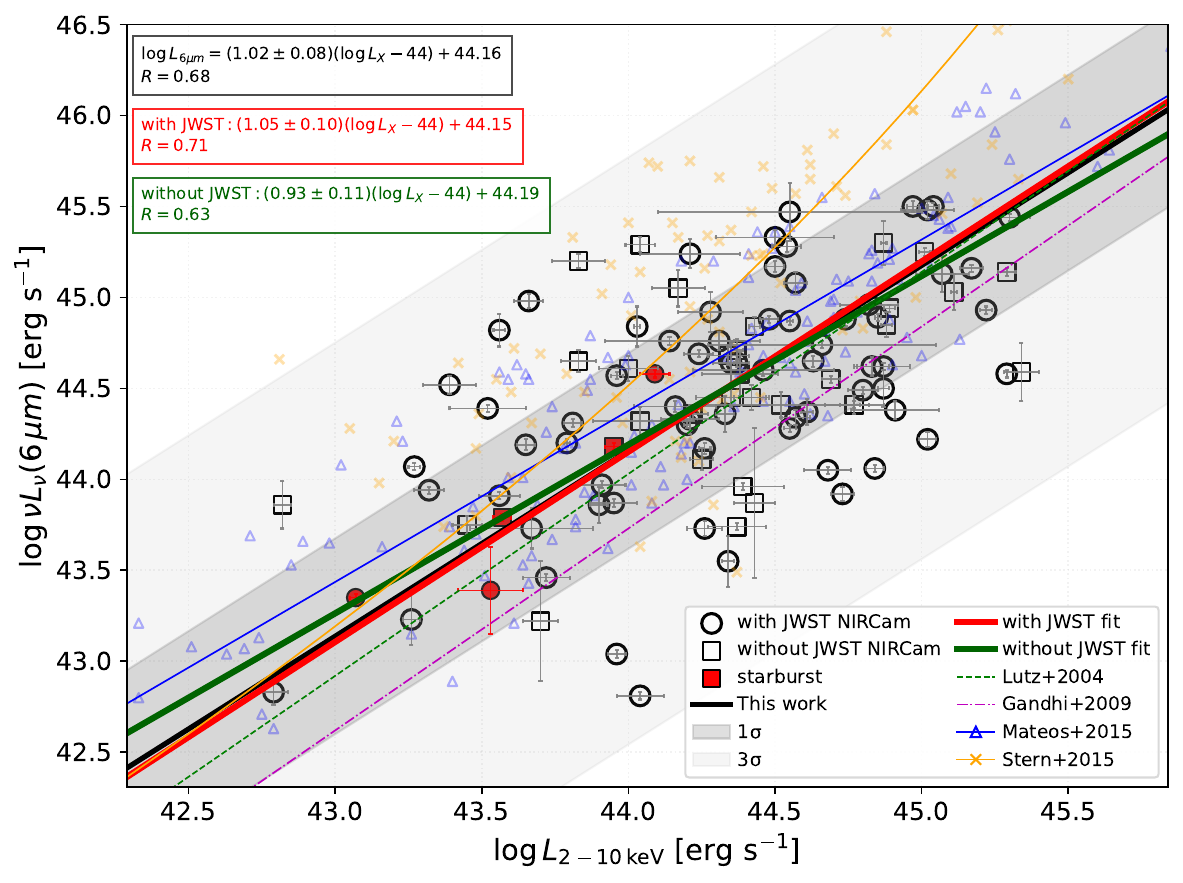}
\caption{Relation between the 2--10 keV X-ray luminosity and the 6 $\mu$m infrared luminosity. Circles and squares indicate sources with and without JWST/NIRCam detections, respectively; red-filled symbols mark starburst broad-line AGNs. The solid black curve shows our best-fit relation for the entire sample, with the 1$\sigma$ (dark gray) and 3$\sigma$ (light gray) confidence bands. The separate fits for the sub-samples with and without JWST detections are represented by the red and green solid lines, respectively. Blue triangles (with blue solid fit line) represent the Mateos et al.\ (2015) sample, while orange cross markers (with orange solid fit line) show the Stern et al.\ (2015) sample. Additional literature relations from Lutz et al.\ (2004) and Gandhi et al.\ (2009) are shown as green dashed and magenta dash--dot lines.}
\label{fig:LxvsL6}
\end{figure*}

\subsection{Deriving the Rest-Frame 2 keV Monochromatic Luminosity} \label{sec:2keVMonochLum}

The rest-frame monochromatic luminosities at 2 $\mathrm{keV}$ were obtained by combining the intrinsic
2--10\,keV luminosities with the photon indices measured from our XMM-Newton spectral modeling in \texttt{XSPEC}. These $L_{2-10\,\mathrm{keV}}$ values are defined in the source's rest-frame, which intrinsically accounts for cosmological bandpass and $(1+z)$ energy-shift corrections.
Each source was modeled with an absorbed power-law continuum, and the best-fit photon index $\Gamma$ was used to reconstruct the intrinsic spectral shape. Following the standard approach  adopted in AGN X-ray/UV studies \citep[e.g.][]{RisalitiLusso2015, Wolf2021}, and assuming a single power-law spectrum of the form $L_{\nu} \propto \nu^{\,1-\Gamma}$, the monochromatic  luminosity at 2 $\mathrm{keV}$ was computed analytically as
\begin{equation}
L_{2\,\mathrm{keV}} =
\frac{L_{2-10\,\mathrm{keV}}}
{\displaystyle \int_{\nu_{2\,\mathrm{keV}}}^{\nu_{10\,\mathrm{keV}}}
\nu^{\,1-\Gamma}\, d\nu}
\, \nu_{2\,\mathrm{keV}}^{\,1-\Gamma}.
\end{equation}

The uncertainties in $L_{2-10}$ and $\Gamma$ were propagated through Monte-Carlo sampling, yielding statistically robust confidence intervals for the derived $2~\mathrm{keV}$ luminosities. \\

\subsubsection{Total AGN} \label{sec:totalAGN}
\begin{figure*}[ht!]
\centering
\includegraphics[width=0.99\textwidth]{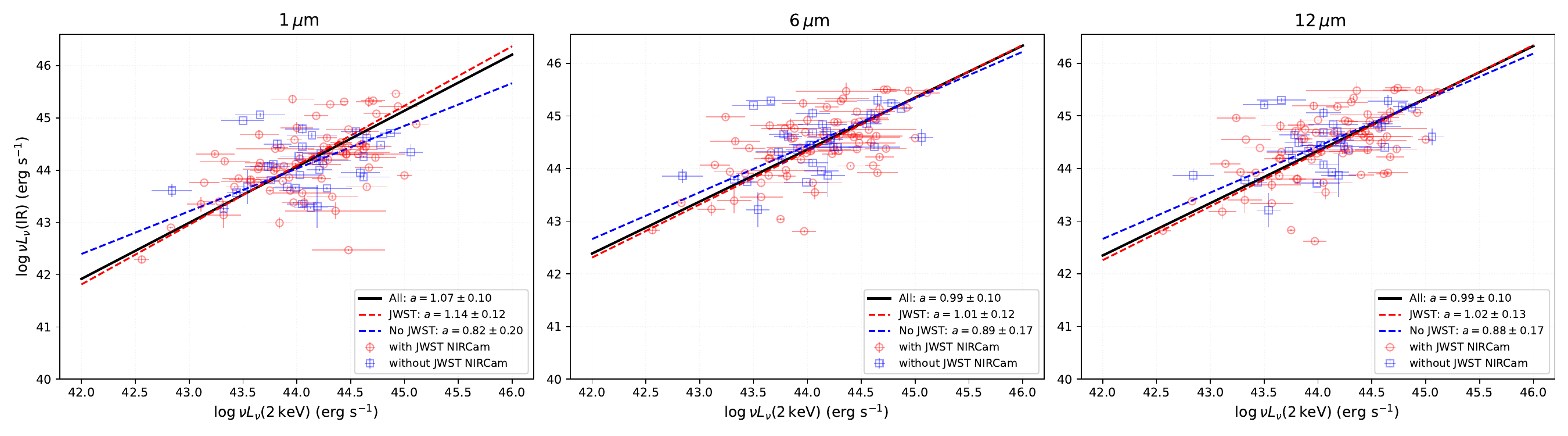}
\includegraphics[width=0.99\textwidth]{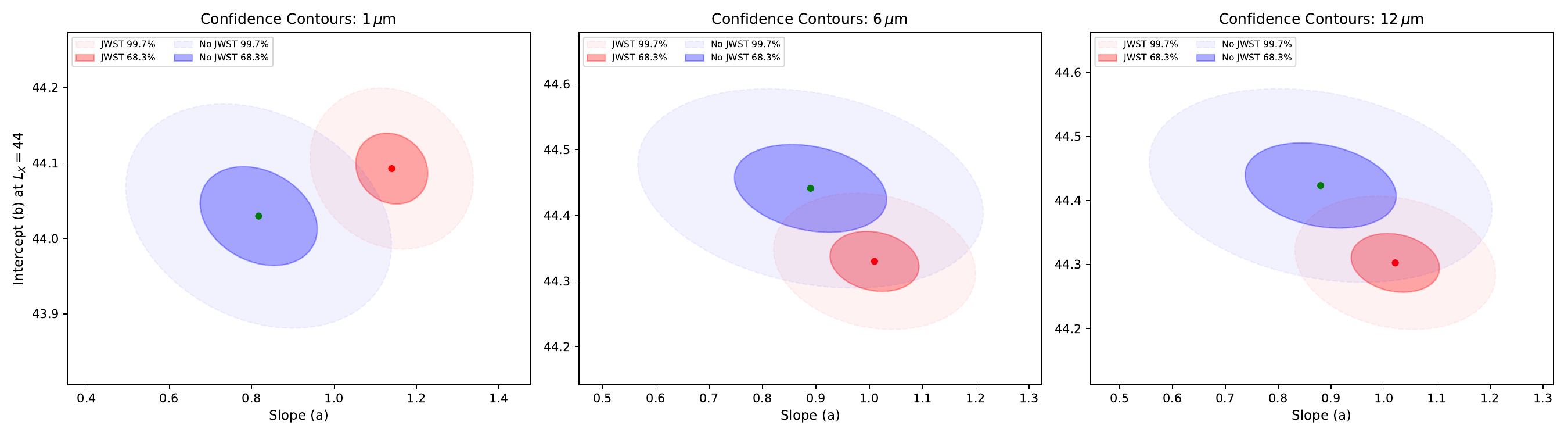}
\caption{$\log \nu L_{\nu}(2\,\mathrm{keV})$ and the AGN infrared luminosities at
$1\,\mu$m (left), $6\,\mu$m (middle), and $12\,\mu$m (right).
Sources with JWST/NIRCam coverage are shown as open circles, while sources without JWST data are shown as open squares. Solid lines indicate the best-fit relations for the full sample (black), the JWST-detected subsample (red), and the subsample without JWST data (blue). Bottom panels: Confidence contours in the slope--intercept ($a$--$b$) parameter space for the corresponding relations shown above, derived from the fits. The inner and outer contours represent the $68.3\%$ ($1\sigma$) and $99.7\%$ ($3\sigma$) confidence levels for the JWST-detected subsample (red) and the subsample without JWST data (blue). While the $68.3\%$ ($1\sigma$) confidence contours remain separated across all three wavelengths, the distinction is strongest at $1\,\mu\mathrm{m}$. At $6\,\mu\mathrm{m}$ and $12\,\mu\mathrm{m}$, the $1\sigma$ contours lie significantly closer to one another and their $3\sigma$ regions overlap extensively, reflecting increased consistency in scaling relations at longer wavelengths.}
\label{fig:Lx2vsLumIR}
\end{figure*}

We investigate the scaling relations between the monochromatic X-ray luminosity, $\log \nu L_{\nu}(2~\mathrm{keV})$, and the infrared luminosities at $1~\mu\mathrm{m}$, $6~\mu\mathrm{m}$, and $12~\mu\mathrm{m}$. To quantify the dependence of the infrared emission on the intrinsic AGN power, we model the relations in the form
\begin{equation}
\log \nu L_{\nu,\mathrm{IR}} = a \cdot  \log \nu L_{\nu}(2~\mathrm{keV}) + b ,
\end{equation}
where the best-fit parameters are derived via the ODR method to account for uncertainties in both variables. The best-fit parameters for the full sample and for the two sub-samples—sources with JWST detections and those without—are summarized in Figure~\ref{fig:Lx2vsLumIR}, which shows a plot of the monochromatic X-ray luminosity, $\log \nu L_{\nu}(2~\mathrm{keV})$, and the infrared luminosities at $1~\mu\mathrm{m}$, $6~\mu\mathrm{m}$, and $12~\mu\mathrm{m}$ respectively.

At 1 $\mu$m, the JWST-detected sources show a tendency toward a steeper relation ($a=1.14\pm0.12$, $R=0.66$) than sources without JWST coverage ($a=0.82\pm0.20$, $R=0.40$), with limited overlap of the 1$\sigma$ confidence contours. The weaker correlation in the non-JWST subsample is consistent with greater sensitivity to host-galaxy contamination and host--AGN decomposition uncertainties when high-spatial-resolution, small-aperture IR photometry is unavailable to constrain the stellar continuum. Rather than being driven by statistical anomalies, the steeper intrinsic scaling in the JWST sample is robustly uncovered thanks to the unprecedented depth and high spatial resolution of JWST. By effectively isolating the nuclear emission from host-galaxy contamination, JWST reveals a more reliable intrinsic dependence on the X-ray luminosity down to the sample limits.

At mid-infrared wavelengths (6 and 12 $\mu$m), however, the scaling relations derived for the two subsamples are much more consistent with each other. At $6\,\mu\mathrm{m}$, the full sample yields a near-unity slope of $a = 0.99 \pm 0.10$ with a strong correlation ($R = 0.68$). The JWST-detected and non-JWST sources produce statistically consistent slopes of $a = 1.01\pm0.12$ ($R = 0.70$) and $a = 0.89 \pm 0.17$ ($R = 0.61$), respectively. Although their $1\sigma$ ($68.3\%$) confidence contours remain distinct due to subtle shifts in slope and intercept, they draw substantially closer compared to $1\,\mu\mathrm{m}$ and show extensive overlap at the $3\sigma$ ($99.7\%$) level, confirming statistical consistency across the two sub-samples. A virtually identical behavior is maintained at $12\,\mu\mathrm{m}$, where the full sample shows a slope of $a = 1.00 \pm 0.10$ ($R = 0.66$), while the individual sub-samples remain fully consistent within $1\sigma$ uncertainties ($a = 1.02 \pm 0.13$, $R = 0.68$ for JWST; $a = 0.88 \pm 0.17$, $R = 0.60$ for non-JWST). These near-linear slopes ($a\simeq1.0$) at 6 and 12 $\mu$m indicate that the torus-dominated mid-infrared emission scales approximately proportionally with the intrinsic X-ray luminosity, with substantially less sensitivity to host-galaxy contamination than at 1 $\mu$m.

\subsubsection{Disk + Torus} \label{sec:disk_torus}
\begin{figure*}[ht!]
\centering
\includegraphics[width=0.99\textwidth]{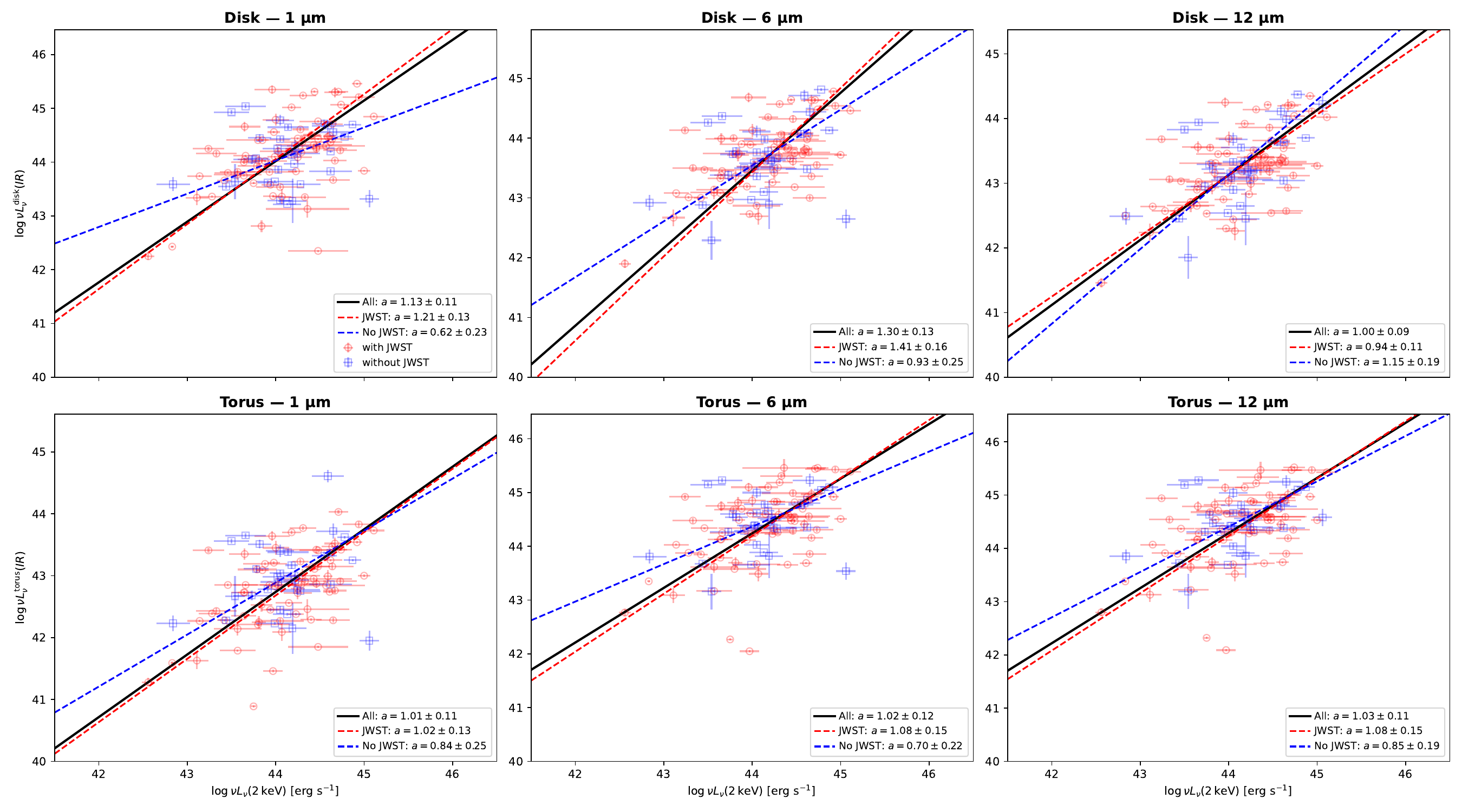}
\caption{Scaling relations between the monochromatic X-ray luminosity, $\nu L_{\nu}(2,\mathrm{keV})$, and the infrared luminosities at 1, 6, and 12~$\mu$m for the disk (upper panels) and torus (lower panels) components derived from SED decomposition. Open circles indicate sources with JWST NIRCam coverage, while open squares denote sources without JWST data. Solid lines represent the best-fit relations for the full sample (black), whereas dashed lines correspond to the JWST-detected sub-sample (red) and the non-JWST sub-sample (blue). The fitted slopes are displayed in the lower right of each panel.}
\label{fig:Lx2vsLumIR_disk_torus}
\end{figure*}

From the SED fits, we can isolate the contributions from the disk and the torus, respectively, and investigate their relation, at 1, 6, and 12~$\mu$m, with the 2~$\mathrm{keV}$ monochromatic luminosity ($\nu L_{\nu}(2~\mathrm{keV})$).  We display these separate contributions in the six panels in Figure~\ref{fig:Lx2vsLumIR_disk_torus} (disk: 3 upper panels and the torus: 3 lower panels).

We perform analyses separately for sources with JWST NIRCam coverage and for those without JWST data. The best-fit parameters derived using ODR are reported in the panels of Figure~\ref{fig:Lx2vsLumIR_disk_torus}. For the accretion disk component in the JWST-detected sample, we obtain slopes of $a = 1.21 \pm 0.13$, $1.41 \pm 0.16$, and $0.94 \pm 0.11$ at 1, 6, and 12~$\mu$m, respectively. The corresponding torus relations in the JWST sample yield nearly linear slopes across all bands: $a = 1.02 \pm 0.13$, $1.08 \pm 0.15$, and $1.08 \pm 0.15$ at 1, 6, and 12~$\mu$m, respectively.

To further validate our results, we performed a non-parametric Spearman rank correlation analysis across all components and wavelengths. We find highly significant intrinsic relationships ($p \ll 10^{-6}$) in all cases. The accretion disk component yields Spearman coefficients of $\rho = 0.46$ at $1\,\mu\text{m}$, $\rho = 0.54$ at $6\,\mu\text{m}$, and $\rho = 0.59$ at $12\,\mu\text{m}$. For the torus component, we obtain $\rho = 0.50$ at $1\,\mu\text{m}$, $\rho = 0.50$ at $6\,\mu\text{m}$, and $\rho = 0.52$ at $12\,\mu\text{m}$. Furthermore, the Spearman correlation strength increases systematically toward the mid-infrared, indicating that the longer-wavelength AGN emission is more tightly coupled to the intrinsic X-ray luminosity. This trend is consistent with reduced sensitivity to host-galaxy starlight contamination at longer wavelengths. Together, these results demonstrate robust IR--X-ray correlations that are consistent with the expected connection between the accretion-powered central engine and circumnuclear dust emission. However, it is critical to emphasize that while \texttt{CIGALE} mathematically decouples the accretion disk and torus components across all wavelengths, the intrinsic emission from the AGN accretion disk is physically confined to the rest-frame UV and optical regimes. Beyond $\sim 1\,\mu\text{m}$, direct emission from the hot accretion disk drops sharply, and the total AGN output becomes heavily dominated by dust reprocessing within the surrounding torus structure. Therefore, the disk emission inferred by the models at 6 and $12\,\mu\text{m}$ does not represent direct observational constraints in the infrared but rather reflects the mathematical extrapolation of the best-fitting UV/optical templates. Consequently, these mid-infrared disk contributions should be interpreted strictly as model-dependent extrapolations rather than independent detections, and caution should be exercised when assigning independent physical significance to the disk component at these longer wavelengths.

The largest discrepancy between the JWST-detected and non-JWST samples is observed for the disk component at $1\,\mu\text{m}$. At these wavelengths, the JWST-detected sources exhibit a steeper relation ($a=1.21\pm0.13$), while the sample without JWST data shows a substantially flatter slope ($a=0.62\pm0.23$). At longer wavelengths (6 and 12 $\mu$m), the disk relations for the two subsamples become consistent within uncertainties.

For the torus component in the JWST-detected sample, the scaling relations show remarkably consistent slopes close to unity ($a\simeq1.02$--$1.08$) across the mid-infrared bands. This near-linear scaling indicates that the mid-infrared torus emission scales approximately proportionally with the intrinsic power of the central engine, with little evidence for a strong luminosity dependence over the range probed here. The similarity of the slopes across bands is also consistent with relatively modest luminosity-dependent changes in the shape of the torus MIR SED.

At mid-infrared wavelengths ($6$ and $12,\mu\mathrm{m}$), the disk component shows a slightly stronger correlation with $\nu L_{\nu}(2~\mathrm{keV})$ than the torus component, while at $1,\mu\mathrm{m}$ the two components exhibit comparable correlation strengths. Among the infrared quantities considered, the total AGN infrared luminosity---combining the disk, torus, and polar-dust components---shows the strongest overall correlation with $\nu L_{\nu}(2~\mathrm{keV})$, indicating that the integrated AGN infrared emission provides the most robust tracer of intrinsic AGN power in our sample.

\section{Discussion} \label{sec:Discussion}

Our results show that the infrared emission of Type~1 AGNs follows a close-to-linear relation with the intrinsic X-ray luminosity when the AGN and host-galaxy contributions are separated through SED decomposition. The dependence on wavelength is particularly informative. At $1~\mu\mathrm{m}$, the relation between $\nu L_{\nu}(2~\mathrm{keV})$ and the AGN infrared luminosity is steeper for sources with JWST/NIRCam coverage than for those without it, with slopes of $1.14\pm0.12$ and $0.82\pm0.20$, respectively. This difference indicates that the near-infrared relation remains sensitive to host-galaxy contamination. Stellar emission from the host can increase the measured infrared luminosity, particularly for lower-luminosity AGNs, and thereby flatten the observed relation. The availability of spatially resolved JWST data provides an important constraint on this contribution and improves the separation of the nuclear emission from the host galaxy.

At $6~\mu\mathrm{m}$ and $12~\mu\mathrm{m}$, the scaling relations are substantially more uniform. For the full sample, we find slopes of $0.99\pm0.10$ at both wavelengths. The corresponding slopes for the JWST and non-JWST subsamples are $1.01\pm0.12$ and $0.89\pm0.17$ at $6~\mu\mathrm{m}$, and $1.02\pm0.13$ and $0.88\pm0.17$ at $12~\mu\mathrm{m}$, respectively. These relations are consistent with a linear scaling within the uncertainties. The smaller dependence on JWST coverage at MIR wavelengths indicates that the nuclear AGN contribution is more reliably recovered in this regime, where the relative contribution of stellar host emission is reduced. The resulting near-linear relation is consistent with the established connection between intrinsic X-ray emission from the corona and hot-dust emission from the circumnuclear environment \citep[e.g.,][]{Gandhi2009, mateos2015}.

The choice of X-ray luminosity tracer also provides a consistent picture. Using the integrated $2\text{--}10~\mathrm{keV}$ luminosity, we obtain a relation with the $6~\mu\mathrm{m}$ AGN luminosity of $L_{6\mu\mathrm{m}} \propto L_{\mathrm{X}}^{1.02\pm0.08}$. The JWST and non-JWST subsamples give slopes of $1.05\pm0.10$ and $0.93\pm0.11$, respectively. Thus, the X-ray--MIR relation is consistent with proportional scaling over the luminosity range covered by our sample. The use of $\nu L_{\nu}(2~\mathrm{keV})$ provides a complementary monochromatic tracer of the coronal emission and yields the same overall behavior. The agreement between the broadband and monochromatic X-ray relations indicates that the observed near-linear IR--X-ray connection is not driven by the specific choice of X-ray luminosity definition.

The component-resolved relations provide further constraints on the origin of the infrared emission. The disk component shows slopes of $1.13\pm0.11$, $1.30\pm0.13$, and $1.00\pm0.09$ at $1$, $6$, and $12~\mu\mathrm{m}$, respectively. At $1~\mu\mathrm{m}$, the JWST and non-JWST subsamples have slopes of $1.21\pm0.13$ and $0.62\pm0.23$, while at $6~\mu\mathrm{m}$ the corresponding slopes are $1.41\pm0.16$ and $0.93\pm0.25$. These differences indicate that the inferred disk contribution is sensitive to the quality of the nuclear infrared constraints, particularly at wavelengths where host emission can contribute substantially. At $12~\mu\mathrm{m}$, the disk relations are consistent with a linear scaling, with slopes of $0.94\pm0.11$ and $1.15\pm0.19$ for the JWST and non-JWST subsamples, respectively.

The interpretation of the disk component at wavelengths beyond $\sim1~\mu\mathrm{m}$ requires particular care. The physical emission from the accretion disk is expected to be concentrated primarily at UV and optical wavelengths, whereas the infrared emission is increasingly dominated by dust reprocessing. The disk luminosities inferred by the SED decomposition at $6$ and $12~\mu\mathrm{m}$ therefore represent the contribution of the adopted disk template to the fitted SED rather than a direct measurement of thermal emission from the accretion disk at these wavelengths. The correlations obtained for these components are consequently useful for characterizing the behavior of the SED model, but they should not be interpreted as independent evidence for substantial direct disk emission in the MIR.

The torus component exhibits a more uniform X-ray--infrared scaling across the wavelength range considered. The full-sample slopes are $1.01\pm0.11$, $1.02\pm0.12$, and $1.03\pm0.11$ at $1$, $6$, and $12~\mu\mathrm{m}$, respectively. At $1~\mu\mathrm{m}$, the JWST and non-JWST subsamples give slopes of $1.02\pm0.13$ and $0.84\pm0.25$. At $6~\mu\mathrm{m}$, the corresponding slopes are $1.08\pm0.15$ and $0.70\pm0.22$, while at $12~\mu\mathrm{m}$ they are $1.08\pm0.15$ and $0.85\pm0.19$. The full-sample relations are consistent with a linear scaling at all three wavelengths, while the larger uncertainties in the non-JWST subsample limit the significance of the differences between the two groups. These results support the interpretation that the torus emission provides a direct infrared tracer of the power of the central engine through reprocessing by circumnuclear dust.

The approximately linear torus--X-ray relation has a natural physical interpretation in terms of radiative reprocessing. The coronal X-ray emission traces the energetic output of the central accretion flow, while the dusty torus absorbs and re-emits a fraction of the accretion-powered radiation in the infrared. The resulting infrared luminosity therefore depends on both the incident AGN radiation field and the efficiency with which the surrounding dust intercepts and reprocesses it. Variations in the dust covering factor, geometry, and characteristic reprocessing radius can introduce scatter in the relation without necessarily producing a strong deviation from linearity. The near-unity slopes measured here indicate that, over the luminosity range sampled, these effects do not dominate the overall X-ray--MIR scaling. This behavior is consistent with the empirical X-ray--MIR relations established from high-angular-resolution observations of AGNs \citep[e.g.,][]{Gandhi2009, mateos2015}.

The behavior of the total AGN emission and its individual components also demonstrates the importance of treating the near- and mid-infrared regimes separately. The strongest dependence on JWST coverage occurs at $1~\mu\mathrm{m}$, where stellar emission from the host can substantially affect the inferred nuclear luminosity. At $6$ and $12~\mu\mathrm{m}$, the total AGN relations remain close to linear and show much smaller differences between the JWST and non-JWST samples. This wavelength dependence indicates that the principal observational limitation in recovering the intrinsic AGN scaling relation is host-galaxy dilution at short infrared wavelengths rather than a fundamental change in the underlying X-ray--infrared connection.

The sources located near the lower X-ray luminosity boundary also require consideration when interpreting the component-specific relations. Objects at the sensitivity limit have relatively large leverage on the fitted slopes and can affect the inferred behavior of the low-luminosity end of the correlations. In particular, the presence of intrinsically X-ray-weak sources can increase the dispersion relative to the main population. For XID~49 and XID~236, the X-ray spectra have photon indices of $\Gamma\approx1.7\text{--}2.0$ and do not show evidence for significant intrinsic absorption. Their low X-ray luminosities therefore cannot be attributed to strong X-ray obscuration and are consistent with intrinsically weak coronal emission. Such objects represent a small fraction of the sample and do not alter the overall near-linear X-ray--infrared relations, but they demonstrate that intrinsic dispersion in the coronal output is present within the Type~1 AGN population, echoing the properties of intrinsically X-ray--weak quasars identified in large surveys \citep[e.g.,][]{Pu2020}.

At the lowest luminosities, the SED decomposition can also constrain the AGN contribution close to the lower boundary of the model parameter space. When the broadband photometry is dominated by host-galaxy emission, the posterior distribution can place the AGN contribution at or close to the minimum allowed value. A formally small uncertainty in such cases reflects the truncation of the posterior at the model boundary rather than a precise measurement of an intrinsically faint nuclear component. Consequently, component luminosities close to the lower model boundary should be interpreted as upper limits or boundary-constrained estimates rather than as precise measurements of the nuclear emission. This consideration is particularly important when interpreting individual objects at the faint end of the sample and when assessing their influence on component-specific scaling relations.

Overall, our results show that the intrinsic infrared emission of Type~1 AGNs follows an approximately linear relation with X-ray luminosity across the $1\text{--}12~\mu\mathrm{m}$ range when the host-galaxy contribution is explicitly modeled. JWST/NIRCam data are particularly important at $1~\mu\mathrm{m}$, where host-galaxy dilution can significantly modify the inferred scaling relation. At $6$ and $12~\mu\mathrm{m}$, the total AGN and torus luminosities provide stable near-linear tracers of the central-engine power. The component-resolved analysis further shows that the torus relation is robust across the MIR, whereas the disk contribution at wavelengths beyond $\sim1~\mu\mathrm{m}$ remains model-dependent. These results establish a consistent connection between the X-ray-emitting corona and the infrared-emitting dust in Type~1 AGNs and demonstrate the value of combining high-resolution infrared observations with physically motivated SED decomposition.

\section{Conclusions}
\label{sec:conclusion}

In this work, we have presented a comprehensive, component-resolved analysis of the IR--X-ray connection in a sample of 104 spectroscopically confirmed Type~1 AGNs drawn from the X-ray-selected catalog of \citet{Lusso2010}. By combining \xmm, \swift/UVOT, and high-angular-resolution \jwst/NIRCam imaging for a sub-sample of sources, we constructed detailed SEDs using \texttt{CIGALE}, explicitly decomposing the nuclear emission into accretion-disk, torus, and polar-dust components while removing host-galaxy contamination. This approach enabled a clean evaluation of the physical coupling between the accretion flow, the circumnuclear dust structure, and the X-ray corona.

\vspace{1em}\noindent Our main findings are summarized as follows:

\begin{enumerate}
    \item Robustness Across X-ray Tracers and Proportional Scaling: We demonstrate that the $L_{\mathrm{IR}}\text{--}L_{\mathrm{X}}$ scaling relation is robustly linear ($a \approx 1.0$) across nearly three orders of magnitude in luminosity. Both the integrated $2\text{--}10\,\mathrm{keV}$ broadband luminosity ($a = 1.02 \pm 0.08$) and the $2\,\mathrm{keV}$ monochromatic luminosity ($\nu L_{\nu}(2\,\mathrm{keV})$; $a = 0.99 \pm 0.10$ at $6\,\mu\mathrm{m}$) yield proportional scaling. This excellent agreement indicates that the observed near-linear IR--X-ray connection is independent of the choice of X-ray luminosity definition and points to a stable average torus covering factor across the sampled Type~1 AGN population (see Section~\ref{sec:Xray-midIR}).

    \item Wavelength-Dependent Impact of \jwst\ Resolution: High-spatial-resolution \jwst/NIRCam imaging is crucial for mitigating host-galaxy starlight dilution at shorter IR wavelengths. At $1\,\mu\mathrm{m}$, the \jwst-detected subsample reveals a steeper intrinsic relation ($a = 1.14 \pm 0.12$) compared to the flattened slope of the non-\jwst\ sample ($a = 0.82 \pm 0.20$). At mid-infrared wavelengths ($6\,\mu\mathrm{m}$ and $12\,\mu\mathrm{m}$), host contamination becomes negligible, resulting in highly consistent near-linear relations ($a \approx 1.01\text{--}1.02$) and extensive overlap of confidence contours between the two subsamples (see Section~\ref{sec:totalAGN}).

    \item Component-Resolved Behavior and MIR Disk Interpretation: Decomposing the nuclear SED reveals that the torus component maintains a remarkably uniform, near-linear scaling ($a \approx 1.01\text{--}1.08$) across $1\text{--}12\,\mu\mathrm{m}$, confirming that circumnuclear dust reprocessing directly tracks central-engine power. Conversely, while the derived disk component shows tight correlations, its inferred contribution at $6\,\mu\mathrm{m}$ and $12\,\mu\mathrm{m}$ represents a mathematical extrapolation of the UV/optical templates rather than direct thermal emission from the accretion disk, which drops sharply beyond $\sim1\,\mu\mathrm{m}$ (see Section~\ref{sec:disk_torus}).

    \item Intrinsic Dispersion and Parameter Space Boundaries: Sources located near the lower X-ray sensitivity boundary (specifically XID~49 and XID~66) exhibit unabsorbed, standard photon indices ($\Gamma \approx 1.7\text{--}1.9$, $N_{\mathrm{H}} < 10^{22}~\mathrm{cm^{-2}}$), strongly suggesting that they are intrinsically X-ray--weak Type~1 AGNs rather than heavily obscured systems, echoing findings in large optical surveys \citep[e.g.,][]{Pu2020}. Furthermore, for extremely faint sources dominated by host starlight, SED posterior distributions can truncate at model parameter limits, indicating that boundary-constrained estimates should be interpreted with caution as upper limits.
\end{enumerate}

\noindent The strong consistency observed in the mid-infrared between \jwst-detected sources and the broader sample confirms that the underlying $L_{\mathrm{X}}\text{--}L_{\mathrm{MIR}}$ scaling law is fundamentally linear in Type~1 AGNs. High-resolution infrared photometry remains indispensible at near-IR wavelengths ($\sim 1\,\mu\mathrm{m}$) to untangle nuclear emission from host-galaxy light, whereas mid-infrared wavelengths provide stable, host-free tracers of total accretion power.\\
\\
\vspace{1em}\noindent Overall, our results establish a coherent physical framework in which the infrared output of Type~1 AGNs-when isolated from host contamination—scales directly and proportionally with the energetic output of the central engine. The torus acts as a reliable tracer of coronal X-ray radiation, while detailed component-resolved modeling provides necessary safeguards against template extrapolations in the mid-infrared.

\vspace{1em}\noindent Looking ahead, expanding spatially resolved infrared coverage with \jwst\ across wider redshift and luminosity ranges will refine our understanding of dust geometry, covering factor variations, and the faint-end AGN population. The physical relations established here provide a calibrated baseline for multiwavelength accretion studies in the era of high-precision AGN surveys.

\begin{acknowledgments}

\noindent We thank the anonymous referee for constructive comments and suggestions that significantly improved the clarity and quality of our manuscript. The JWST data presented in this article were obtained from the the Mikulski Archive for Space Telescopes (MAST) at the Space Telescope Science Institute. The specific observations analyzed can be accessed via \dataset[doi:  10.17909/44vg-nm41]{https://doi.org/XYZ}.
This research is supported by the Scientific and Technological Research Council of Turkey (T\"UB\.ITAK) through project number 123F272.

\end{acknowledgments}

\vspace{5mm}
\facilities{HST(STIS), Swift(XRT and UVOT), AAVSO, CTIO:1.3m,
CTIO:1.5m,CXO}

\software{astropy \citep{Astropy2013,Astropy2018,astropy:2022},  
 Source Extractor \citep{Bertin1996}
          }

\clearpage
\appendix

\section{Analysis of High-Inclination Sources and AGN Fraction}
\label{app:B}
\noindent To assess the robustness of sources characterized by intermediate-to-edge-on inclination angles ($i \ge 50^\circ$), we performed a detailed inspection of their SED-derived parameters. Although these sources are categorized as Type~1 AGNs based on their X-ray and optical properties, their edge-on torus configurations and host-dominated infrared emission motivate a careful evaluation of whether the inferred geometries reflect physical structures or modeling degeneracies. To ensure the statistical reliability of this subsample, we implemented a strict quality filter and completely excluded sources exhibiting poor goodness-of-fit metrics (reduced $\chi^2 > 10$) from our dataset.

Following this filtering, a final clean subsample of 11 high-inclination sources remains ($i \ge 50^\circ$). This population is characterized by distinct structural regimes rather than clear evolutionary phases. As shown in the left panel of Figure~\ref{fig:appendix_B1}, the vast majority of these high-inclination sources (9 out of 11) cluster tightly within the $50^\circ \le i \le 60^\circ$ range. These targets are heavily supported by high-quality JWST data and likely represent intermediate-type configurations where the central engine remains partially visible through a clumpy or patchy obscuring structure. In contrast, sources at the highest inclination boundaries ($i \ge 70^\circ$) with low AGN fractions ($f_{\rm AGN} \le 0.20$), such as XID24 ($i = 70.00^\circ$, $f_{\rm AGN} = 0.15$) and XID6 ($i = 90.00^\circ$, $f_{\rm AGN} = 0.20$), suffer from increased geometric uncertainty due to host-galaxy light dominance. At these extremes, the edge-on orientation of the host stellar or dust disk can blend with and artificially drive the torus model parameters toward higher viewing angles. Particularly for XID6, which is independently classified as a starburst galaxy in the literature, the intense star-formation-driven dust emission from the host disk likely dominates the infrared, leading the SED model to mimic an extreme edge-on torus configuration while yielding a low AGN fraction. The distribution of these inclinations and their relationships with X-ray photon indices ($\Gamma$) and mid-infrared luminosities are shown in Figure~\ref{fig:appendix_B1}.

\renewcommand{\thefigure}{\arabic{figure}}
\restartappendixnumbering
\begin{figure*}[ht!]
\centering
\includegraphics[width=\textwidth]{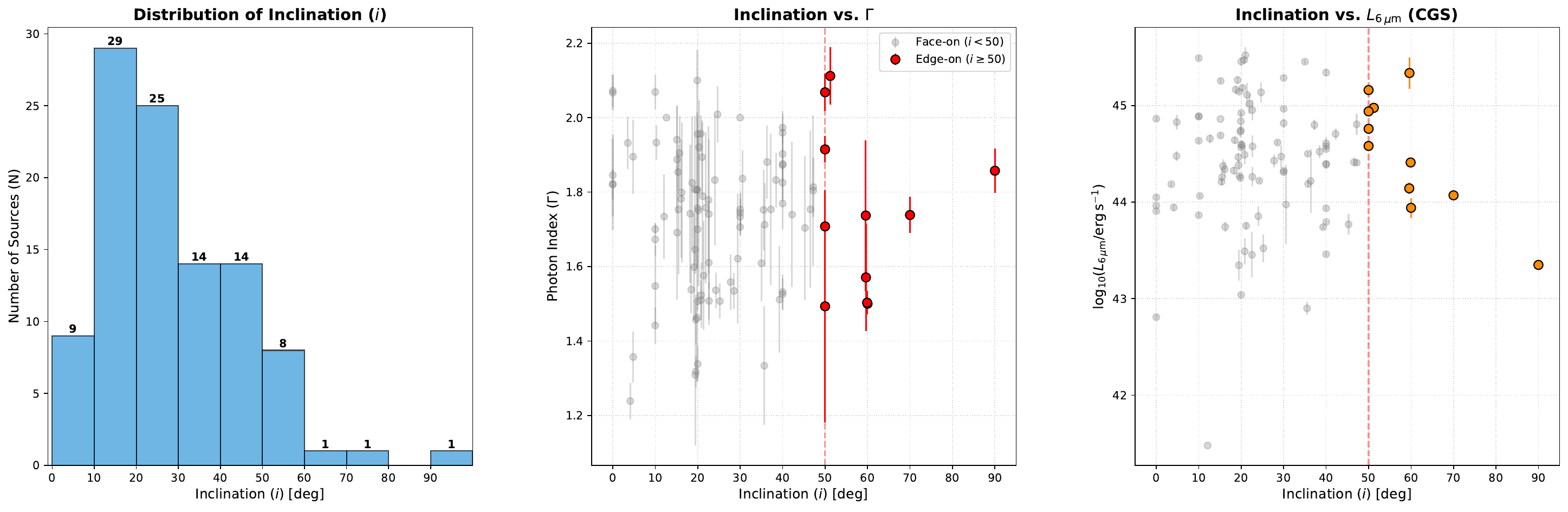}
\caption{Analysis of the intermediate-to-edge-on population ($i \ge 50^\circ$). Left panel: Distribution of inclination angles ($i$) for the sample, highlighting the final clean distribution. Middle panel: Photon index ($\Gamma$) vs. inclination angle, showing no systematic X-ray hardening for edge-on targets. Right panel: Mid-infrared luminosity ($L_{6\,\mu{\rm m}}$) vs. inclination angle, illustrating the isotropic nature of the integrated dust emission. Error bars are included to reflect parameter uncertainties.}
\label{fig:appendix_B1}
\end{figure*}

We find no systematic hardening of the X-ray spectra at high inclinations, with photon indices remaining in the range $\Gamma \approx 1.5$--$2.1$, consistent with unobscured or mildly obscured nuclear emission (middle panel of Figure~\ref{fig:appendix_B1}). This behavior is fully compatible with a clumpy torus scenario, in which X-ray photons can escape through low-density lines of sight even at relatively large viewing angles. Furthermore, as demonstrated in the right panel of Figure~\ref{fig:appendix_B1}, high mid-infrared luminosities ($\log_{10}(L_{6\,\mu{\rm m}} / \mathrm{erg\,s^{-1}}) \sim 43.3$--$45.3$) are sustained across all inclinations, confirming the approximately isotropic nature of the dust emission when integrated over the full SED.

\begin{figure*}[ht!]
\centering
\includegraphics[width=0.8\textwidth]{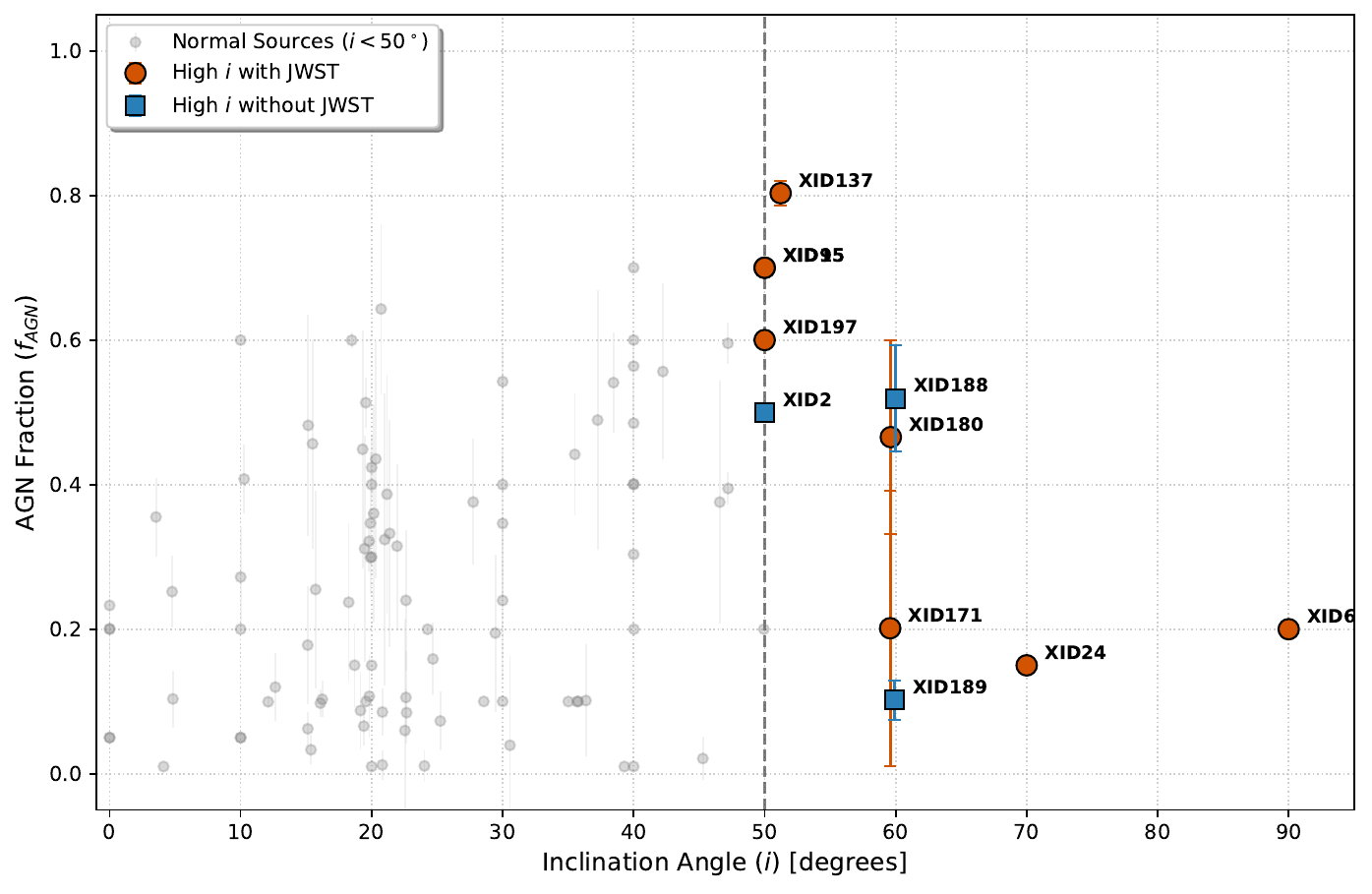}
\caption{AGN Fraction ($f_{\rm AGN}$) as a function of inclination angle ($i$) for the filtered sample. Sources identified with high-quality JWST data are highlighted with orange circles, and those without are marked with blue squares. Vertical error bars represent the formal uncertainties on the inferred quantities. High $f_{\rm AGN}$ values at large angles (e.g., XID137, XID95) validate the reliability of the torus geometry fits against host-galaxy blending.}
\label{fig:appendix_agnfrac}
\end{figure*}

Low AGN fractions ($f_{\rm AGN} \le 0.20$) in these high-inclination systems should therefore be interpreted as signatures of host-galaxy dominance in the optical/IR regime rather than the absence of nuclear activity. This interpretation is illustrated in Figure~\ref{fig:appendix_agnfrac}, which shows the distribution of $f_{\rm AGN}$ as a function of inclination angle for sources with and without JWST coverage. For sources with high AGN fractions ($f_{\rm AGN} \ge 0.50$; e.g., XID137, XID15, XID95, XID197, XID188, XID2), the inferred high inclination is particularly robust because the total SED possesses a strong torus signal to break parameter degeneracies. Conversely, at lower AGN fractions (e.g., XID180, XID171, XID24, XID6, XID189), the geometric parameters carry model-dependent limits and should be interpreted with caution due to the lack of independent spatial constraints.

The physical properties and qualitative interpretations of the clean high-inclination subsample are summarized in Table~\ref{table:appendix_table_sample}. Taken together, the consistency between the SED fits, X-ray properties, and JWST-constrained emission supports a scenario where these sources represent a population observed through a patchy, clumpy dust distribution rather than a homogeneous obscuring medium, with host-galaxy emission dominating the systems exhibiting the lowest AGN fractions.

\begin{table}[ht!]
\centering
\caption{Properties and Interpretations of High-Inclination Sources ($i \ge 50^\circ$)}
\begin{tabular}{lccl}
\hline \hline
Source ID & $i$ [deg] & $f_{\rm AGN}$ & Interpretation / Confidence \\ \hline
XID137 & 51.22 & 0.80 & High Confidence: Strong torus detection with JWST. \\
XID15  & 50.00 & 0.70 & High Confidence: Strong AGN signal supporting high angle. \\
XID95  & 50.00 & 0.70 & High Confidence: Strong torus signature with JWST. \\
XID197 & 50.00 & 0.60 & Reliable fit: Strong torus component with JWST. \\
XID188 & 60.00 & 0.52 & Reliable fit: Moderate AGN contribution (Non-JWST). \\
XID2   & 50.00 & 0.50 & High Confidence: Clear AGN signal (Non-JWST). \\
XID180 & 59.63 & 0.47 & Moderate Confidence: Intermediate AGN strength with JWST. \\
XID171 & 59.58 & 0.20 & Galaxy-dominated: Intermediate angle with low $f_{\rm AGN}$ (JWST). \\
XID6   & 90.00 & 0.20 & Edge-on Host: Angle likely driven by starburst host disk (JWST). \\
XID24  & 70.00 & 0.15 & Host-dominated: High angle carries model uncertainty (JWST). \\
XID189 & 59.90 & 0.10 & Galaxy-dominated: Low $f_{\rm AGN}$, model-dependent limit (Non-JWST). \\ \hline
\end{tabular}
\label{table:appendix_table_sample}
\end{table}

\section{CIGALE Mock Analysis and Parameter Recovery}
\label{app:mock_check}
\begin{figure*}[ht!]
\restartappendixnumbering
\centering
    \begin{minipage}{0.32\textwidth}
        \centering
        \includegraphics[width=\textwidth]{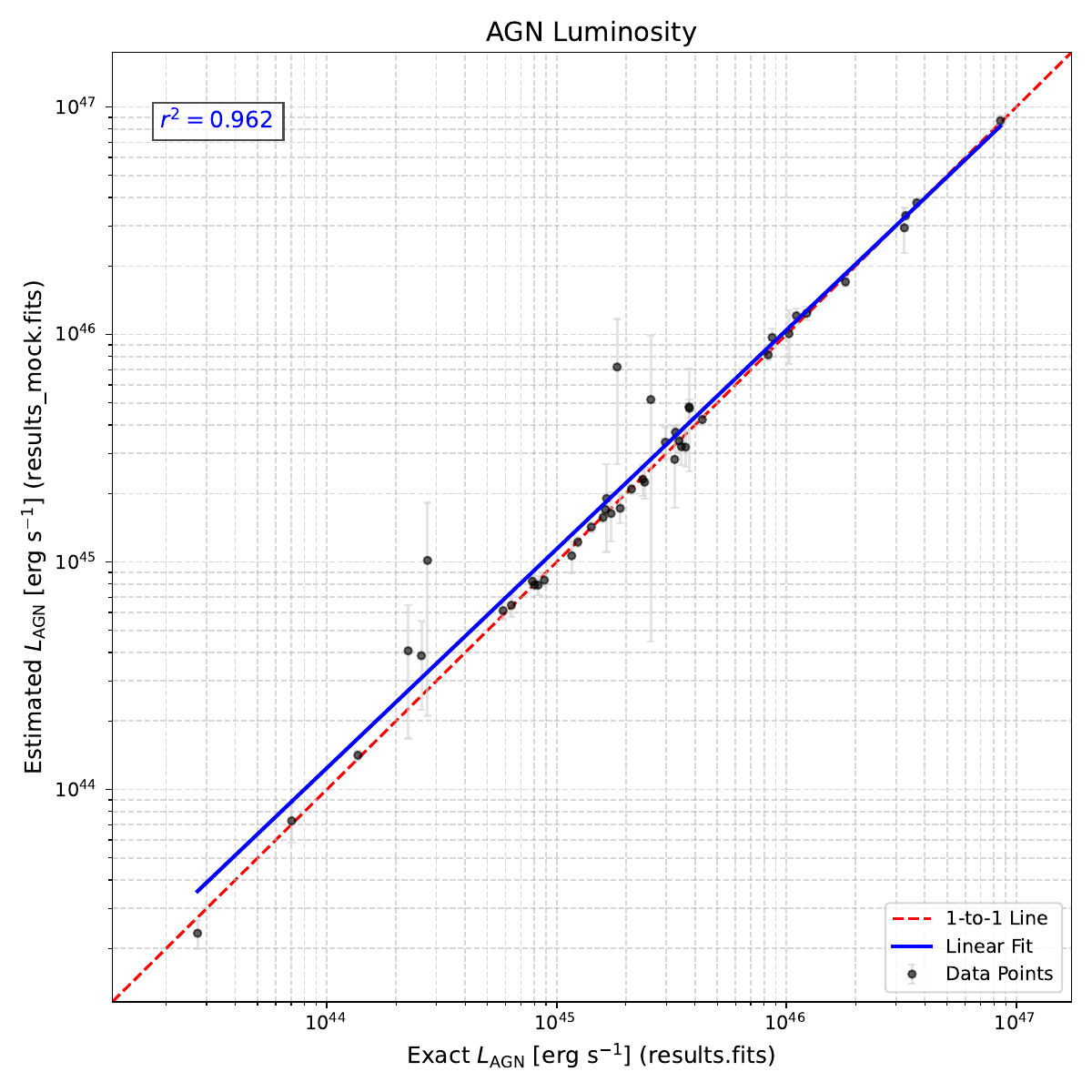}
    \end{minipage}\hfill
    \begin{minipage}{0.32\textwidth}
        \centering
        \includegraphics[width=\textwidth]{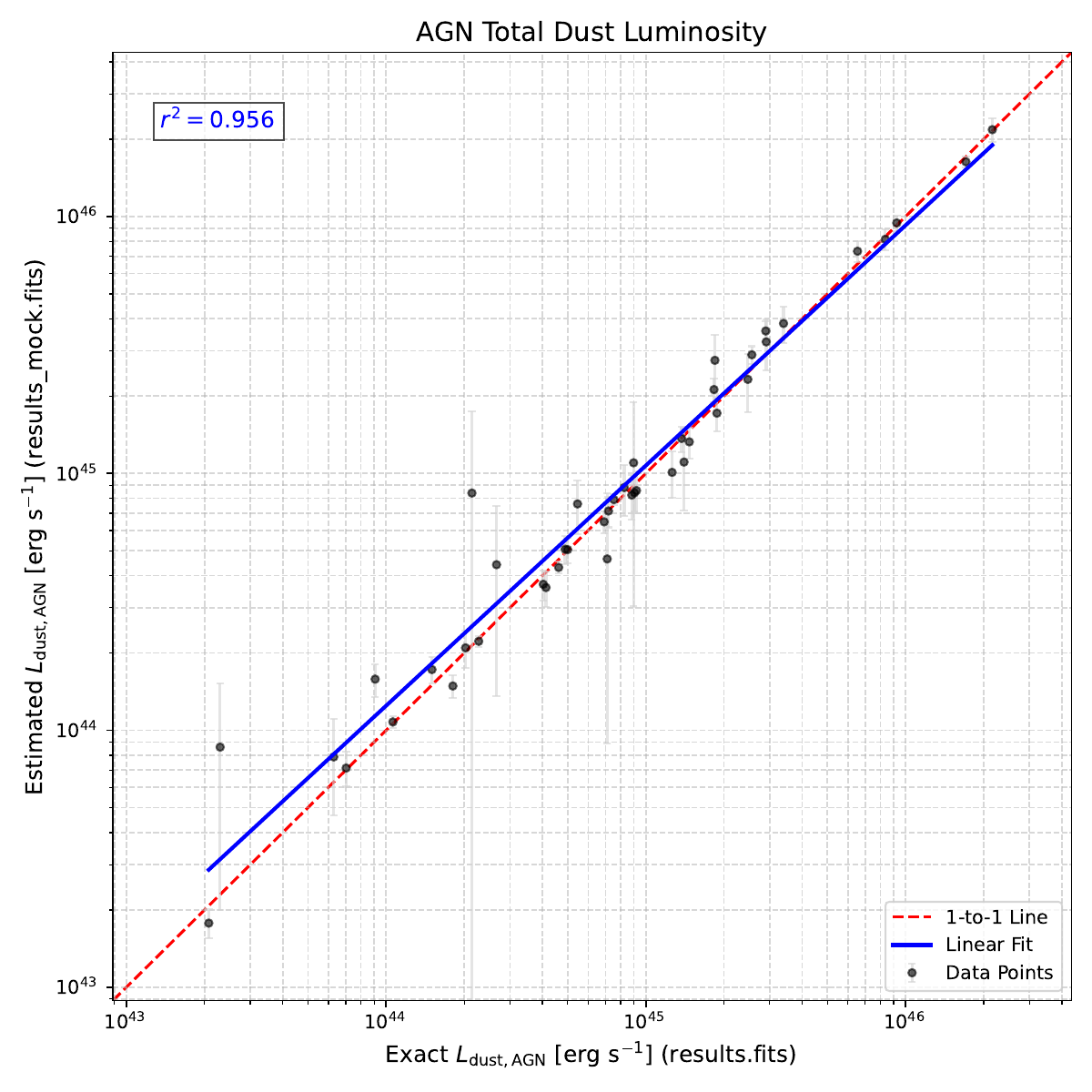}
    \end{minipage}\hfill
    \begin{minipage}{0.32\textwidth}
        \centering
        \includegraphics[width=\textwidth]{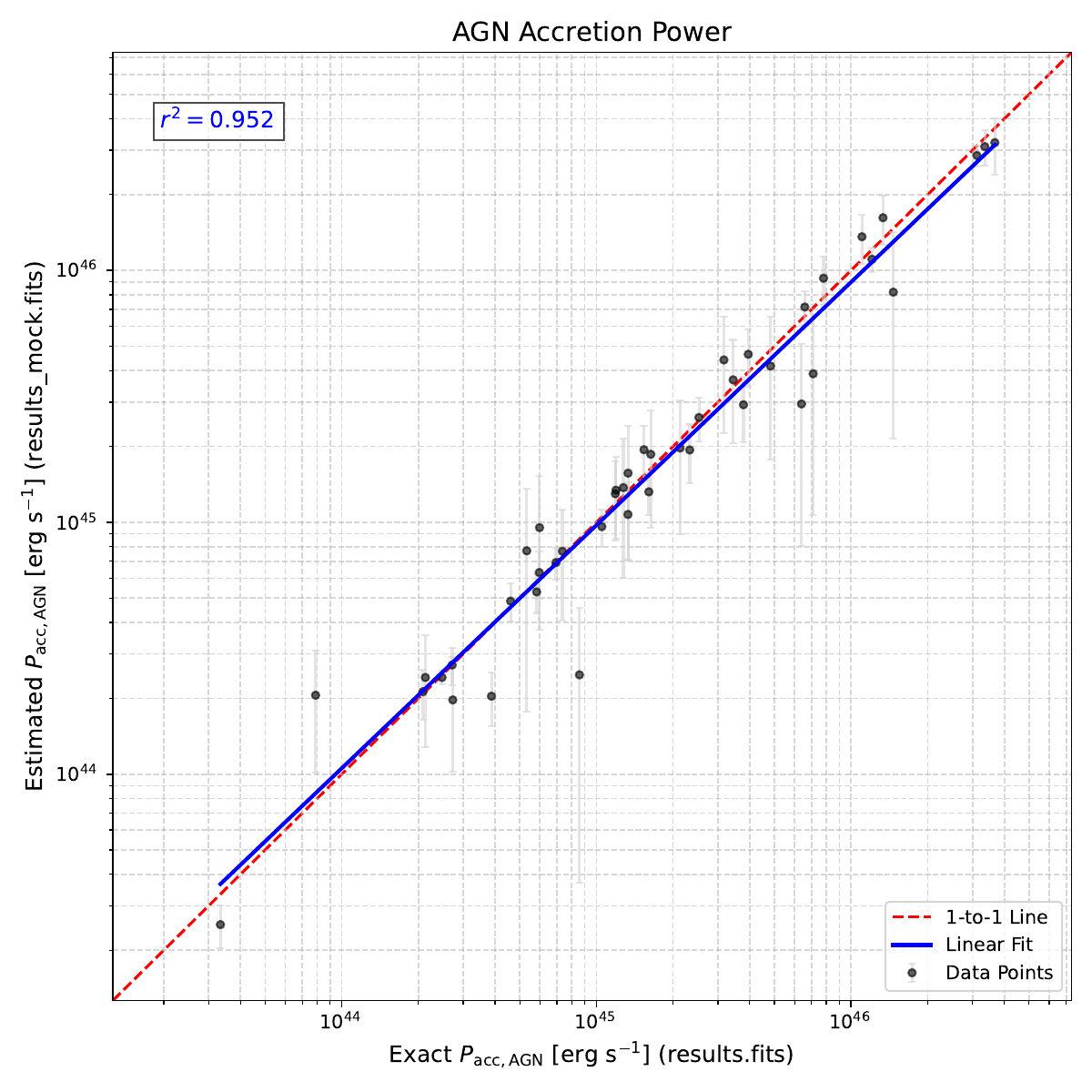}
    \end{minipage}
    
    \vspace{0.3cm} 
    
    \begin{minipage}{0.32\textwidth}
        \centering
        \includegraphics[width=\textwidth]{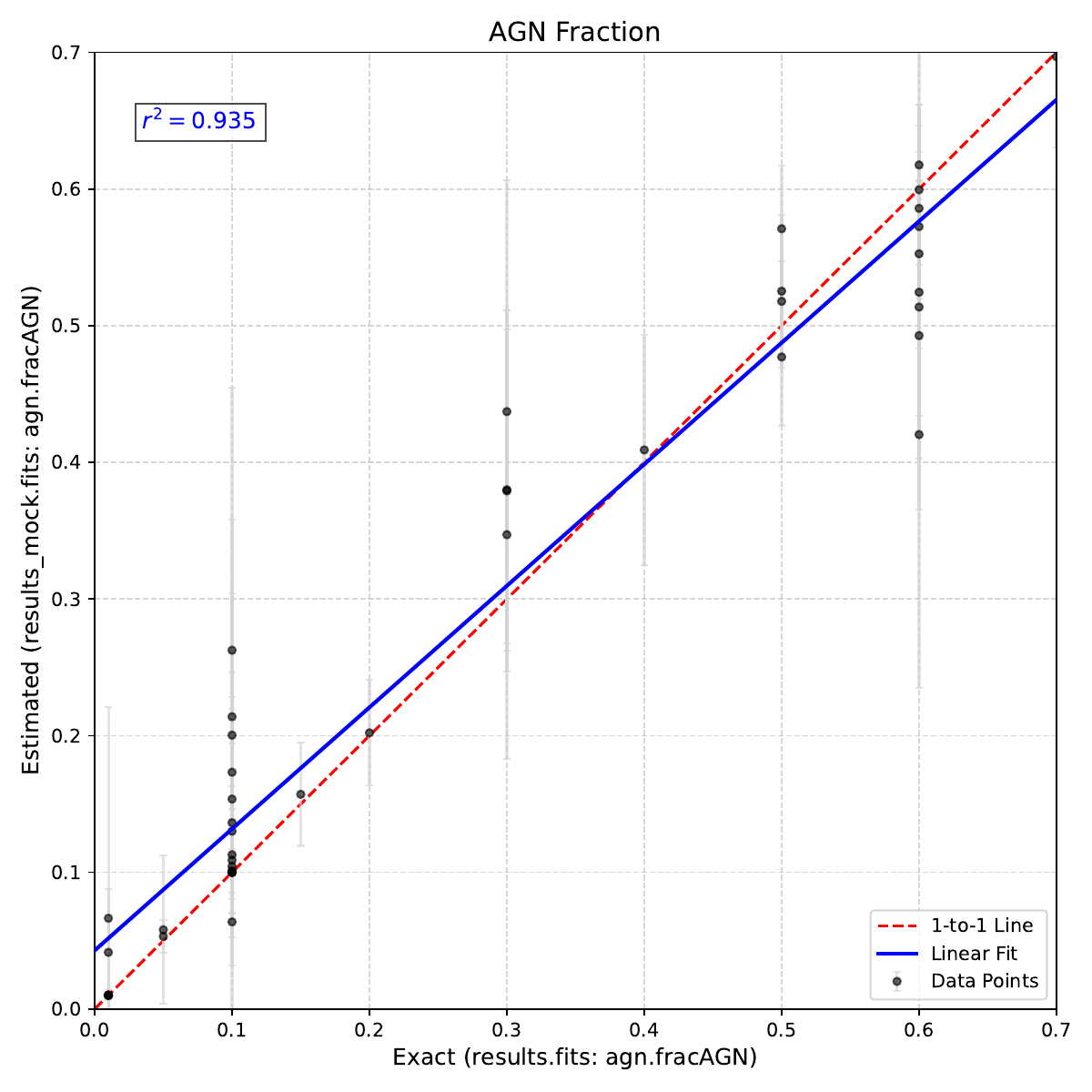}
    \end{minipage}\hfill
    \begin{minipage}{0.32\textwidth}
        \centering
        \includegraphics[width=\textwidth]{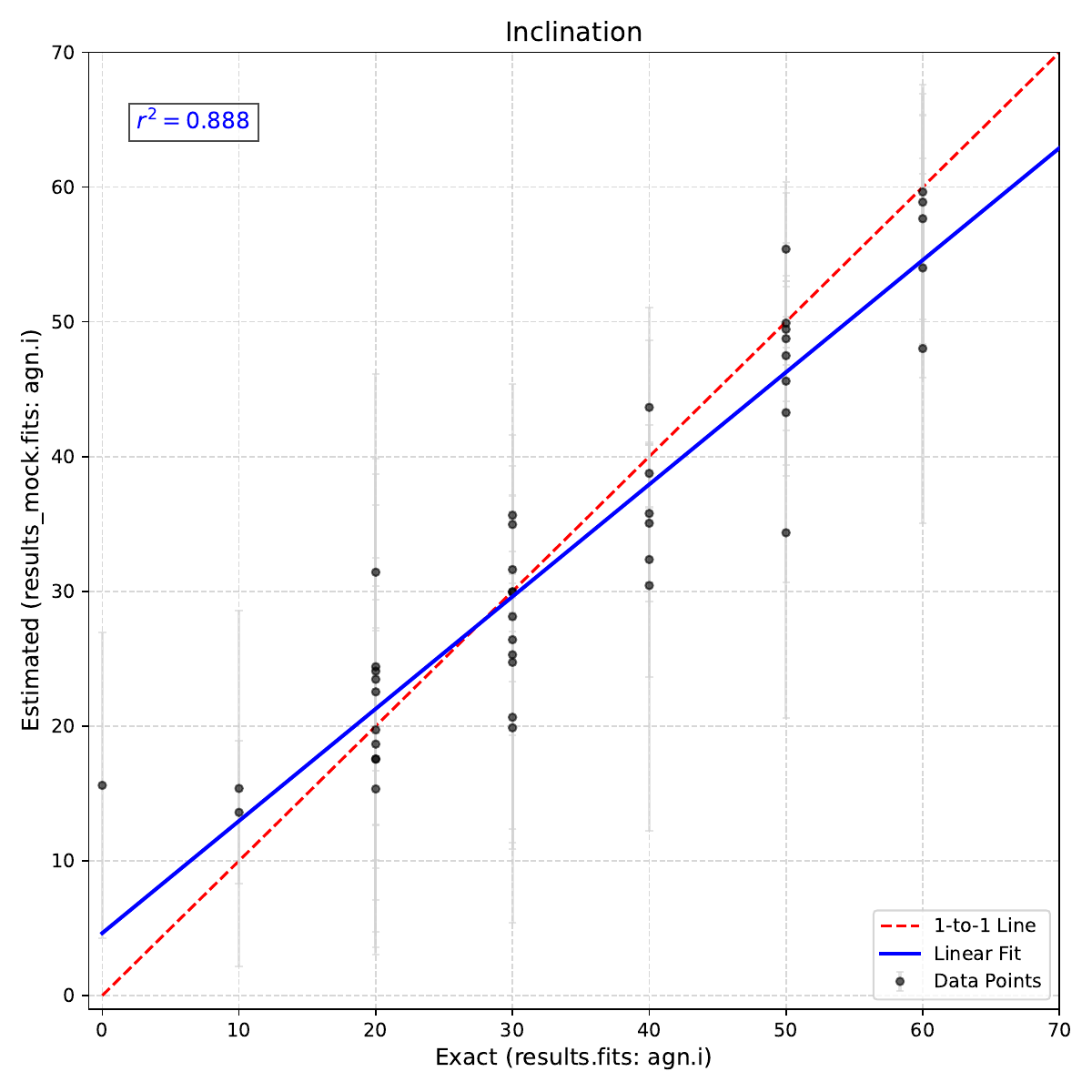}
    \end{minipage}\hfill
    \begin{minipage}{0.32\textwidth}
        \centering
        \includegraphics[width=\textwidth]{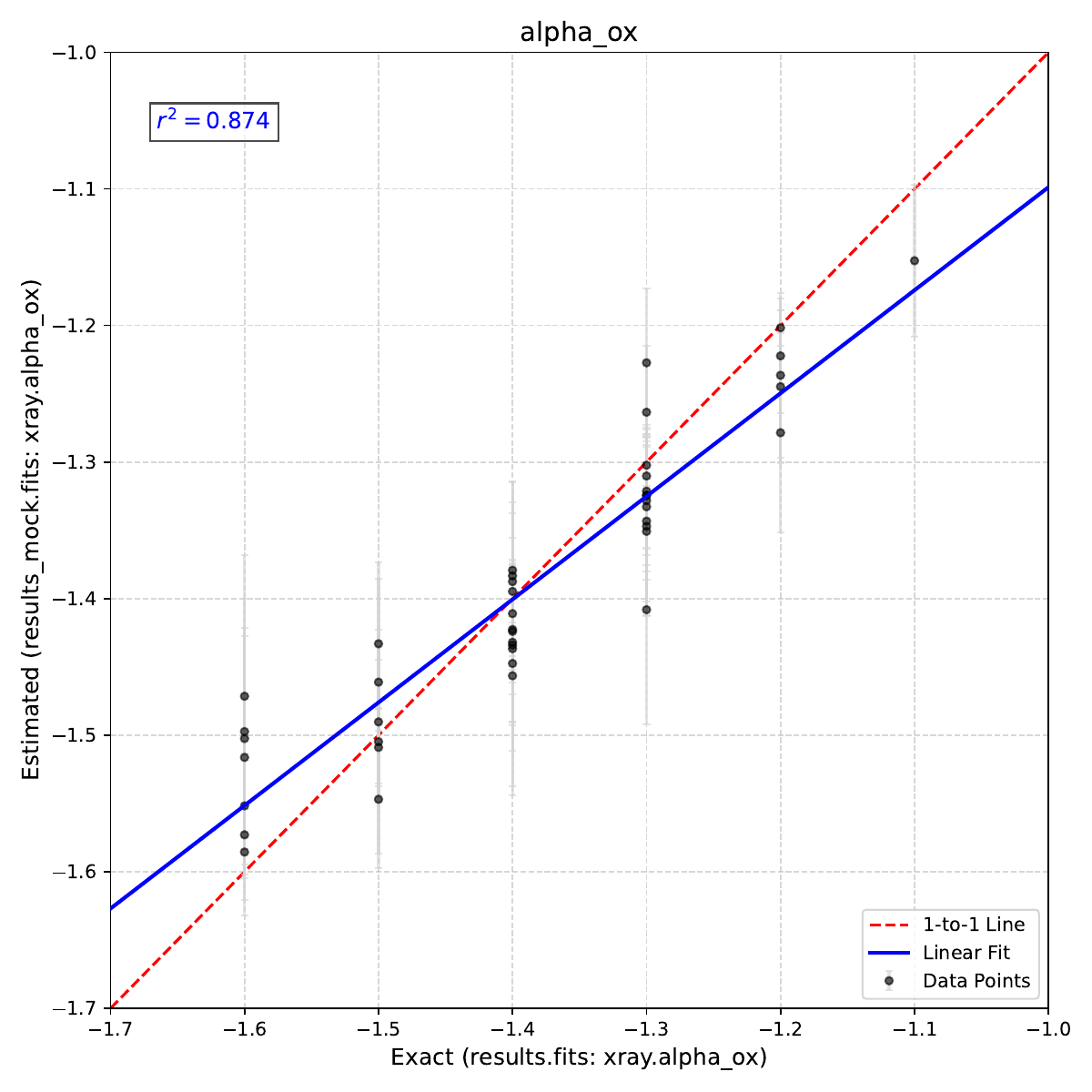}
    \end{minipage}
\caption{CIGALE mock analysis parameter recovery for our control sample ($N=50$). Panels compare exact input values against recovered estimates for bolometric and geometric parameters: AGN luminosity ($L_{\text{AGN}}$), total dust luminosity ($L_{\text{dust, AGN}}$), accretion power ($P_{\text{acc, AGN}}$), AGN fraction ($f_{\text{AGN}}$), inclination ($i$), and X-ray slope ($\alpha_{\text{ox}}$). Red dashed lines indicate the 1:1 identity relation, solid blue lines represent best linear fits, and individual sub-panel text boxes state the resulting coefficient of determination ($r^2$). High correlation across all panels validates the robustness of our SED decomposition procedure.}
\label{fig:mock_recovery_plots}
\end{figure*}

In this appendix, we present the quantitative graphical results of the CIGALE mock analysis performed on our 50-galaxy control sample, as outlined in Section 3.1. Figure~\ref{fig:mock_recovery_plots} displays the direct 1:1 parameter recovery comparison between the synthetic input (exact) values and the post-analysis output (recovered) parameters. Overall, the energy-integrated bolometric properties—specifically the AGN luminosity ($L_{\text{AGN}}$, $r^2 = 0.962$), total dust luminosity ($L_{\text{dust, AGN}}$, $r^2 = 0.956$), and accretion power ($P_{\text{acc, AGN}}$, $r^2 = 0.952$)—exhibit exceptionally tight clustering along the identity line in logarithmic space. Geometric and fractional parameters such as the AGN fraction ($f_{\text{AGN}}$, $r^2 = 0.935$), torus inclination angle ($i$, $r^2 = 0.888$), and optical-to-X-ray spectral index ($\alpha_{\text{ox}}$, $r^2 = 0.874$) also display robust linear fits despite minor parameter degeneracies. These consistently high $r^2$ metrics confirm that CIGALE accurately breaks model degeneracies and reliably disentangles host-galaxy stellar emission from coronal and torus-reprocessed AGN outputs.

\renewcommand{\thefigure}{\Alph{section}\arabic{figure}}
\begin{figure*}
\begin{center}$
\begin{array}{lll}
\includegraphics[scale=0.30]{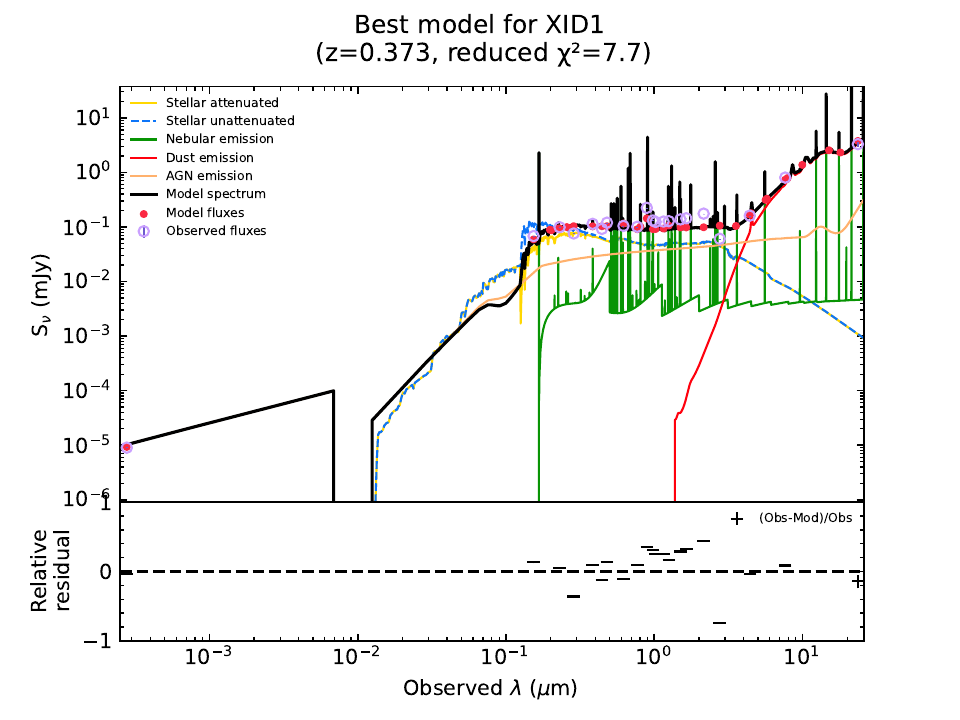}&
\includegraphics[scale=0.30]{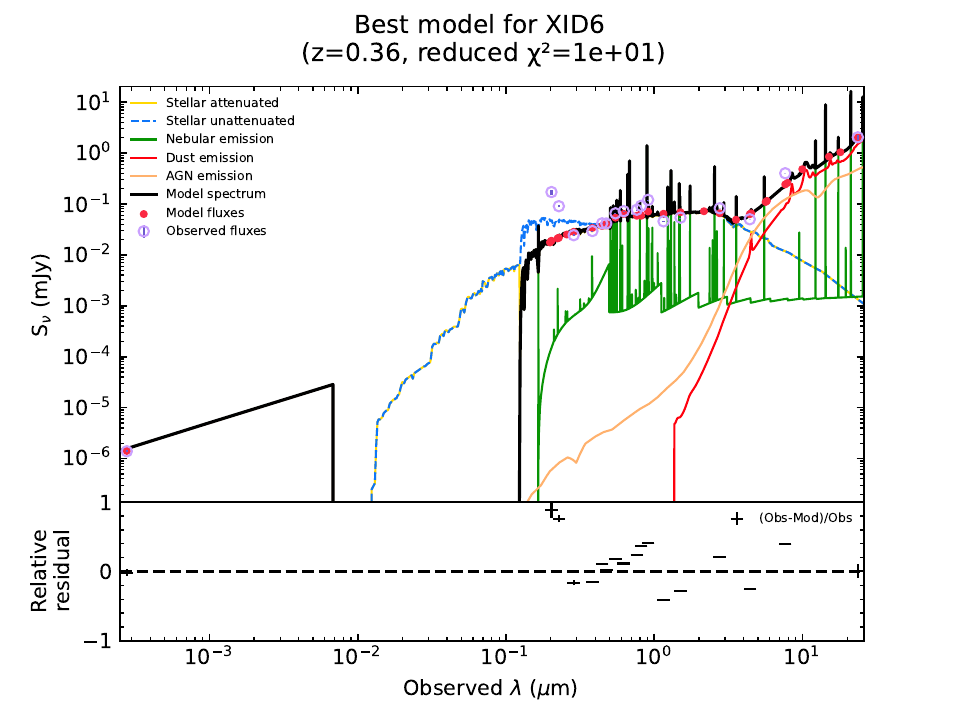}&
\includegraphics[scale=0.30]{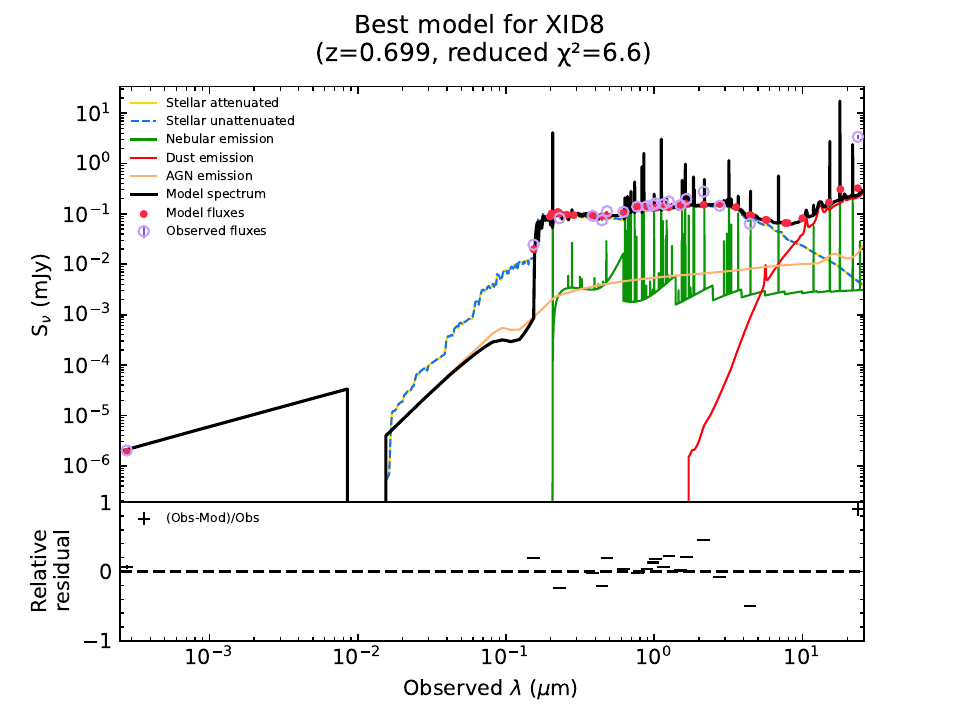}\\
\includegraphics[scale=0.30]{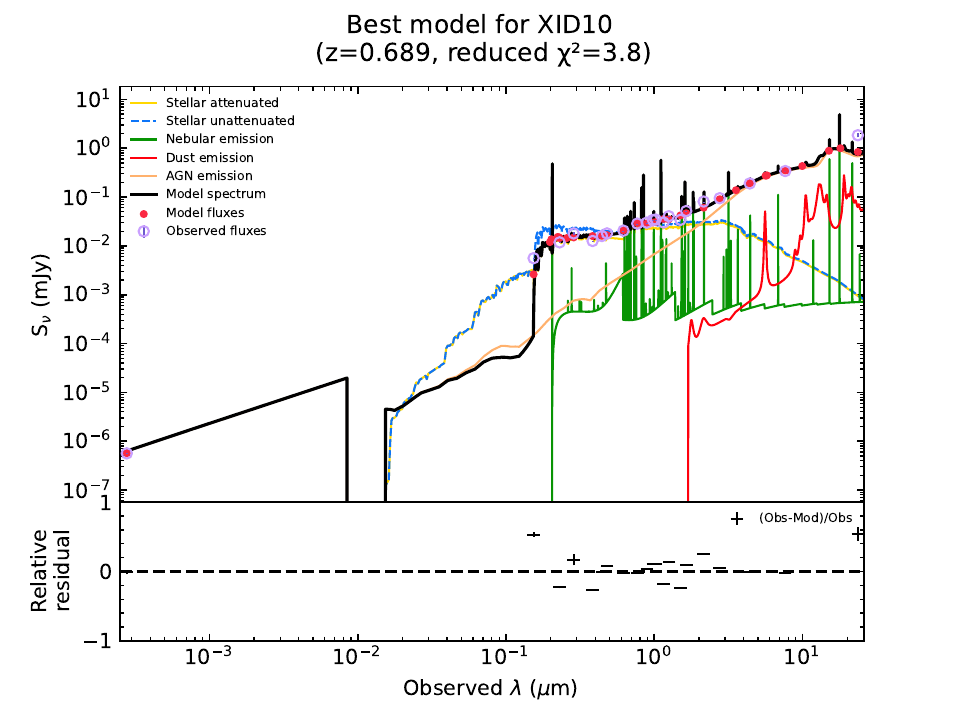}&
\includegraphics[scale=0.30]{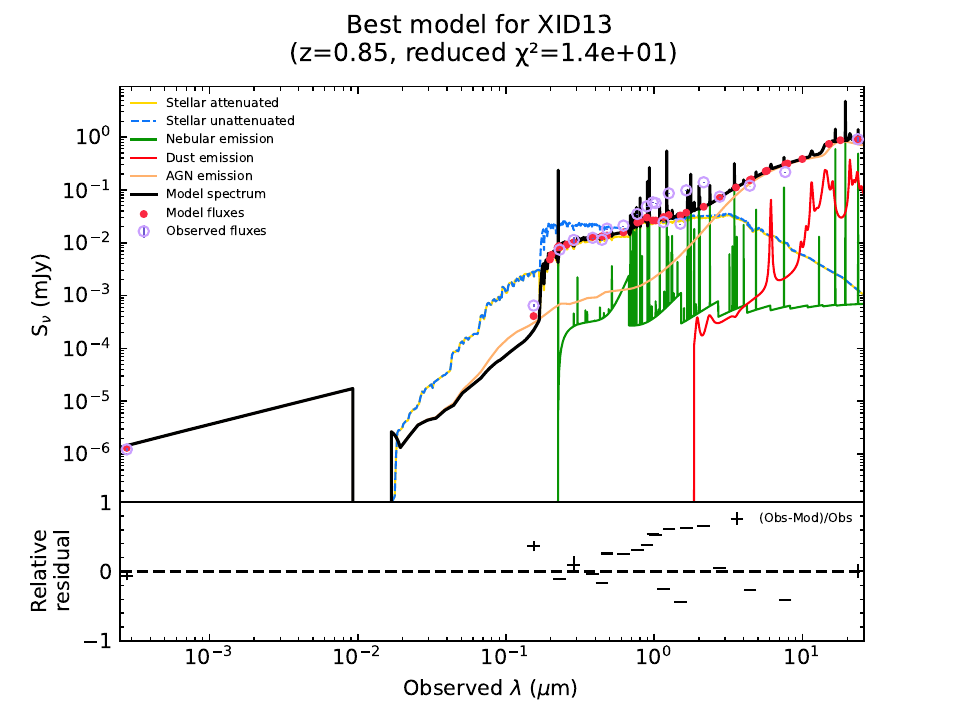}&
\includegraphics[scale=0.30]{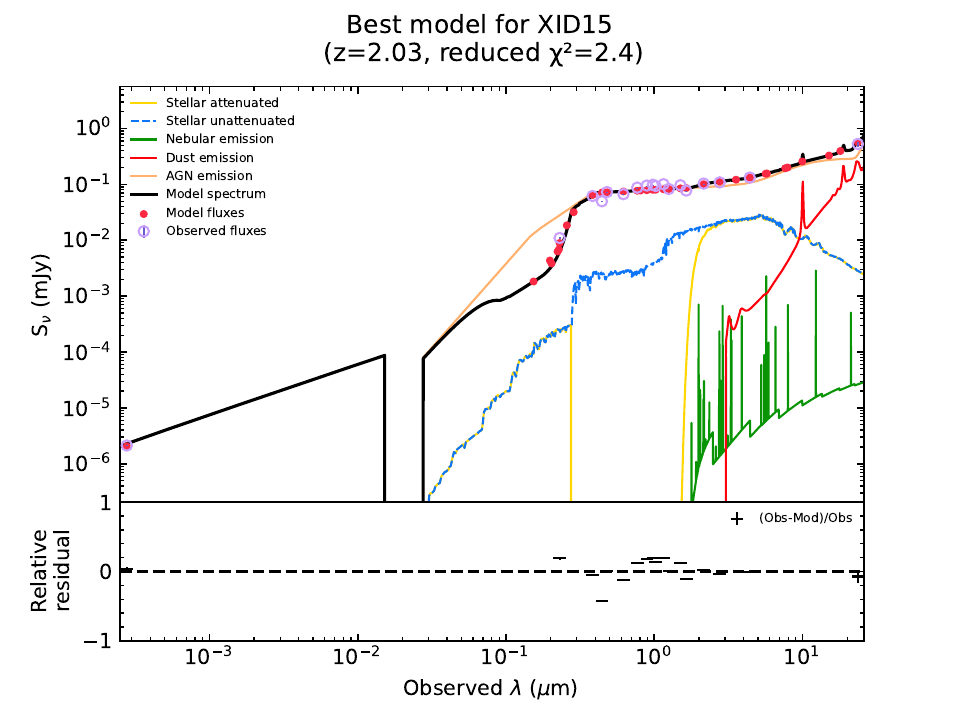}\\
\includegraphics[scale=0.30]{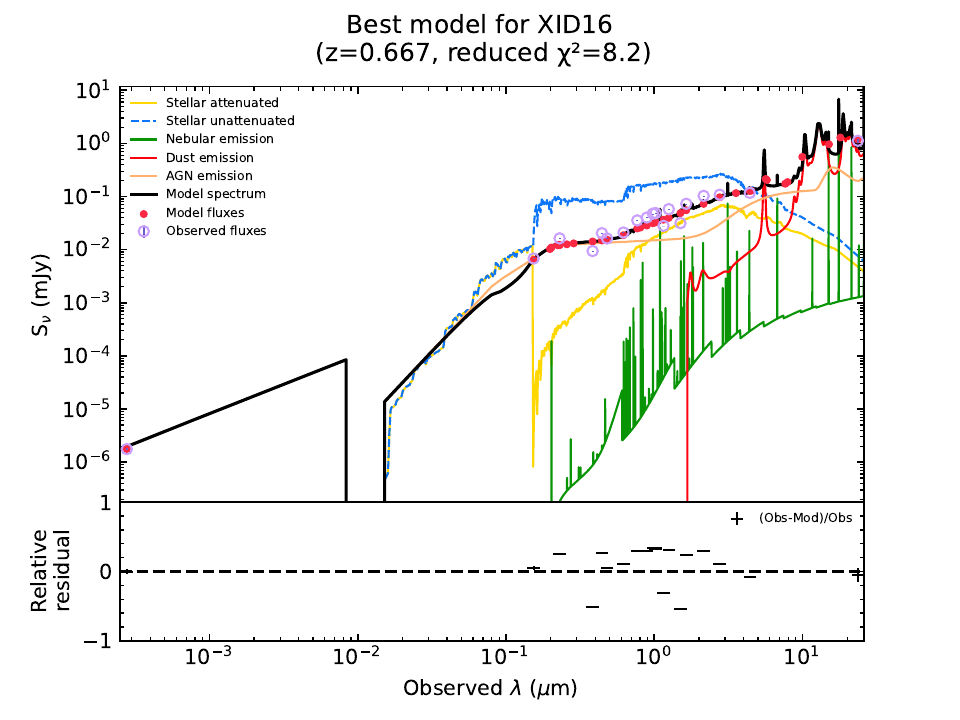}&
\includegraphics[scale=0.30]{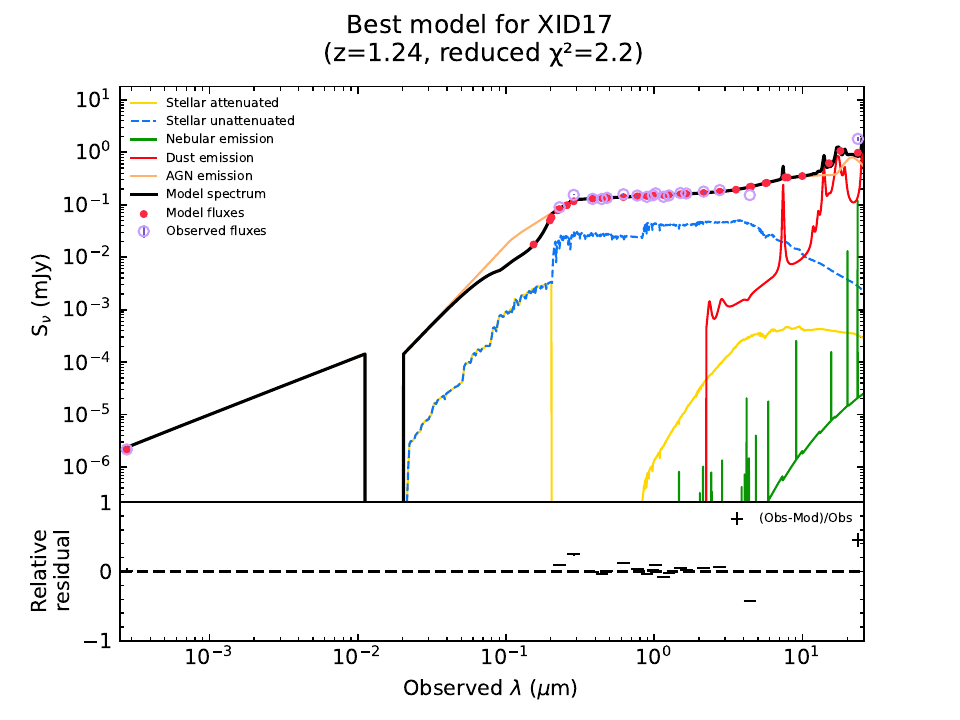}&
\includegraphics[scale=0.30]{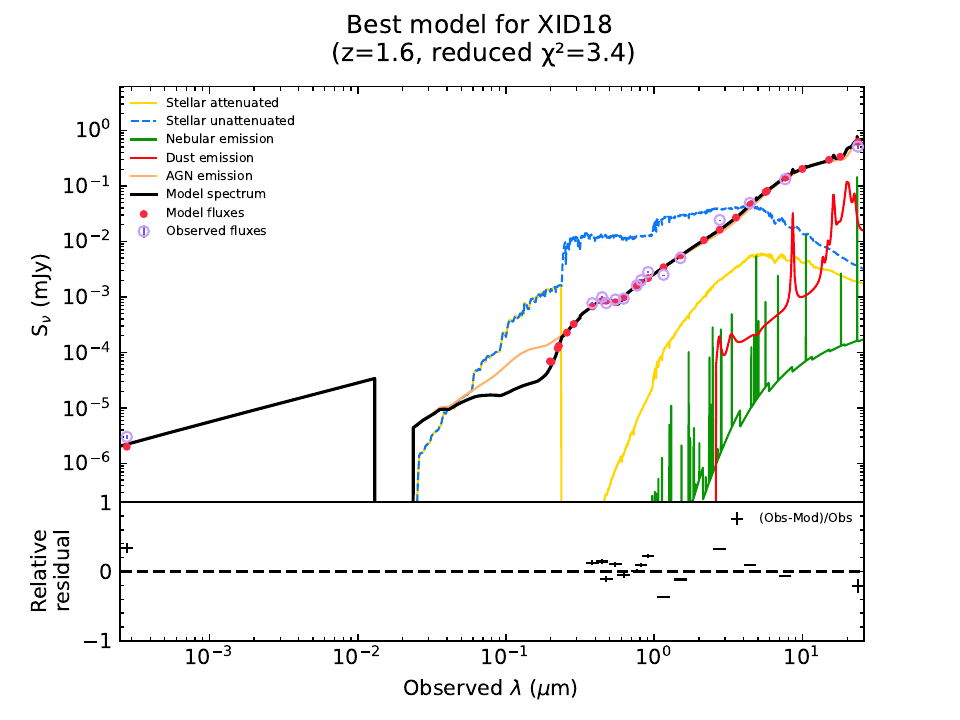}\\
\includegraphics[scale=0.30]{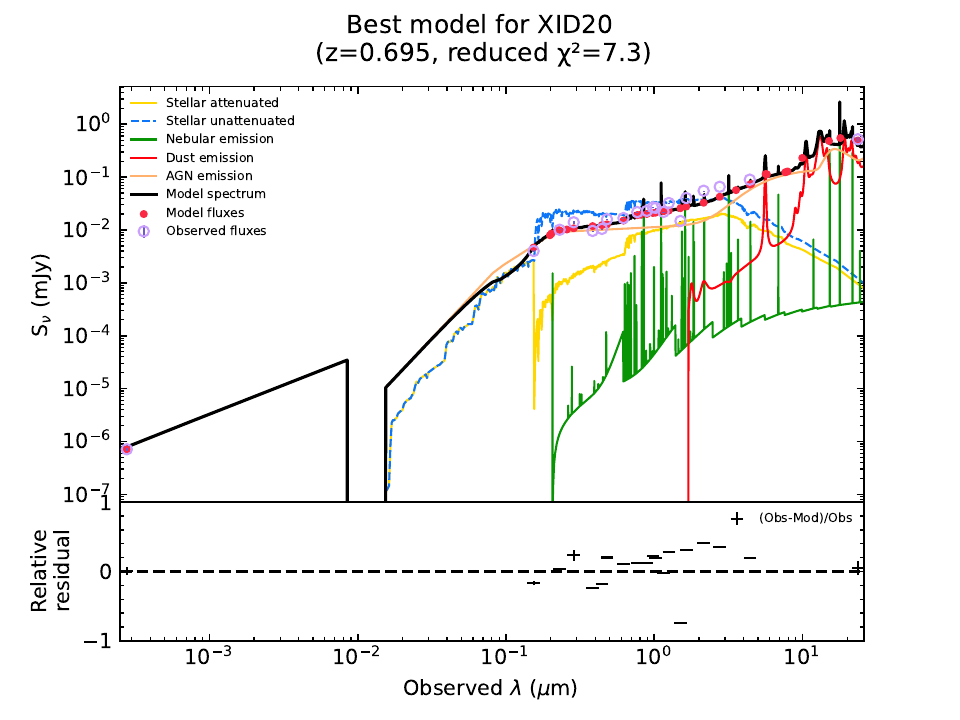}&
\includegraphics[scale=0.30]{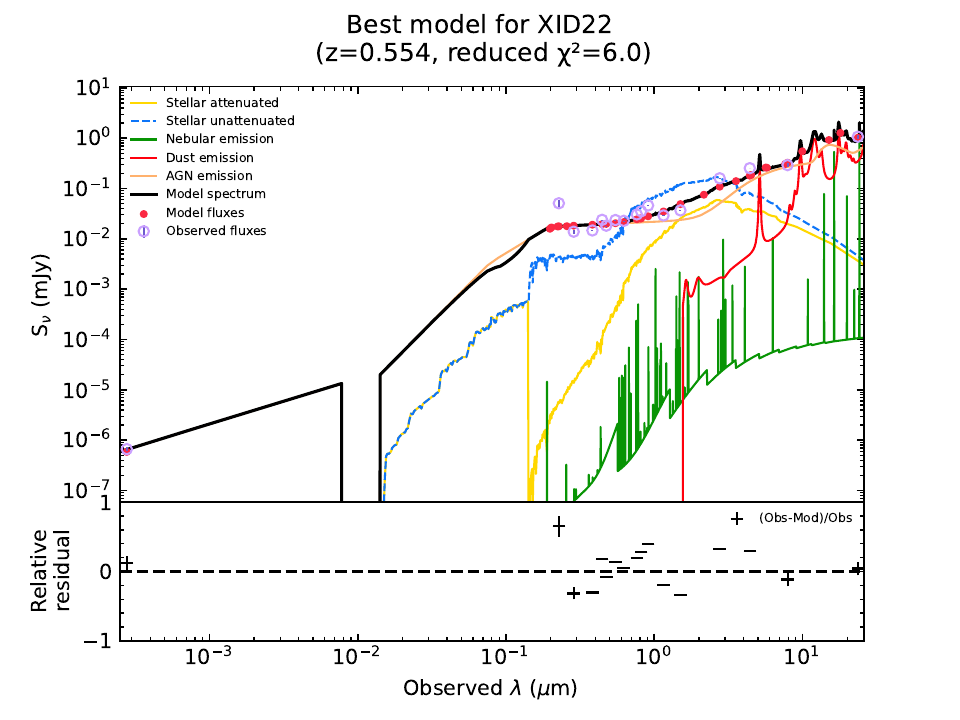}&
\includegraphics[scale=0.30]{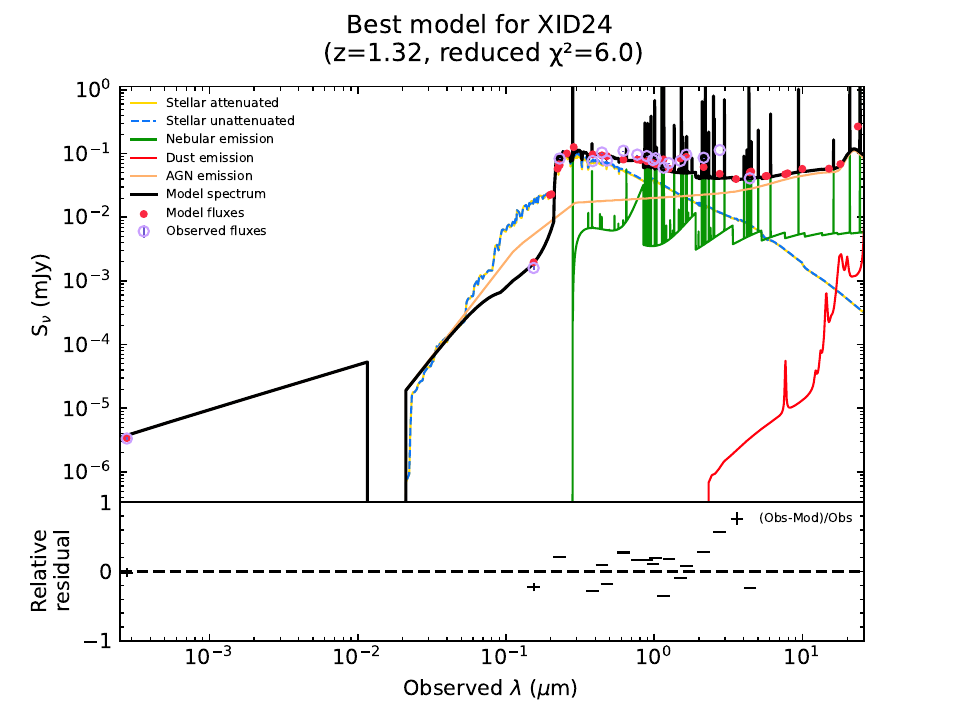}\\
\includegraphics[scale=0.30]{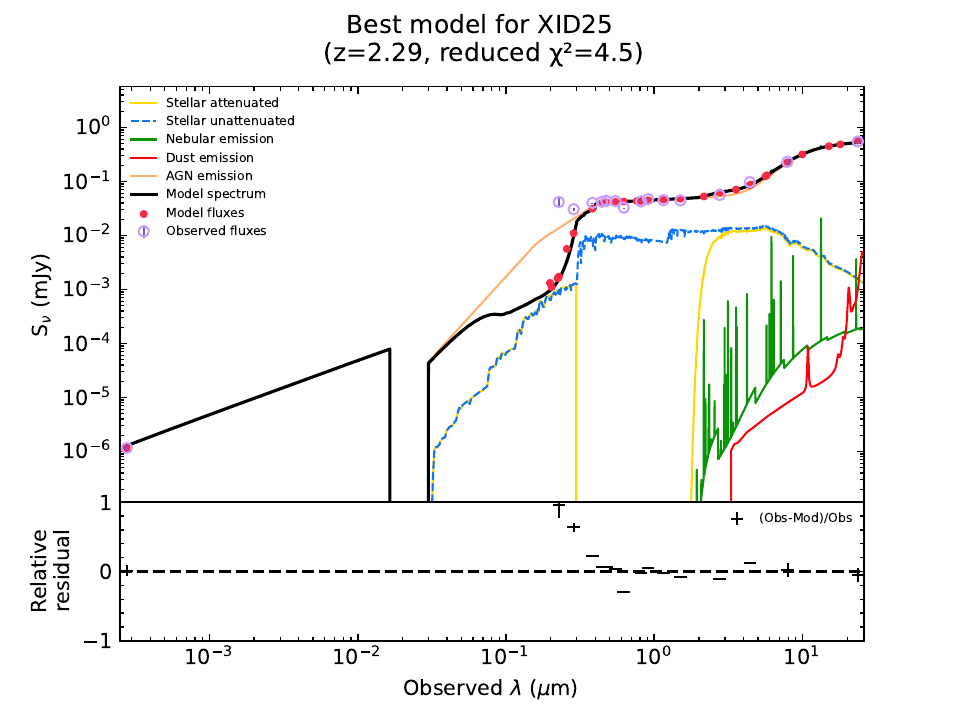}&
\includegraphics[scale=0.30]{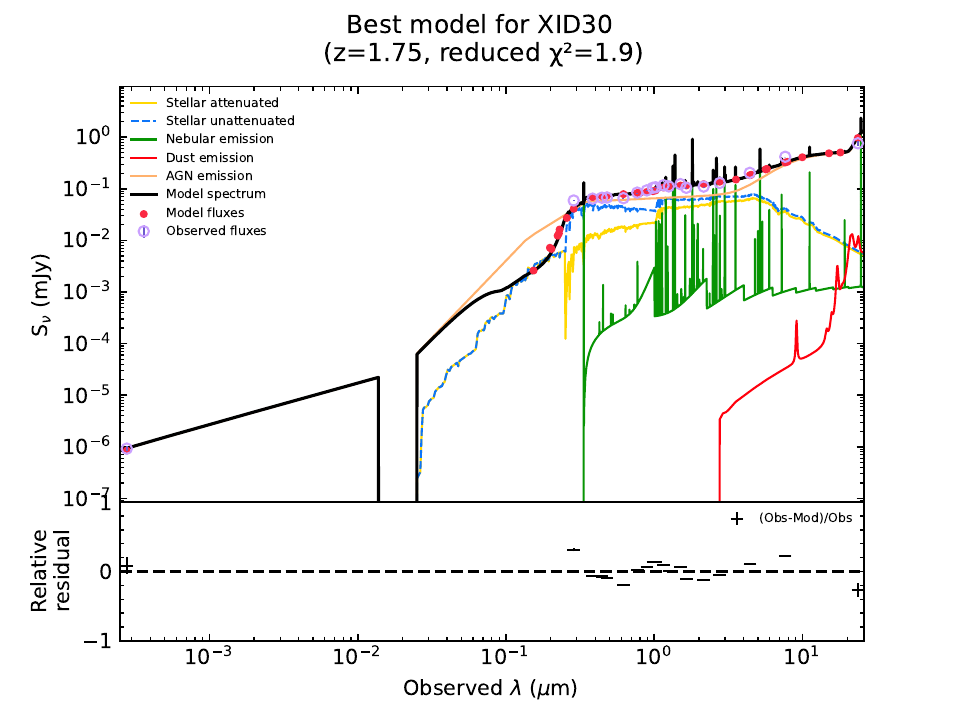}&
\includegraphics[scale=0.30]{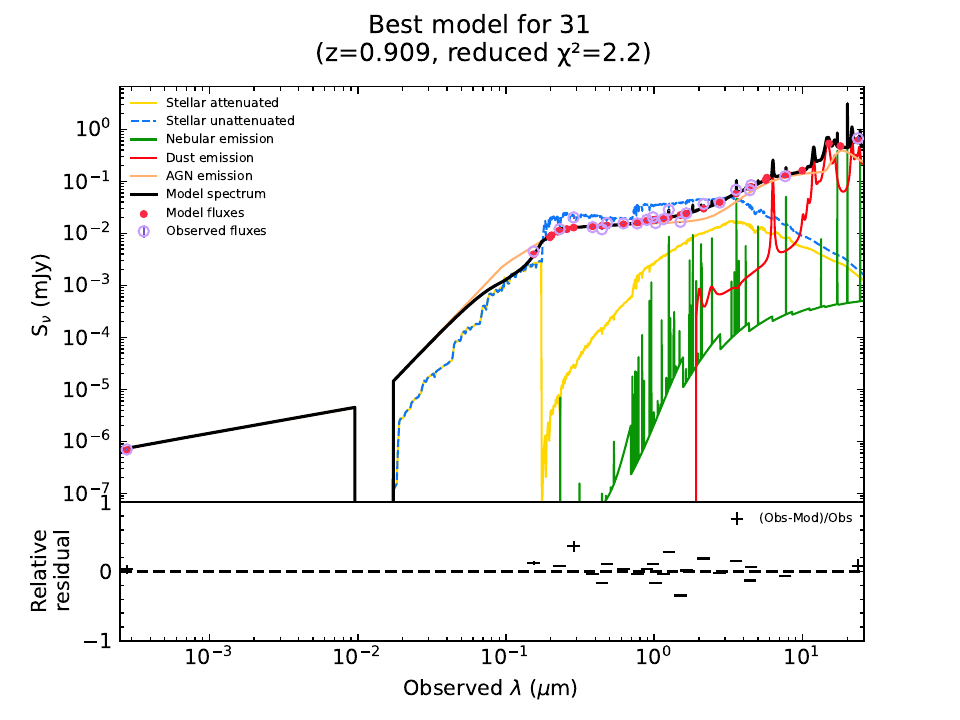}\\
\includegraphics[scale=0.30]{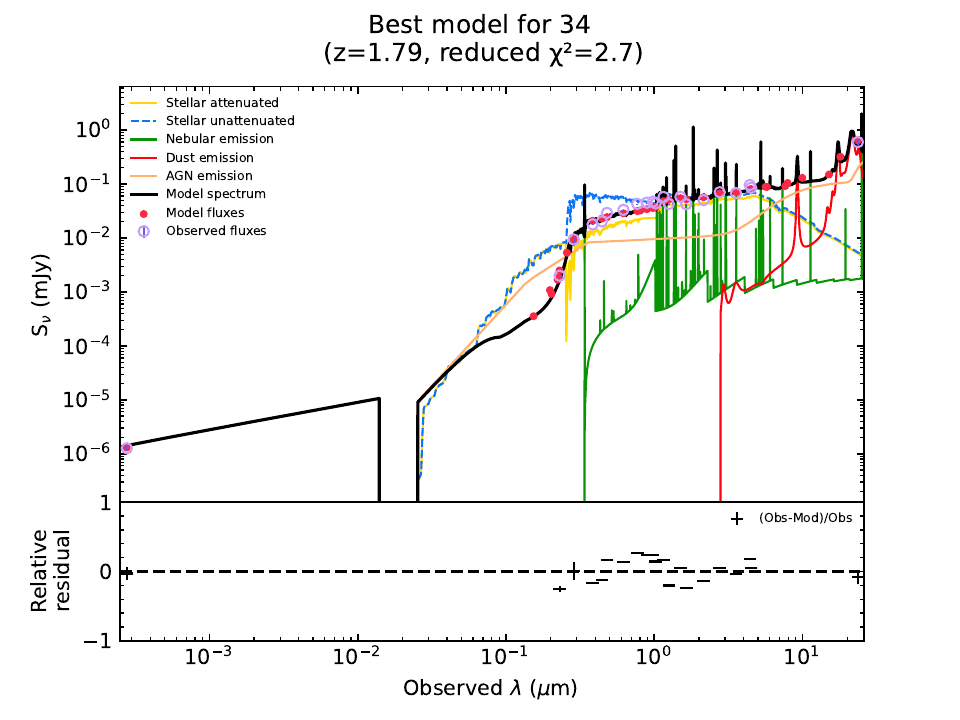}&
\includegraphics[scale=0.30]{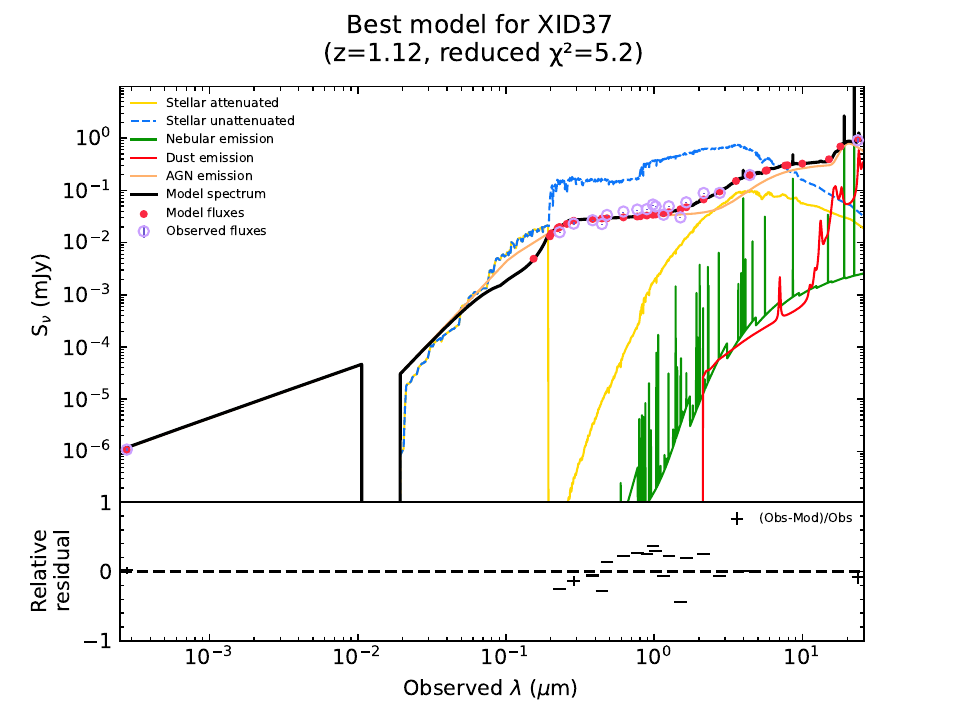}&
\includegraphics[scale=0.30]{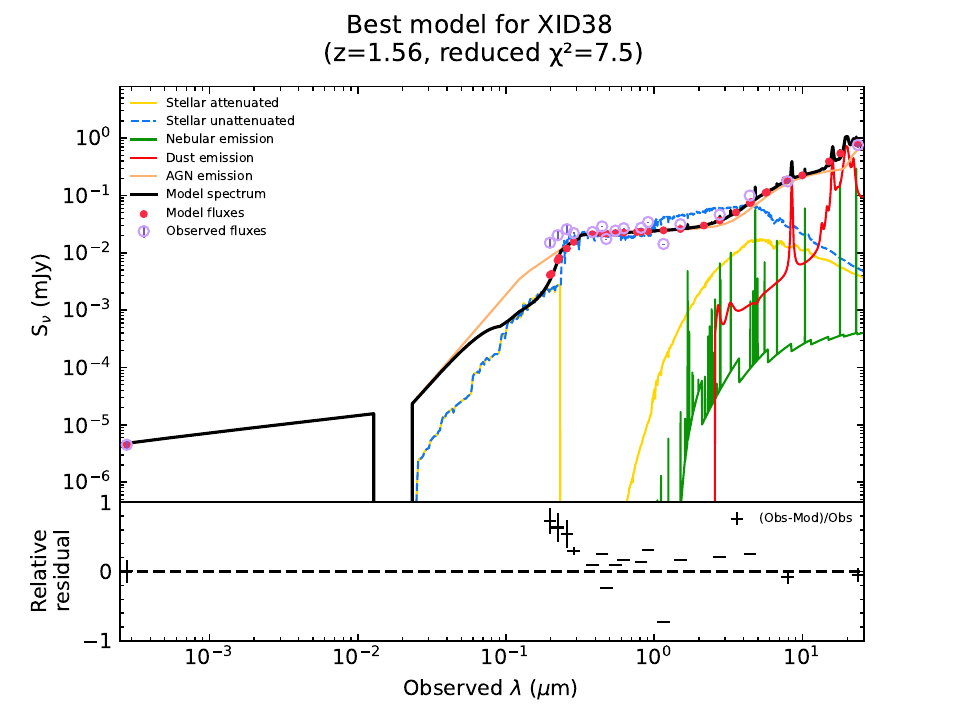}\\
\end{array}$
\end{center}
\caption{The best-fitting SEDs of S1, the AGN with JWST NIRCam coverage. The dust emission is shown by the red line. The orange line shows the AGN model. The blue curve shows the stellar light. The observed data is shown by the violet points. The red circles are the model fluxes in the given bands.}
\label{fig:figB3}
\end{figure*}

\renewcommand{\thefigure}{\Alph{section}\arabic{figure}(Cont.)}
\addtocounter{figure}{-1}
\begin{figure*}
\begin{center}$
\begin{array}{lll}
\includegraphics[scale=0.30]{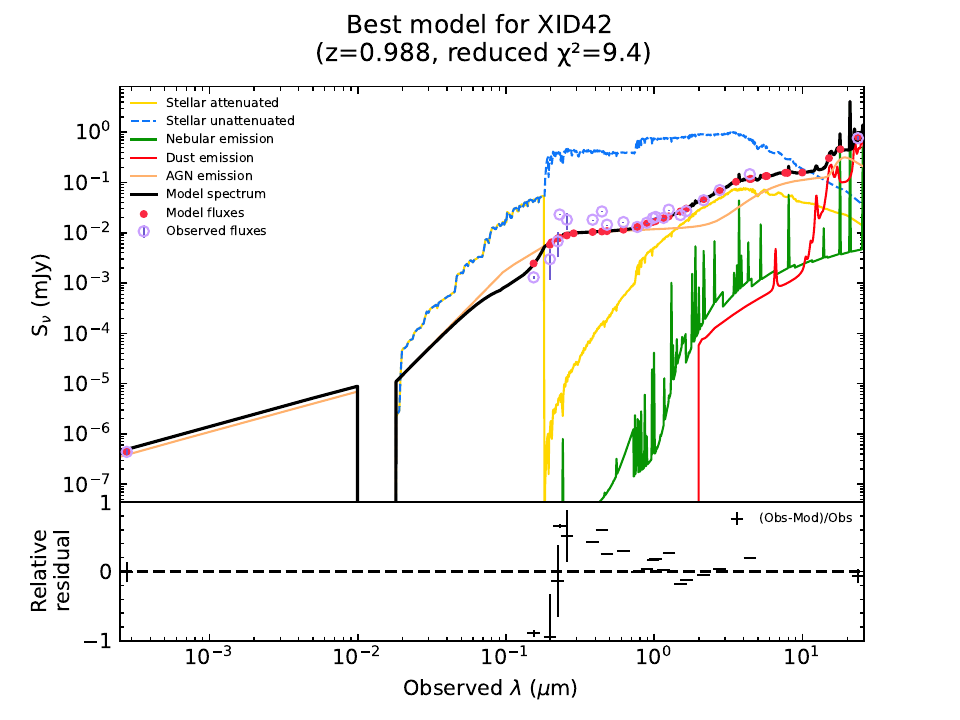}&
\includegraphics[scale=0.30]{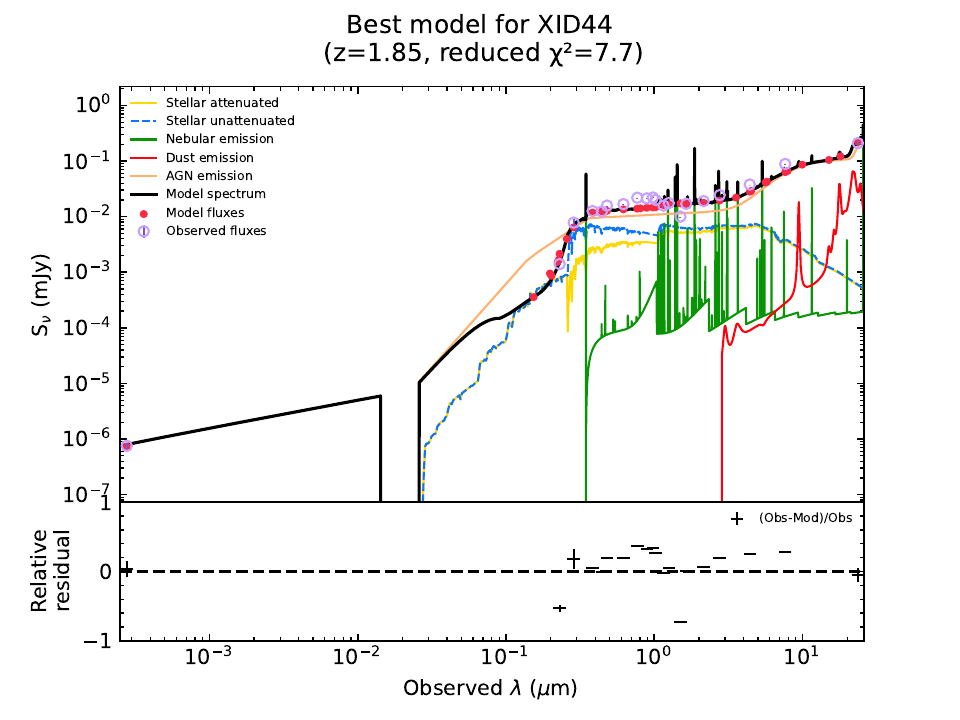}& 
\includegraphics[scale=0.30]{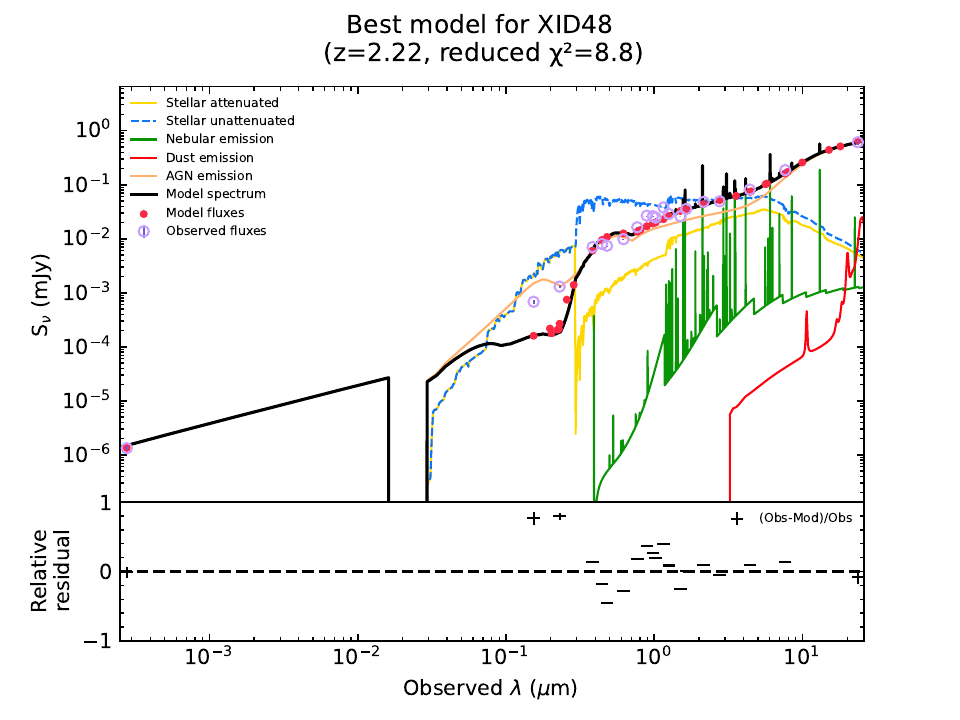}\\ 
\includegraphics[scale=0.30]{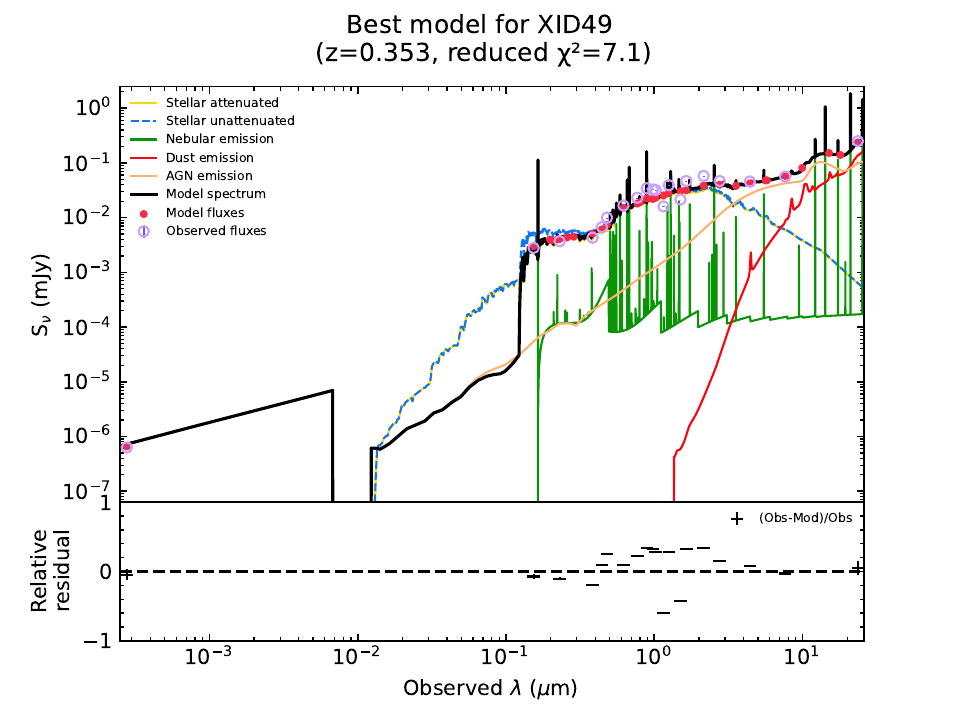}& 
\includegraphics[scale=0.30]{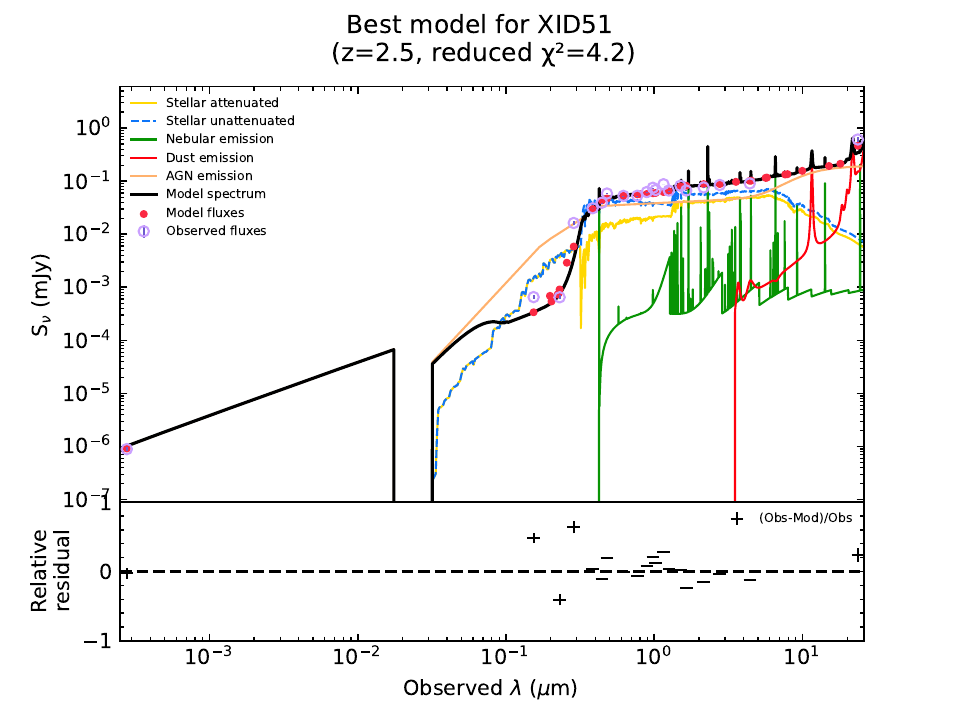}&
\includegraphics[scale=0.30]{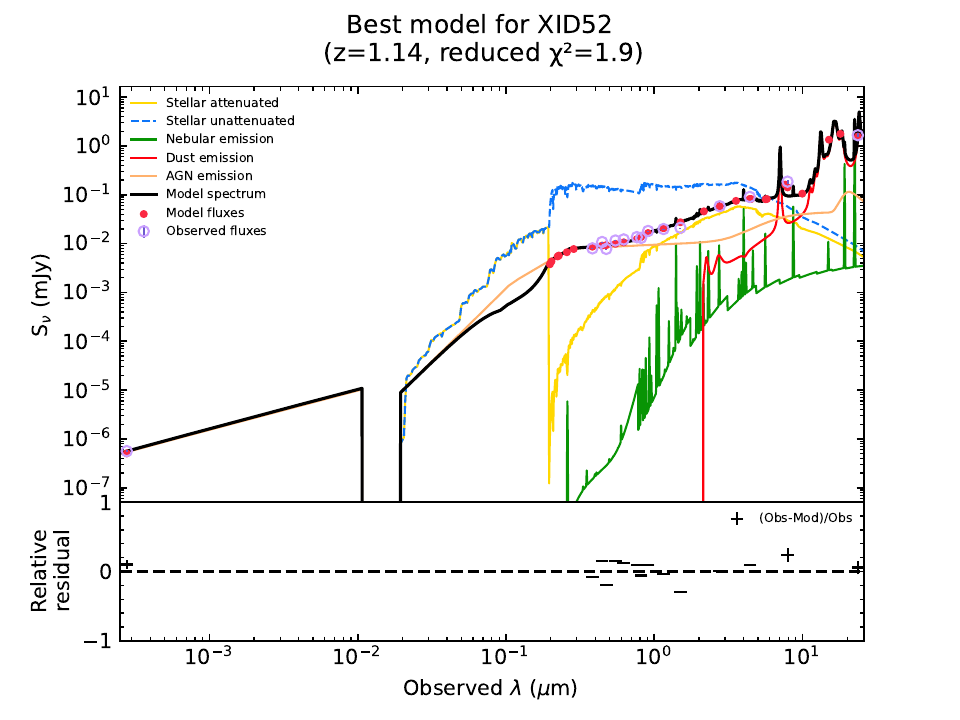}\\
\includegraphics[scale=0.30]{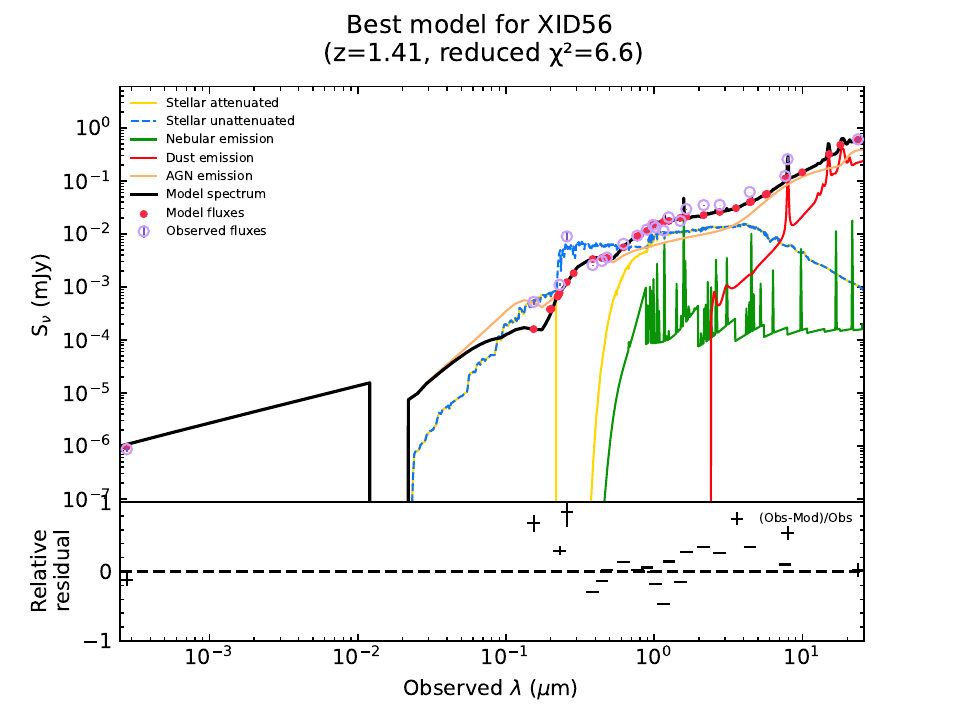}& 
\includegraphics[scale=0.30]{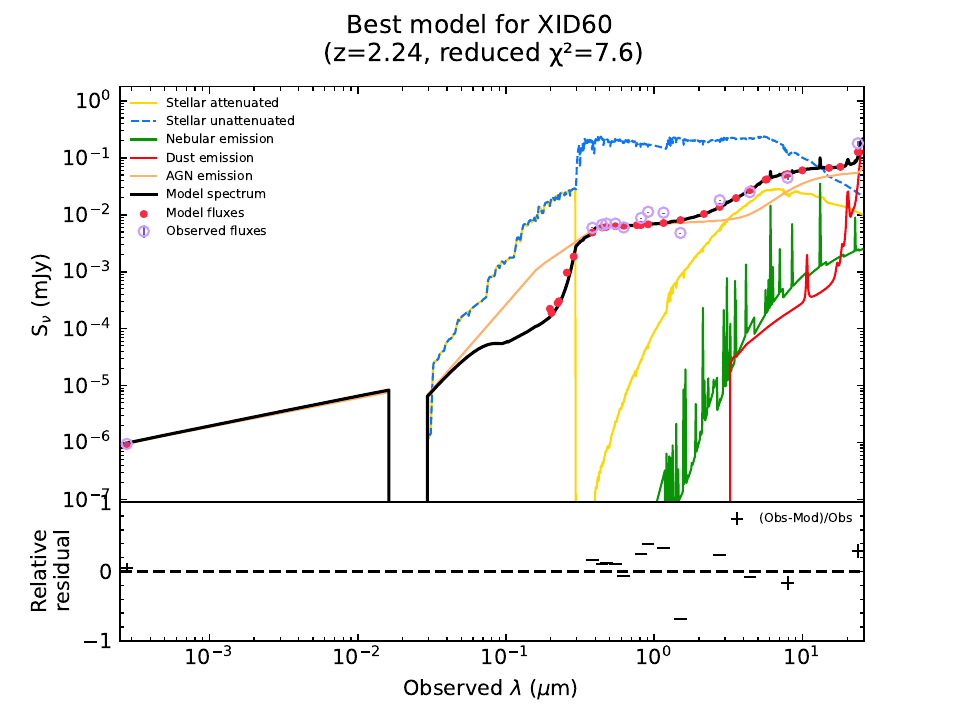}&
\includegraphics[scale=0.30]{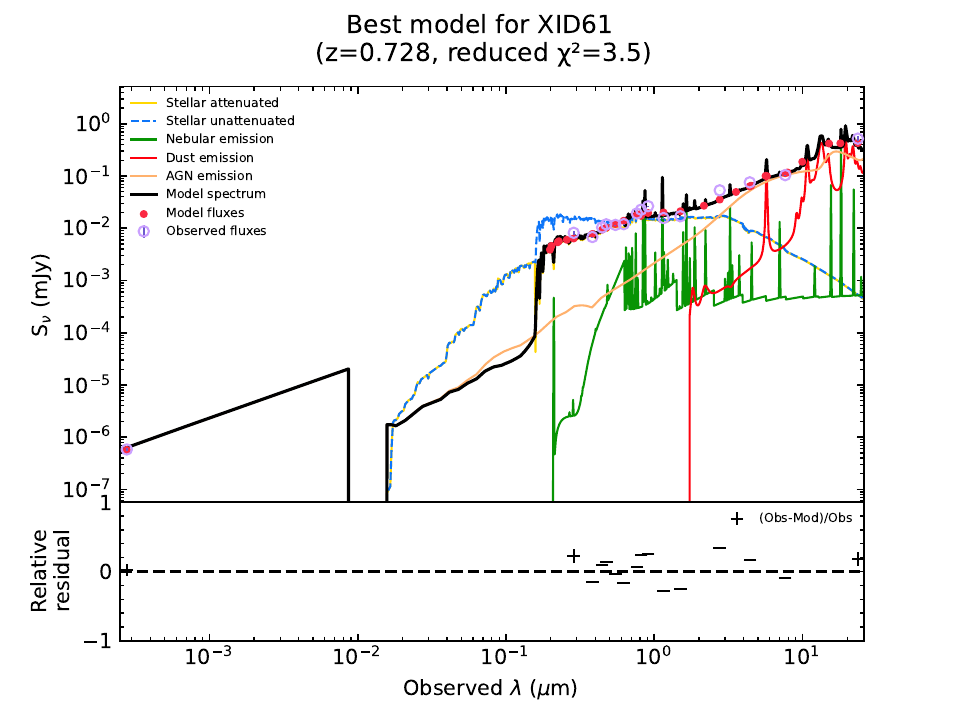}\\
\includegraphics[scale=0.30]{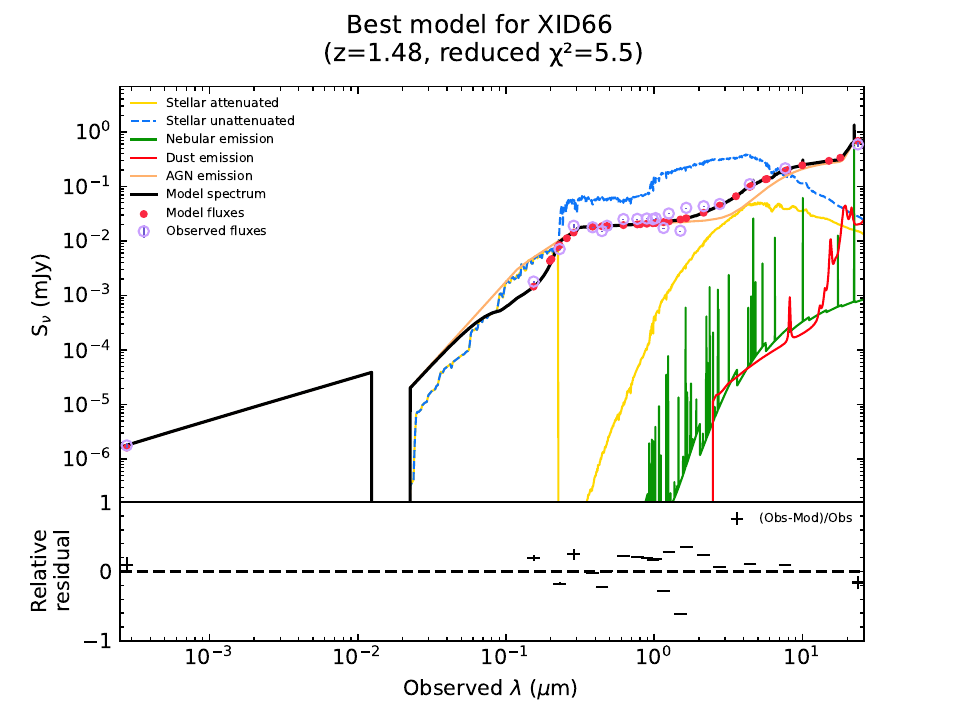}& 
\includegraphics[scale=0.30]{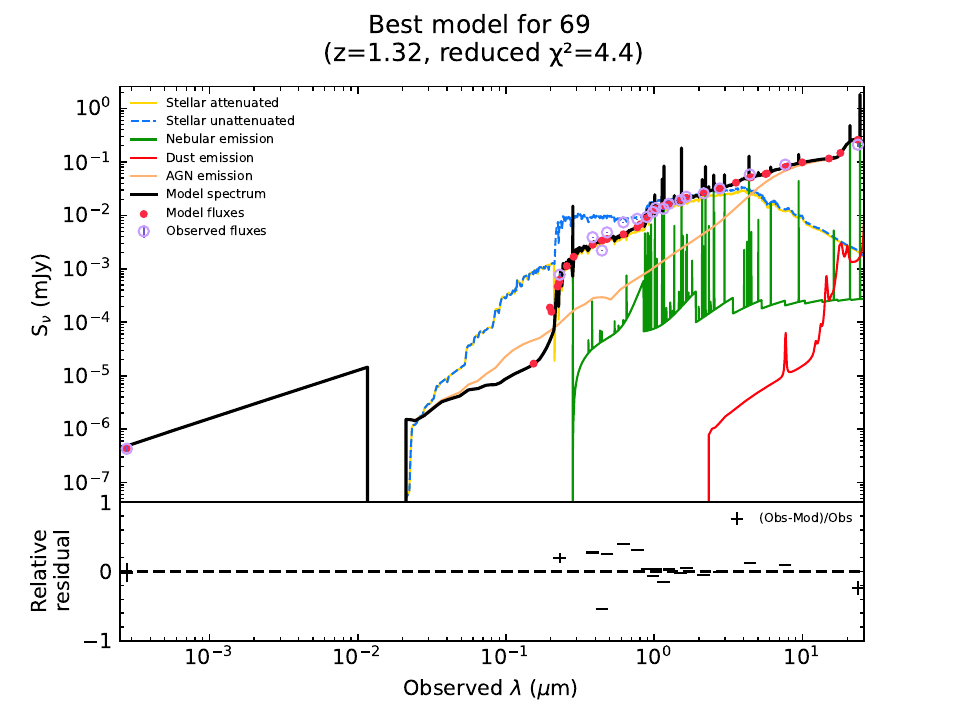}&  
\includegraphics[scale=0.30]{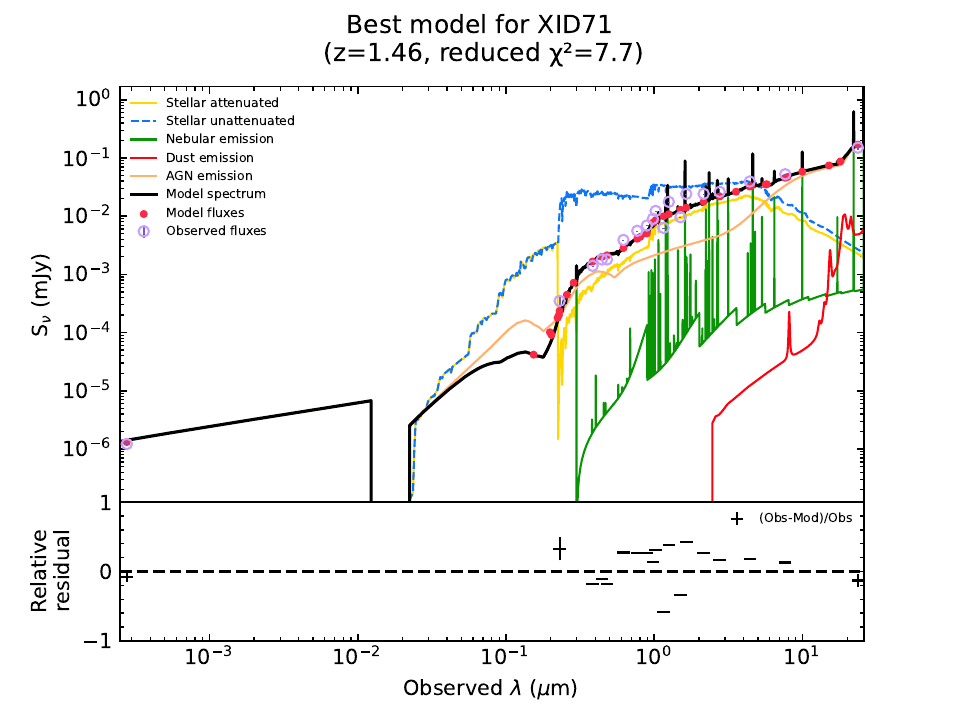}\\  
\includegraphics[scale=0.30]{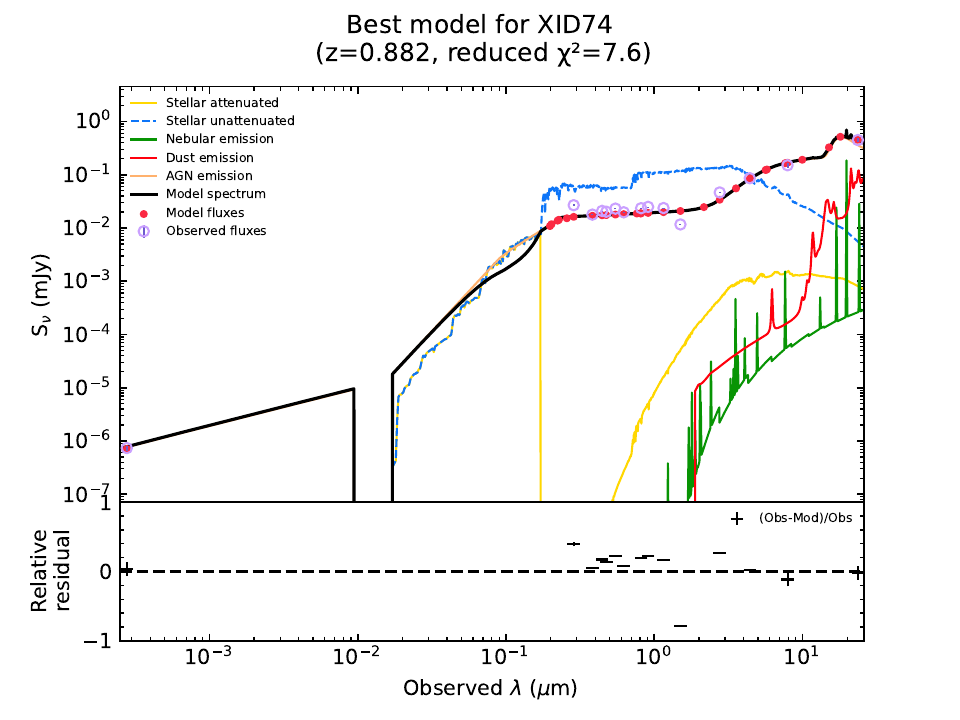}&  
\includegraphics[scale=0.30]{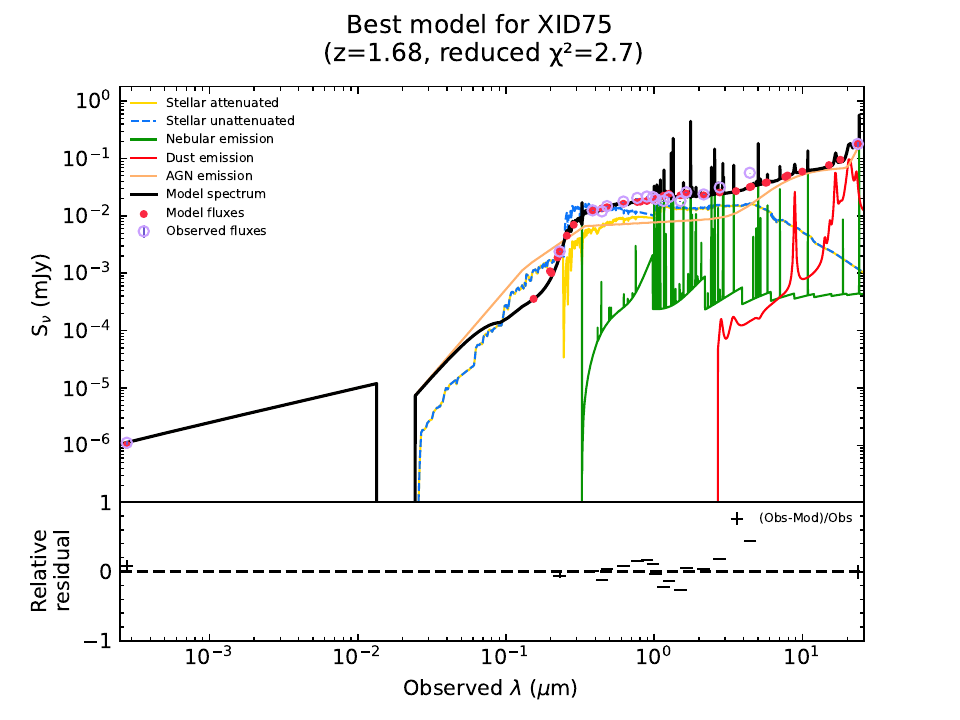}&  
\includegraphics[scale=0.30]{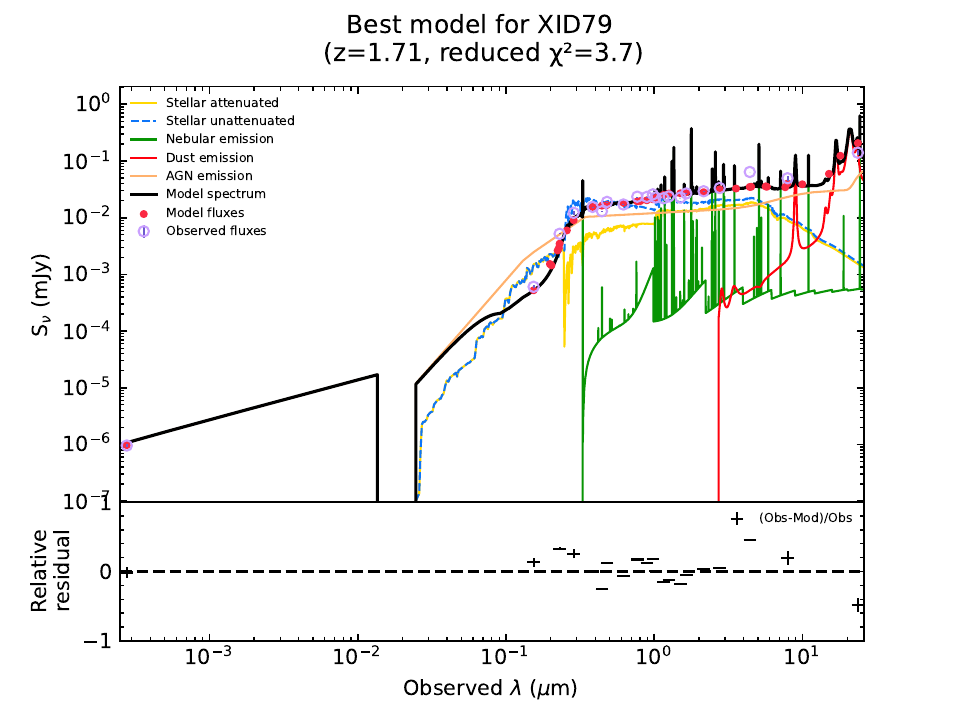}\\ 
\includegraphics[scale=0.30]{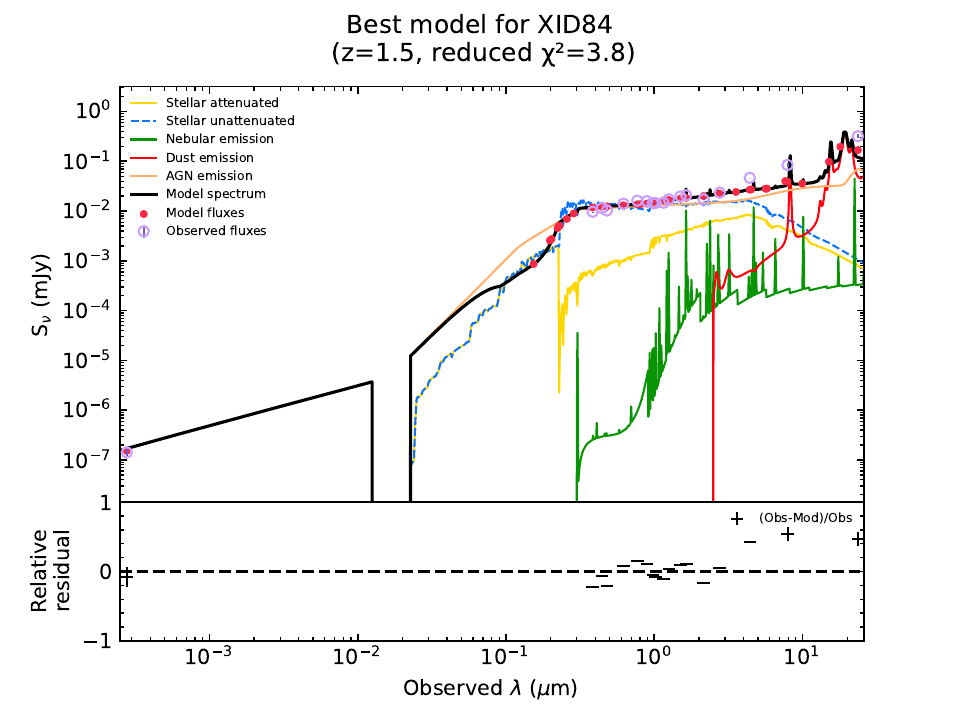}& 
\includegraphics[scale=0.30]{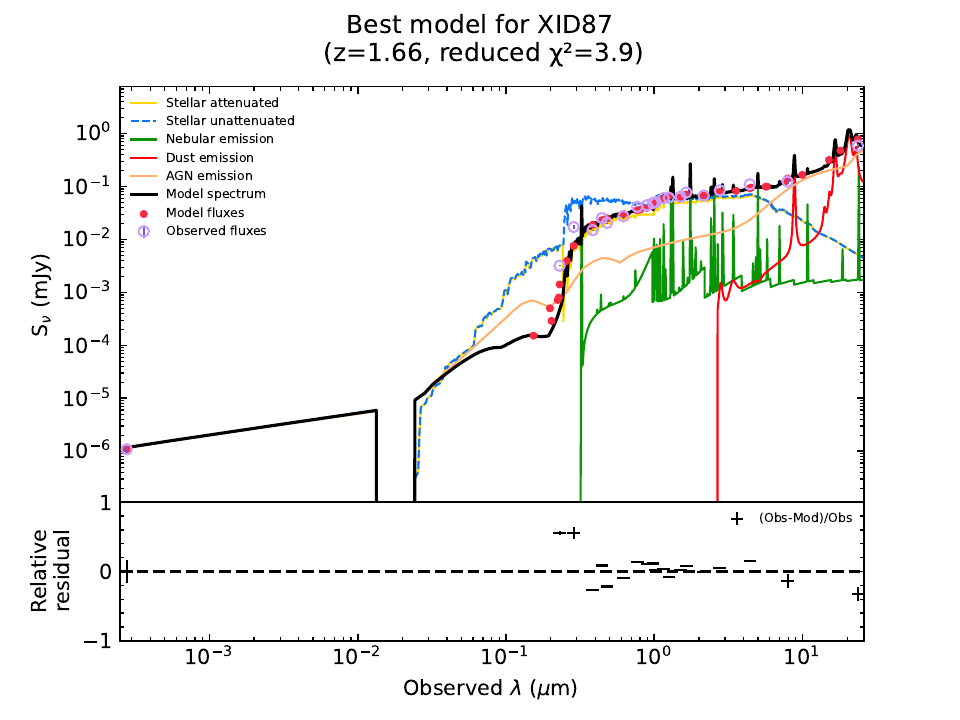}& 
\includegraphics[scale=0.30]{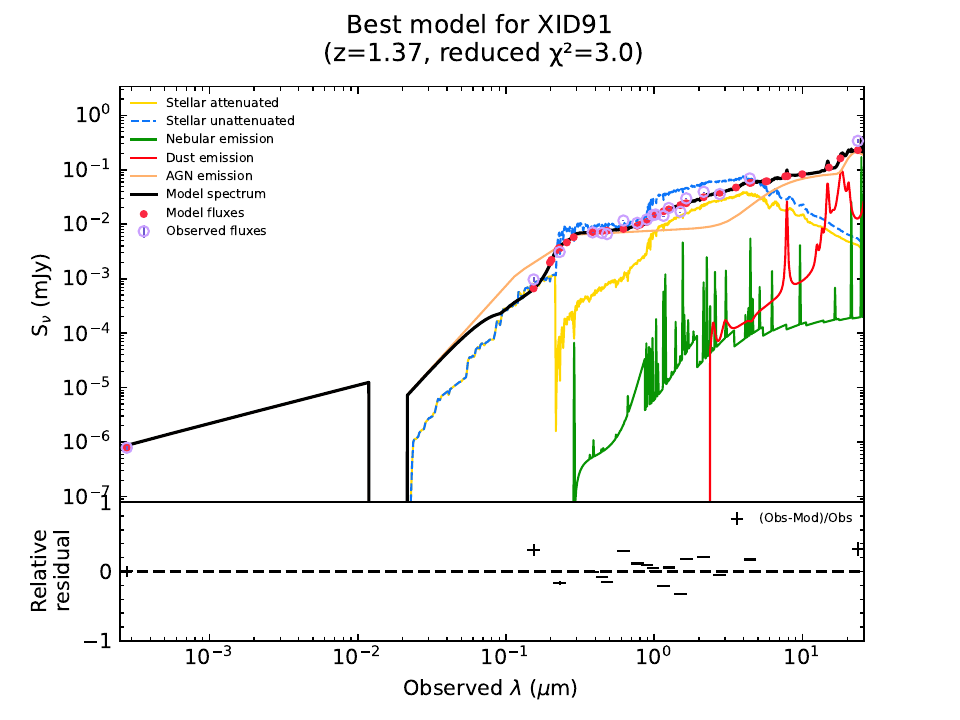}\\ 
\end{array}$
\end{center}
\caption{The best-fitting SEDs of S1, the AGN with JWST NIRCam coverage.}
\end{figure*}

\renewcommand{\thefigure}{\Alph{section}\arabic{figure}(Cont.)}
\addtocounter{figure}{-1}
\begin{figure*}
\begin{center}$
\begin{array}{lll}
\includegraphics[scale=0.30]{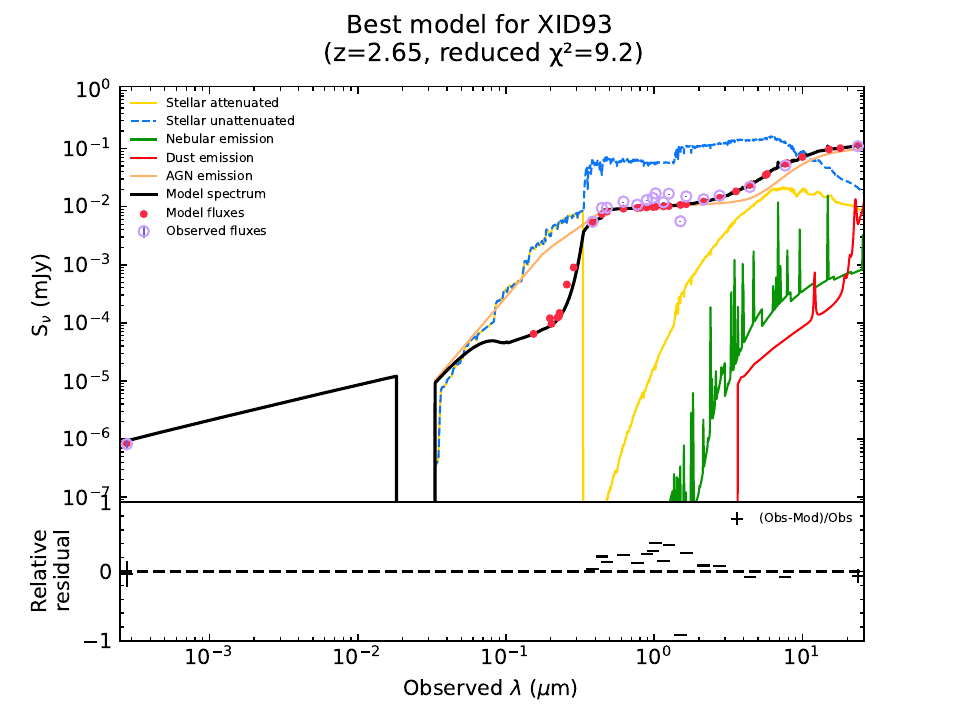}&  
\includegraphics[scale=0.30]{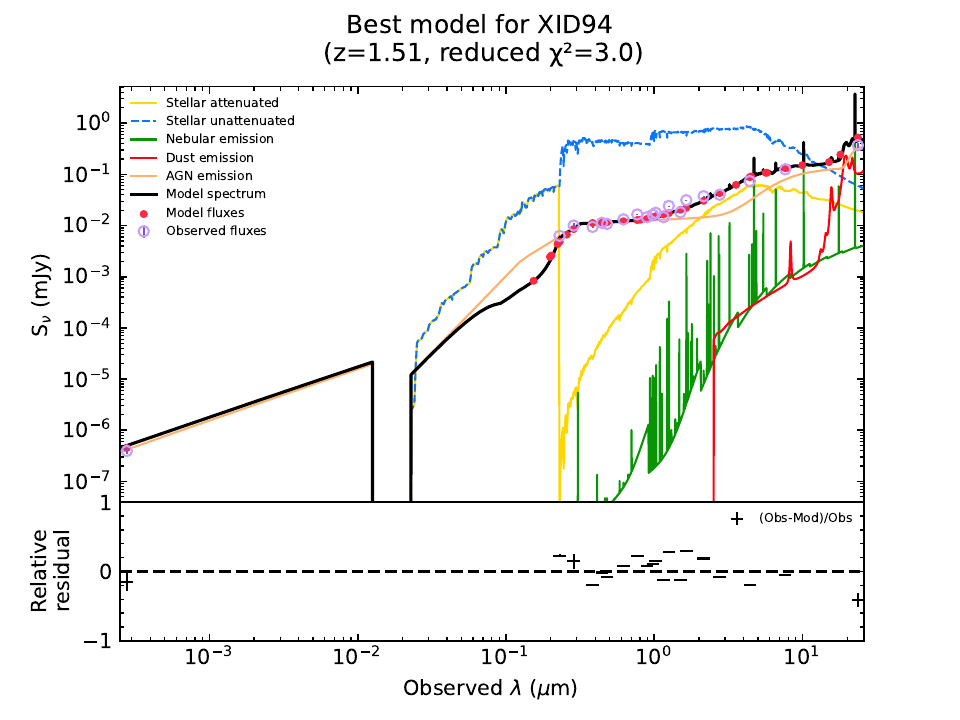}&  
\includegraphics[scale=0.30]{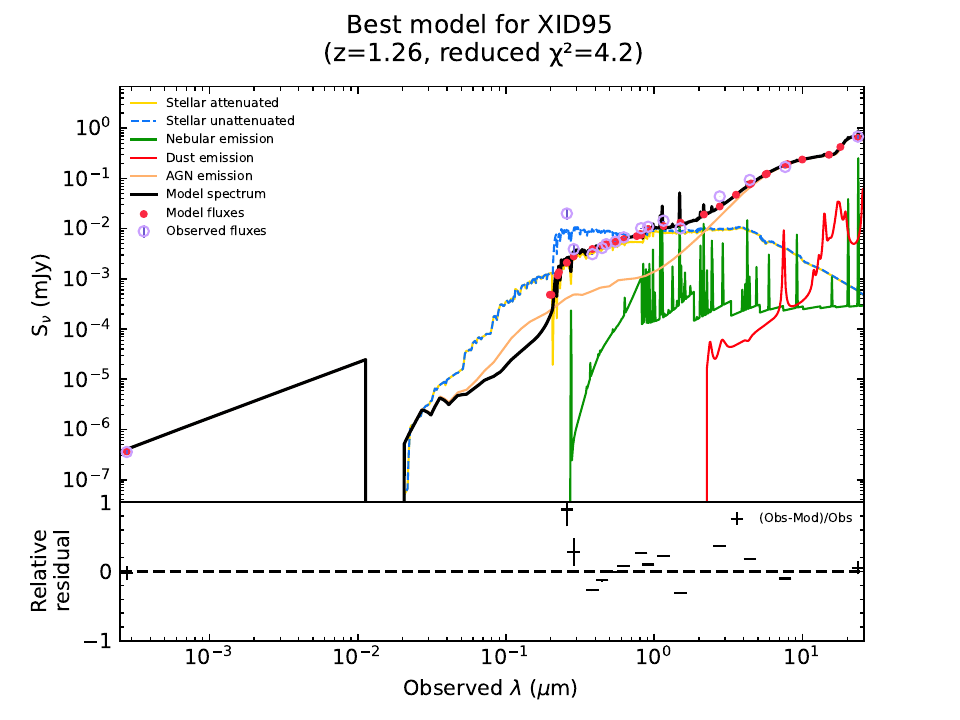}\\  
\includegraphics[scale=0.30]{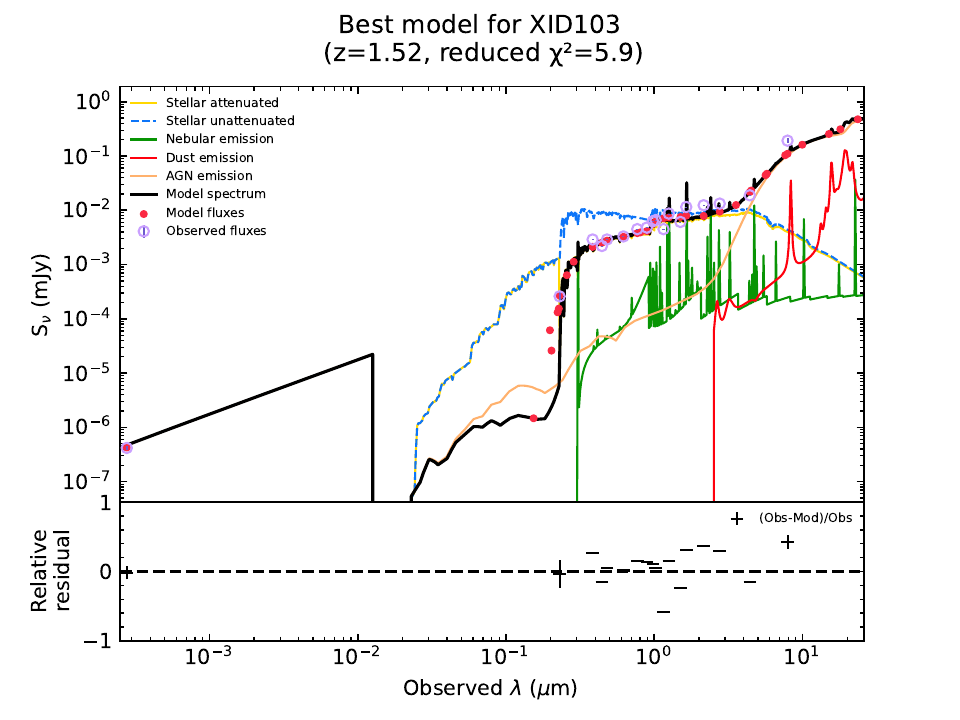}&  
\includegraphics[scale=0.30]{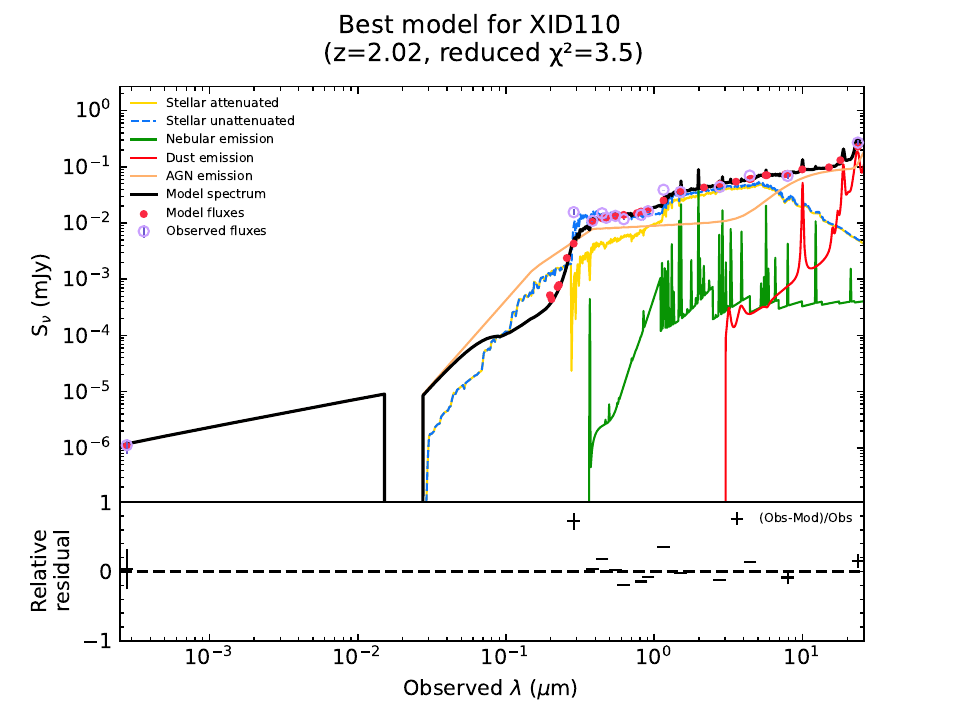}&  
\includegraphics[scale=0.30]{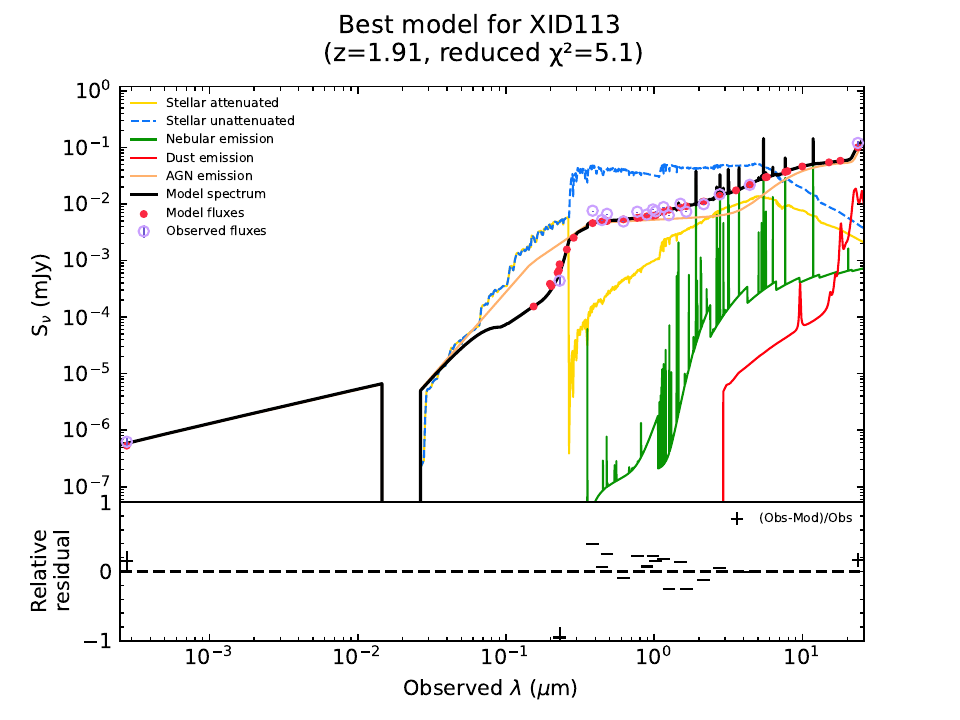}\\  
\includegraphics[scale=0.30]{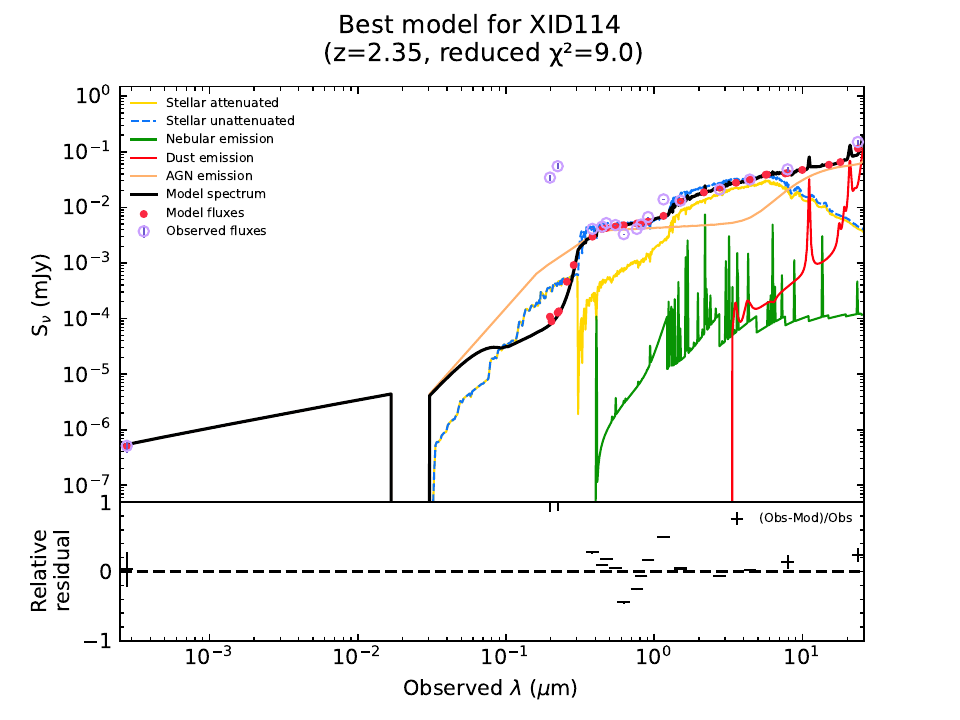}&  
\includegraphics[scale=0.30]{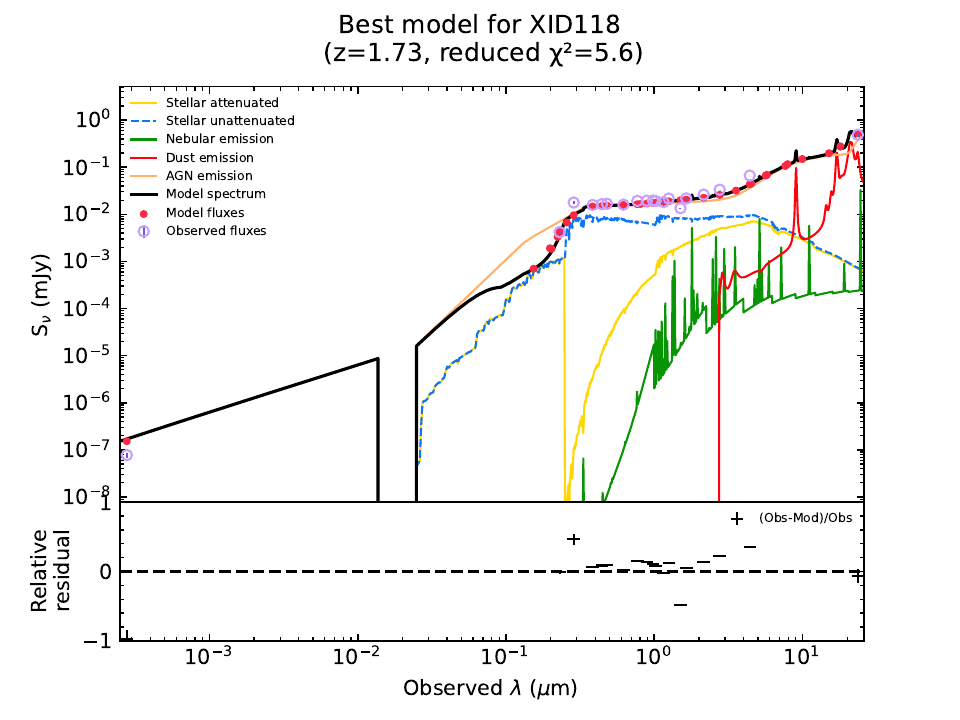}&  
\includegraphics[scale=0.30]{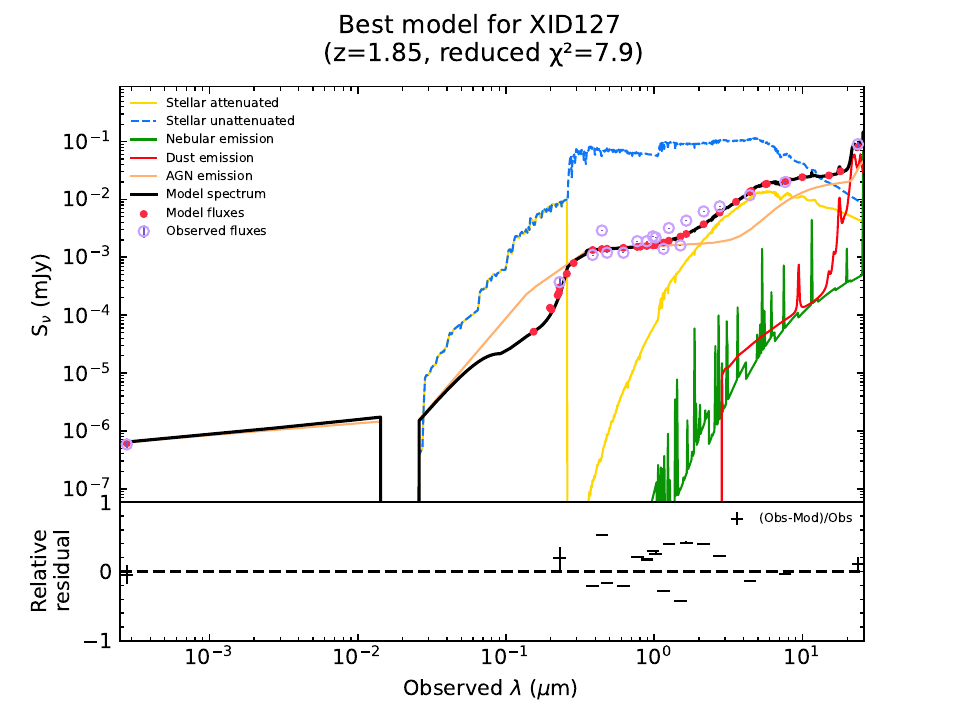}\\  
\includegraphics[scale=0.30]{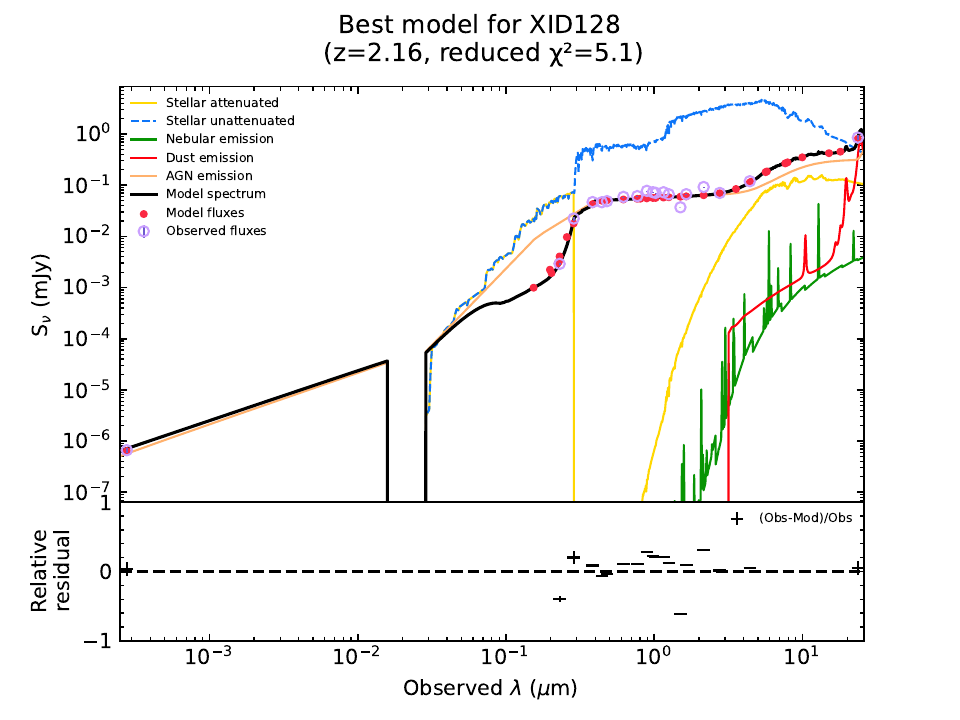}&  
\includegraphics[scale=0.30]{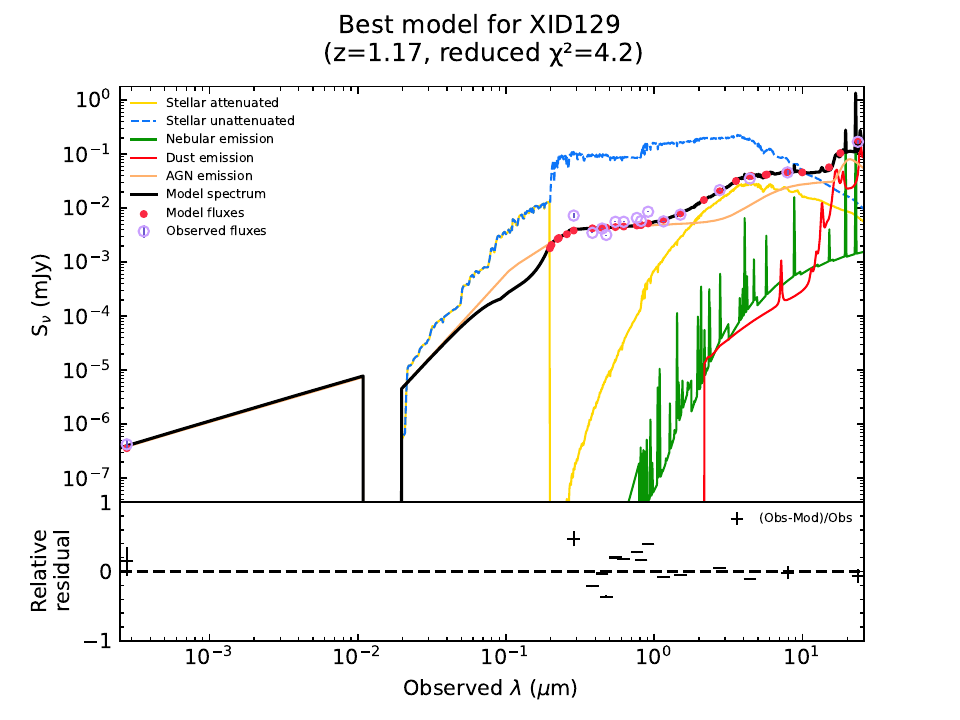}& 
\includegraphics[scale=0.30]{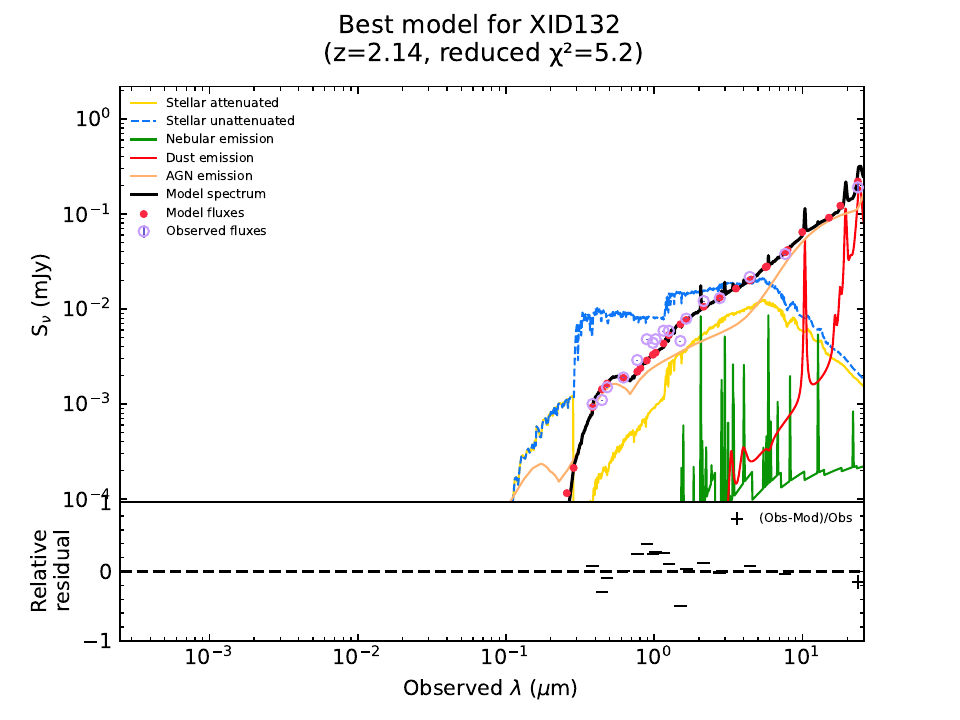}\\ 
\includegraphics[scale=0.30]{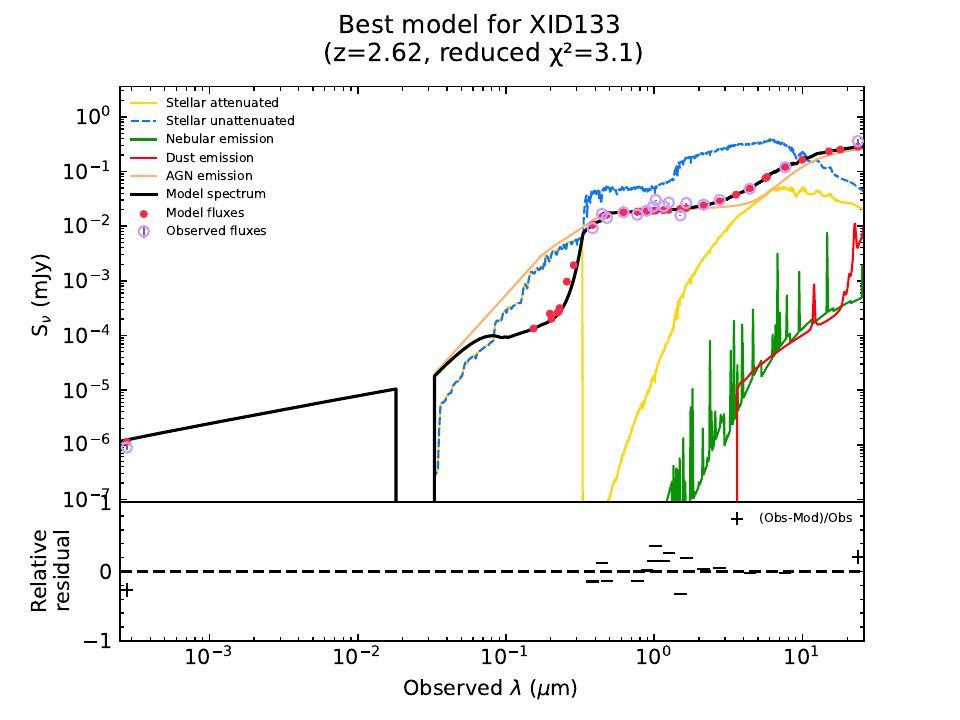}& 
\includegraphics[scale=0.30]{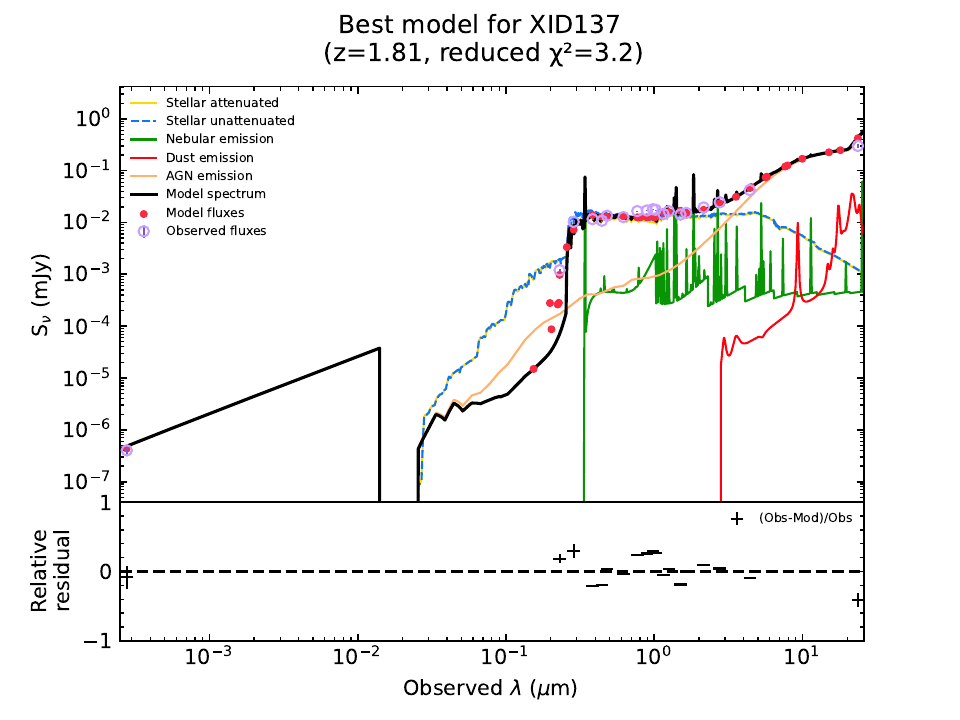}& 
\includegraphics[scale=0.30]{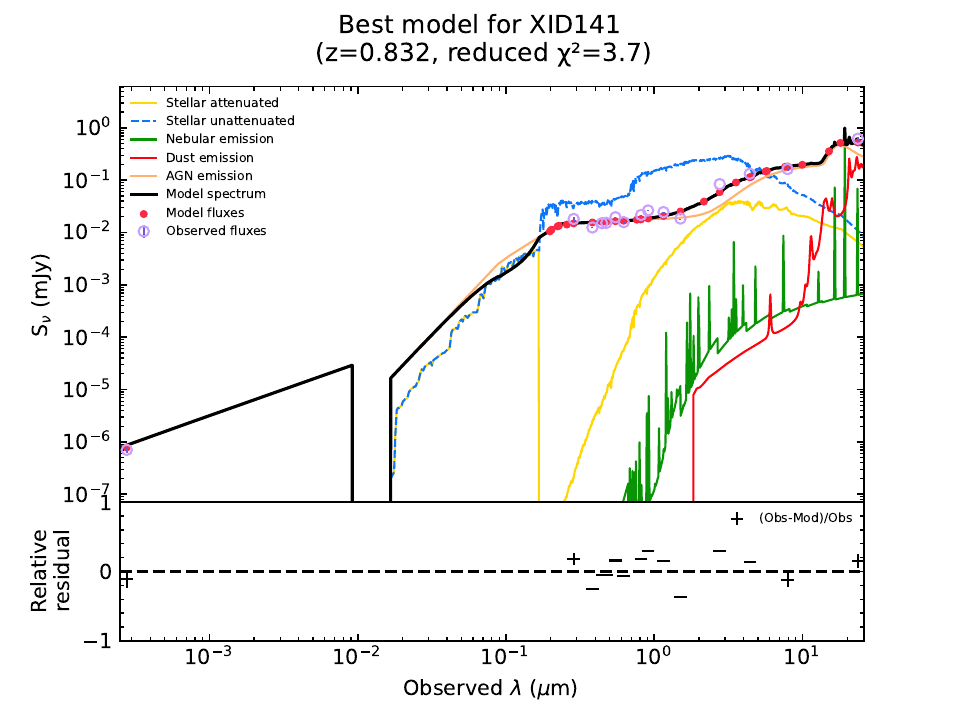}\\ 
\includegraphics[scale=0.30]{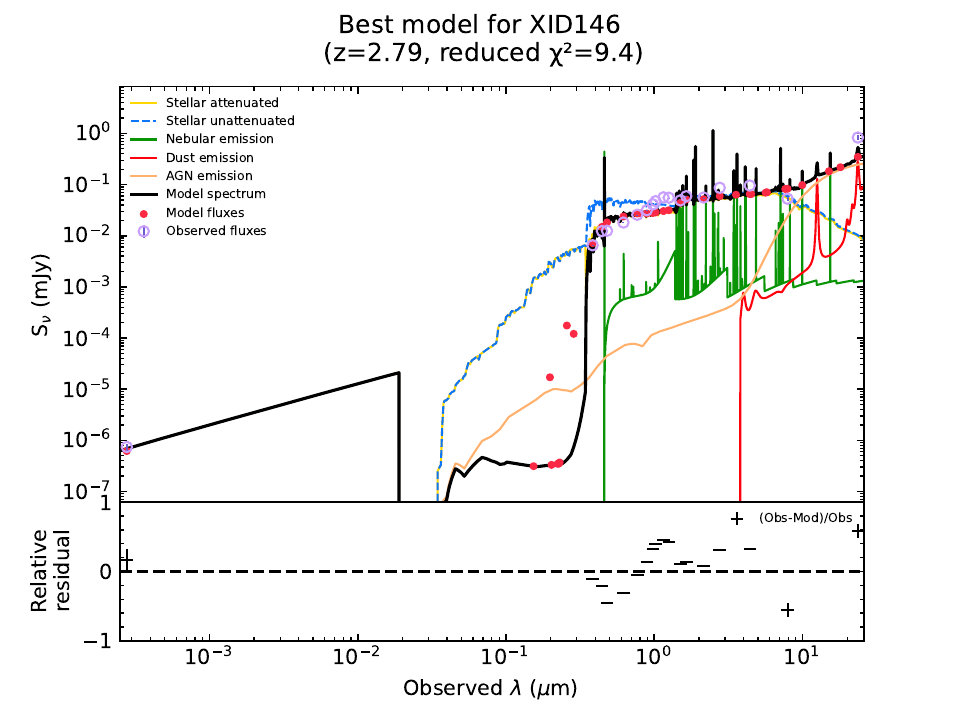}& 
\includegraphics[scale=0.30]{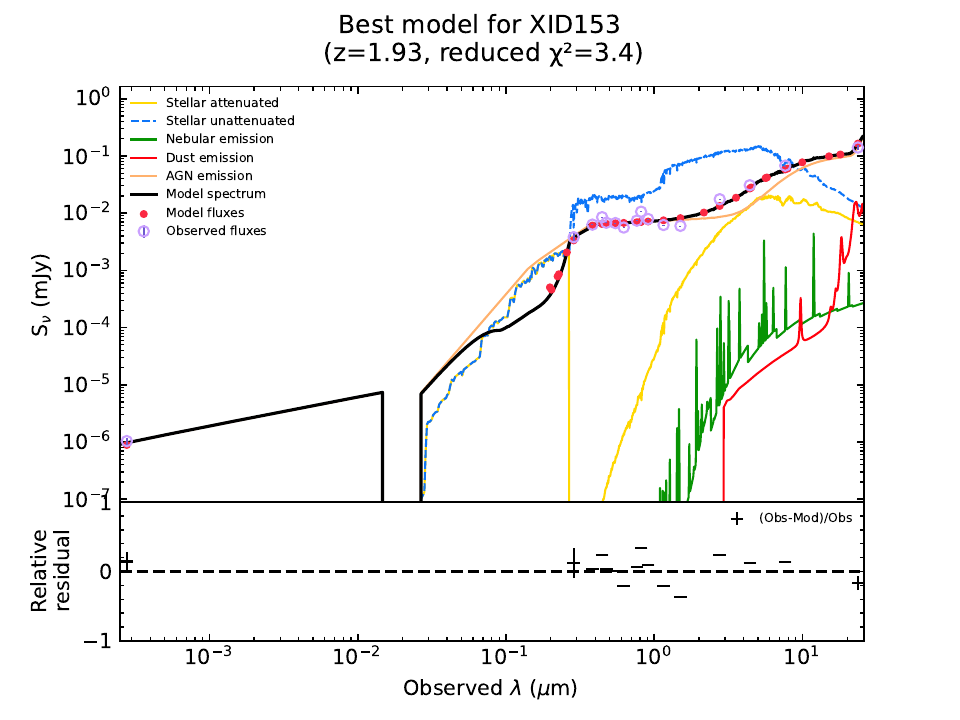}& 
\includegraphics[scale=0.30]{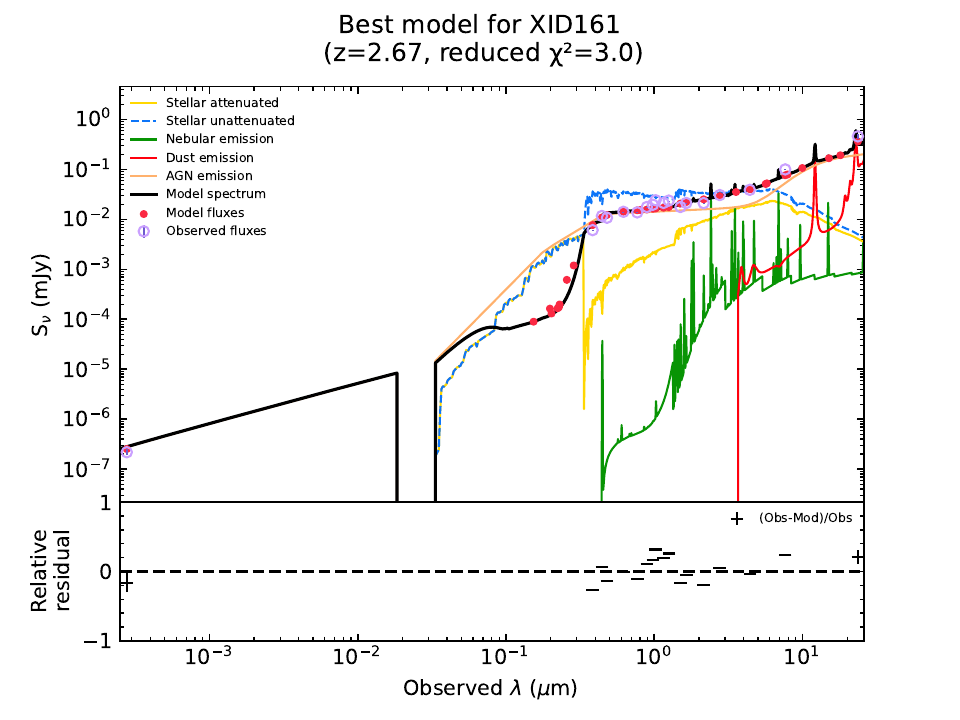}\\ 
\end{array}$
\end{center}
\caption{The best-fitting SEDs of S1, the AGN with JWST NIRCam coverage.}
\end{figure*}

\renewcommand{\thefigure}{\Alph{section}\arabic{figure}(Cont.)}
\addtocounter{figure}{-1}
\begin{figure*}
\begin{center}$
\begin{array}{lll}
\includegraphics[scale=0.30]{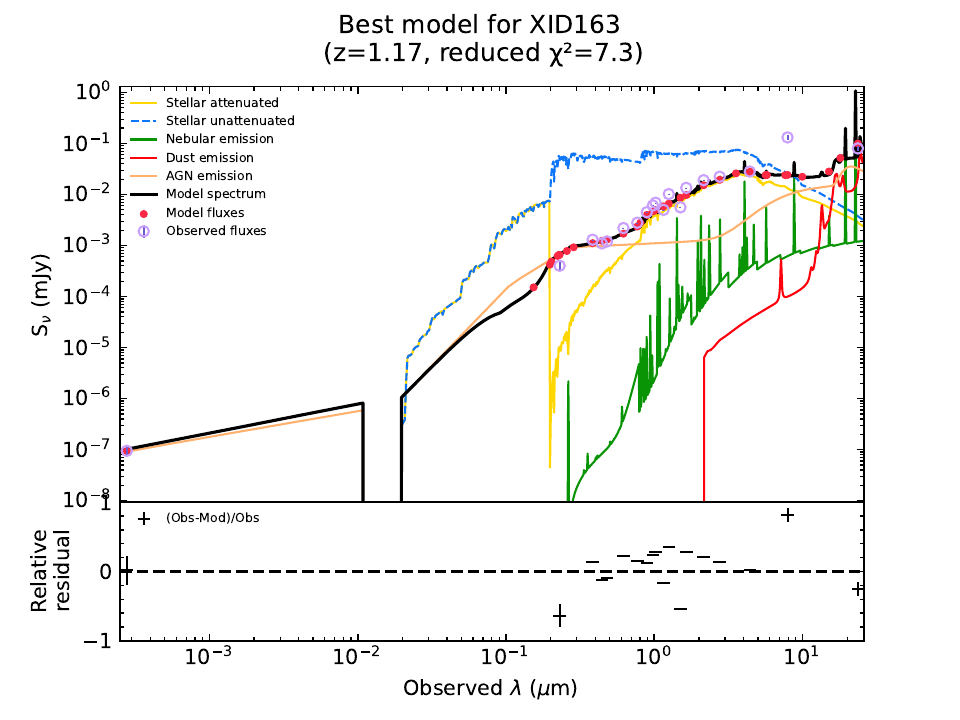}& 
\includegraphics[scale=0.30]{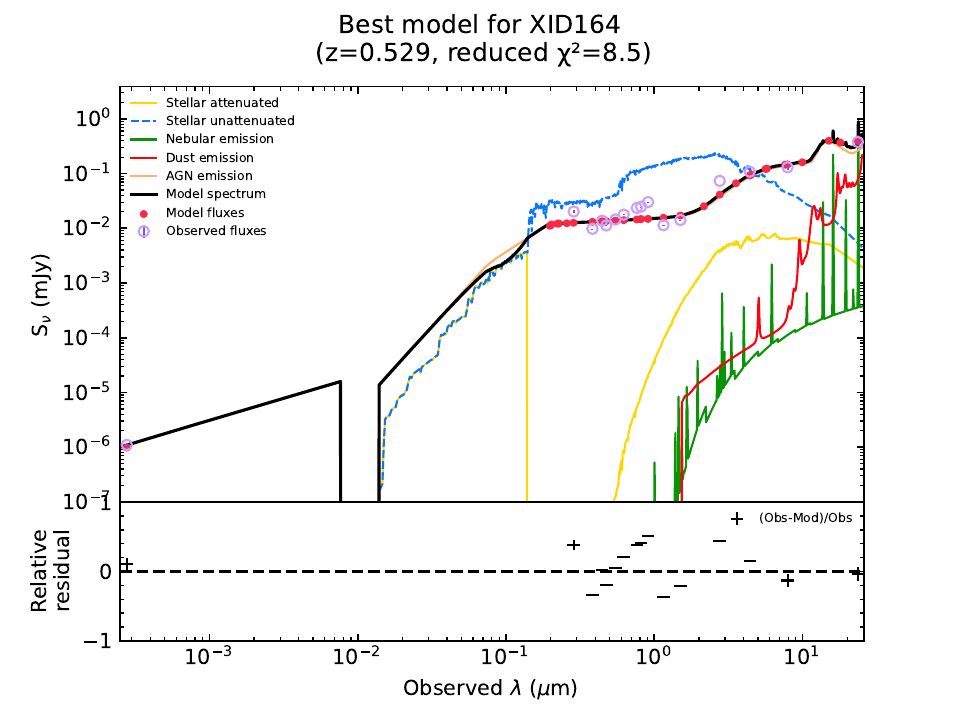}& 
\includegraphics[scale=0.30]{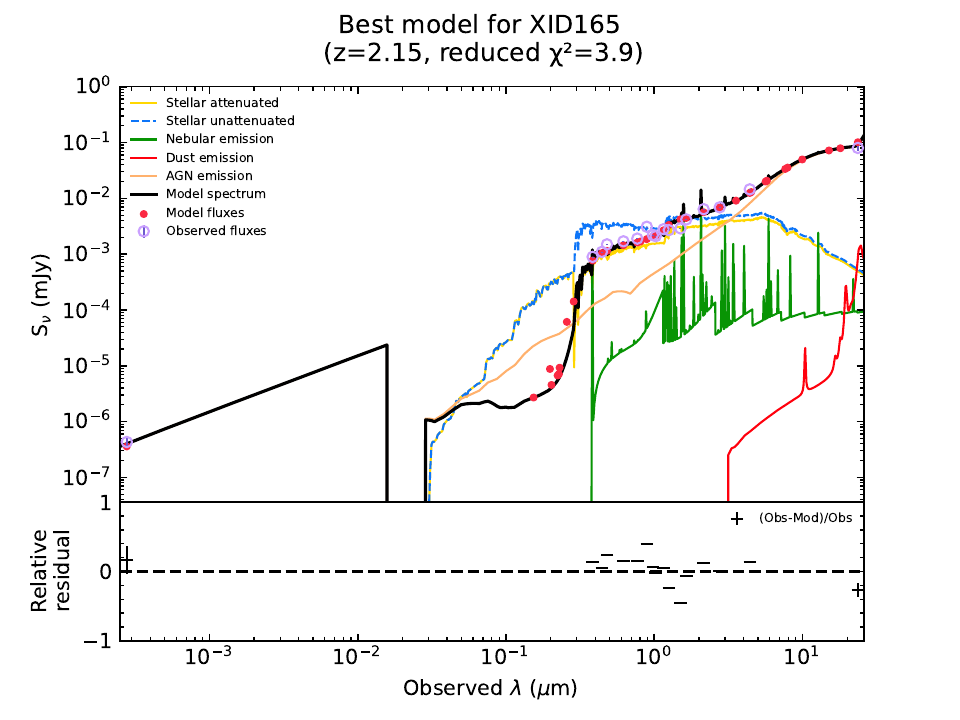}\\ 
\includegraphics[scale=0.30]{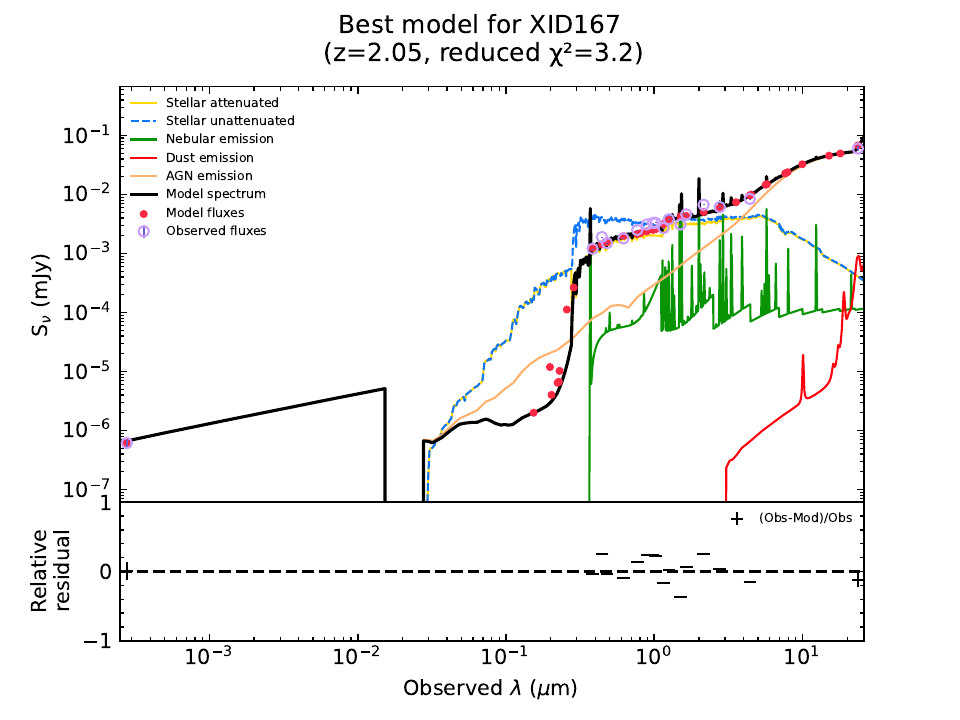}& 
\includegraphics[scale=0.30]{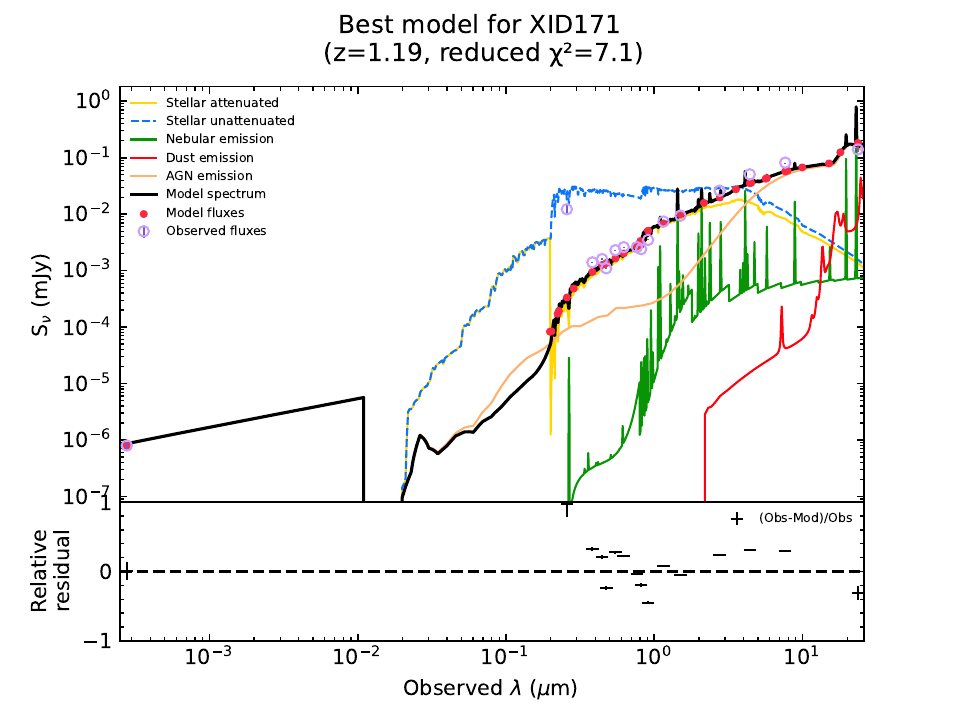}& 
\includegraphics[scale=0.30]{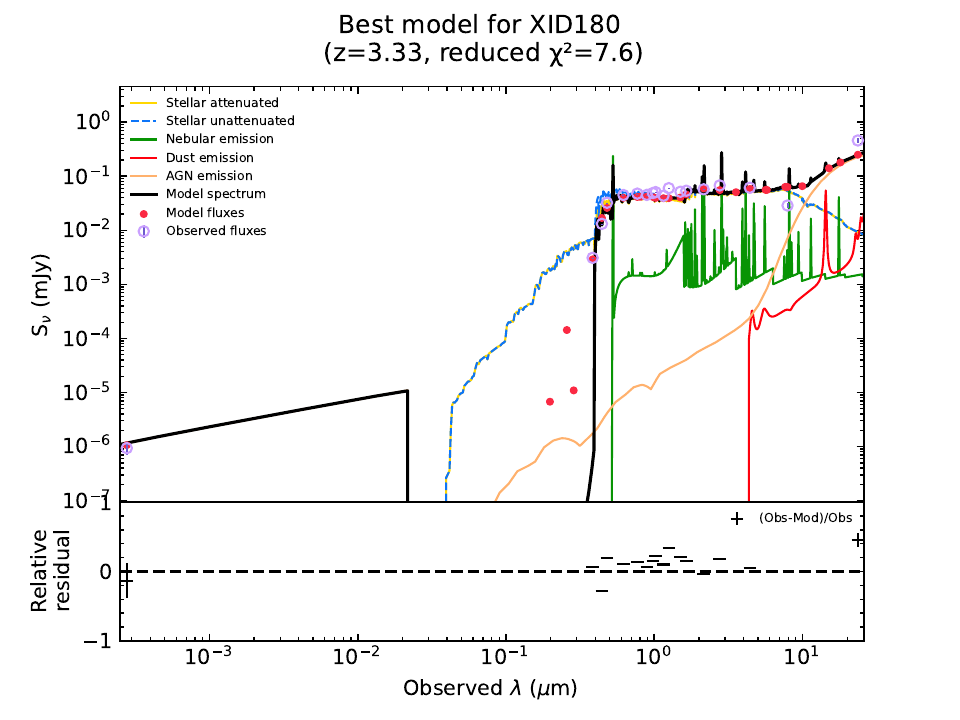}\\ 
\includegraphics[scale=0.30]{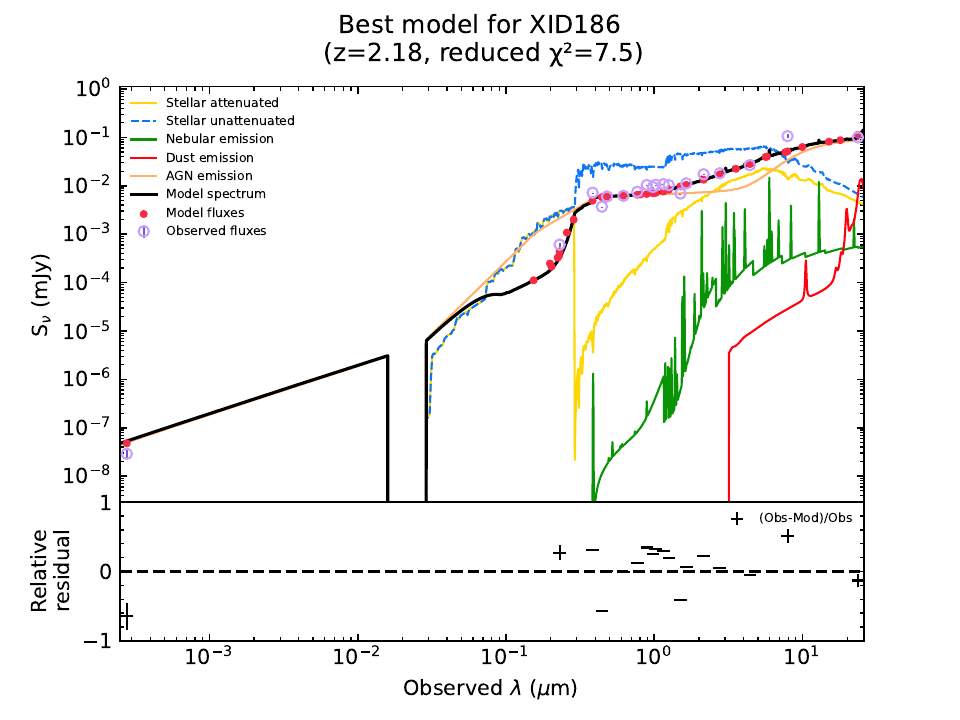}& 
\includegraphics[scale=0.30]{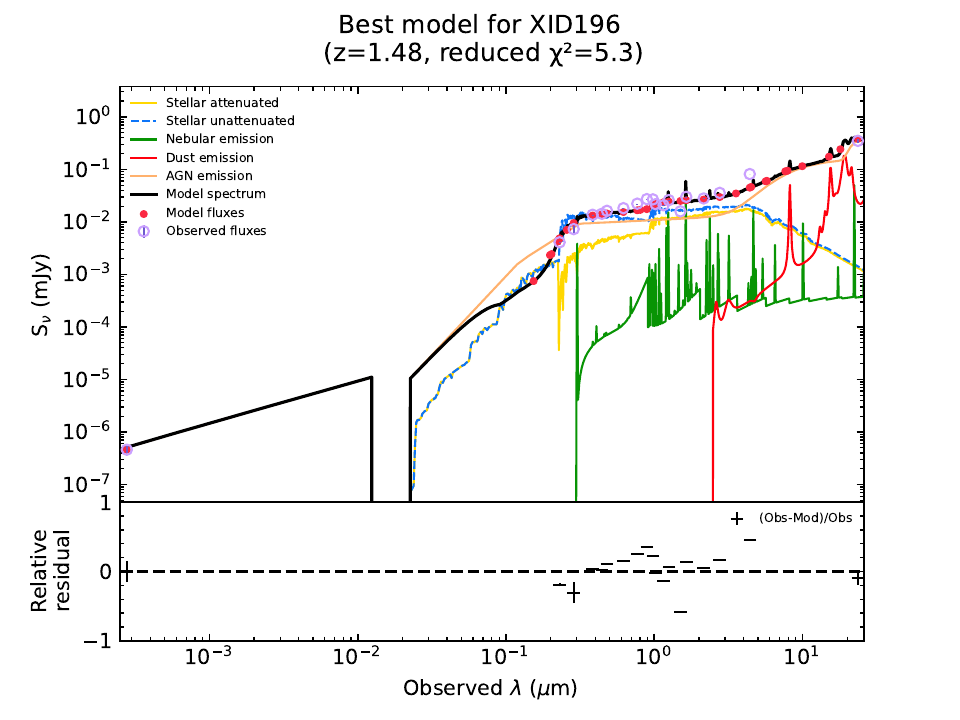}& 
\includegraphics[scale=0.30]{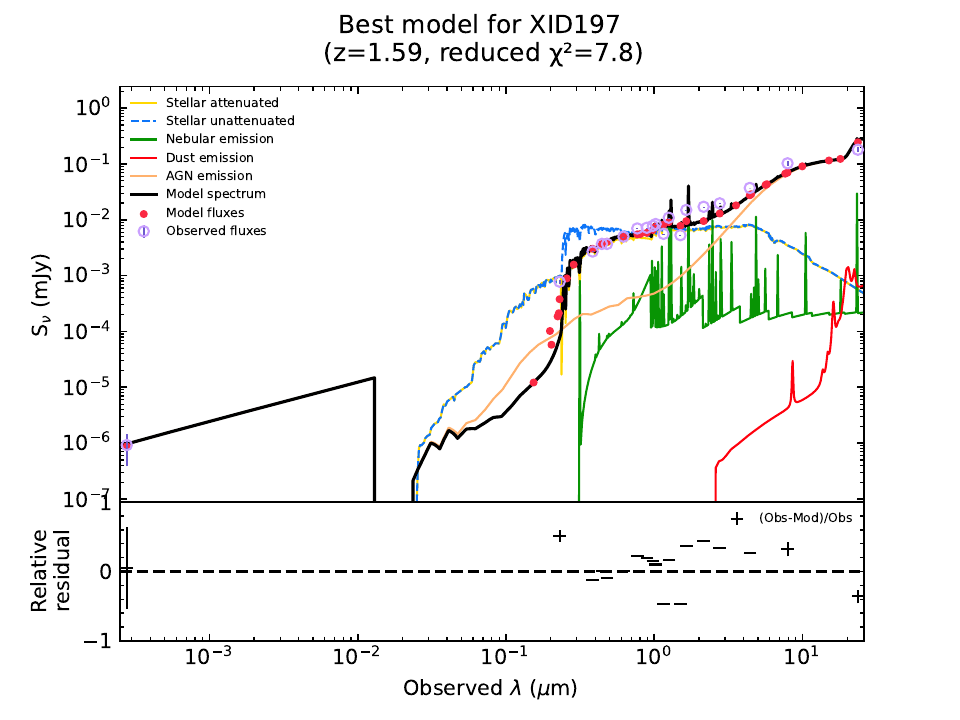}\\ 
\includegraphics[scale=0.30]{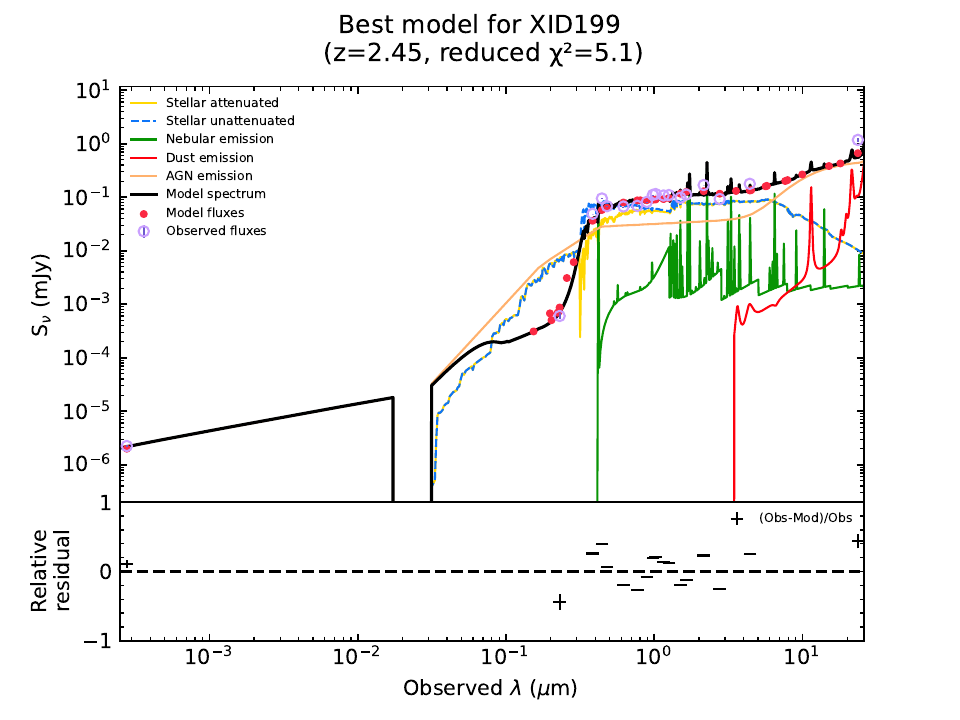}& 
\includegraphics[scale=0.30]{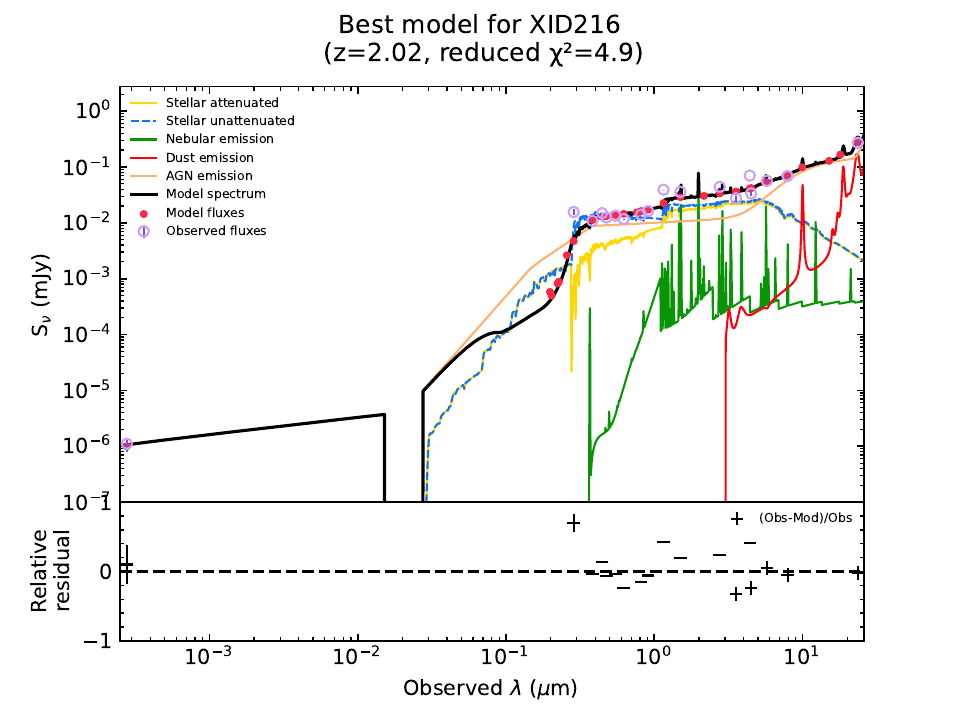}& 
\includegraphics[scale=0.30]{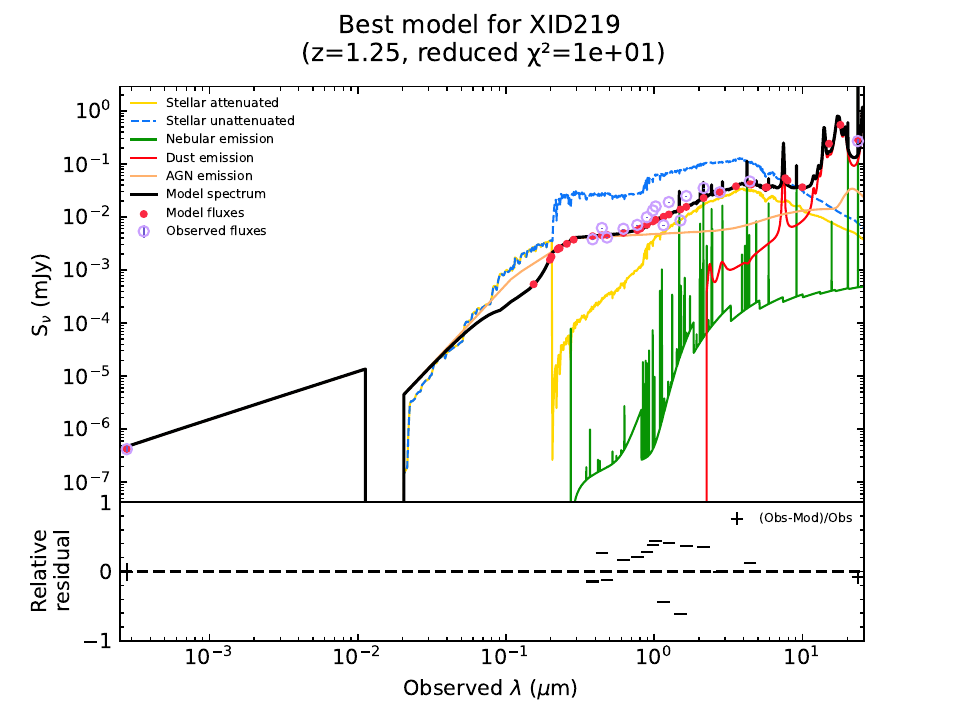}\\ 
\includegraphics[scale=0.30]{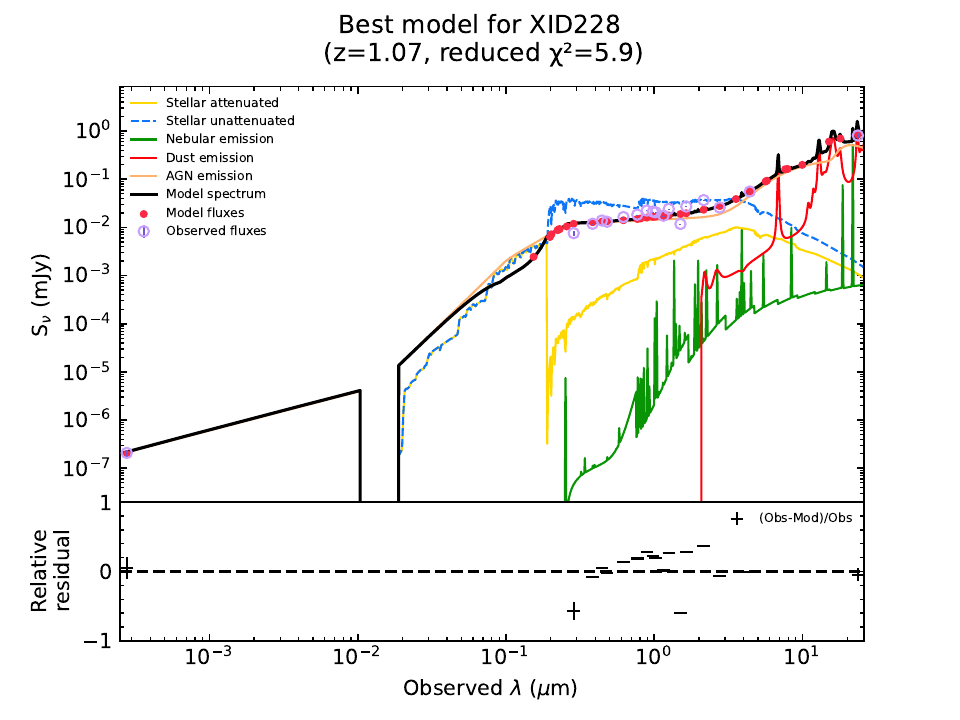}& 
\includegraphics[scale=0.30]{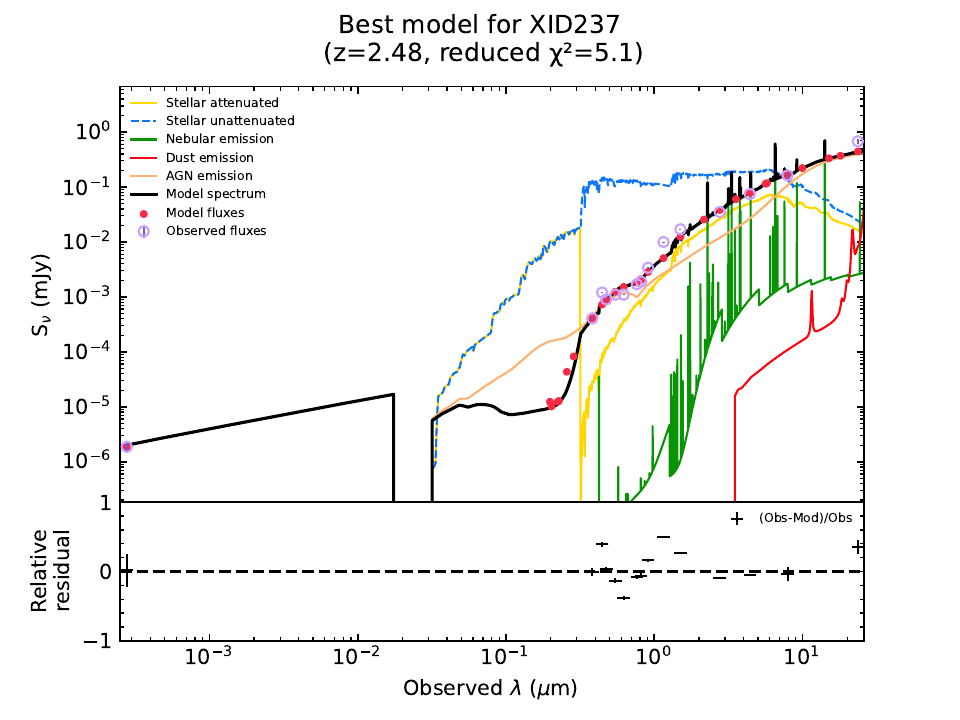}& 
\includegraphics[scale=0.30]{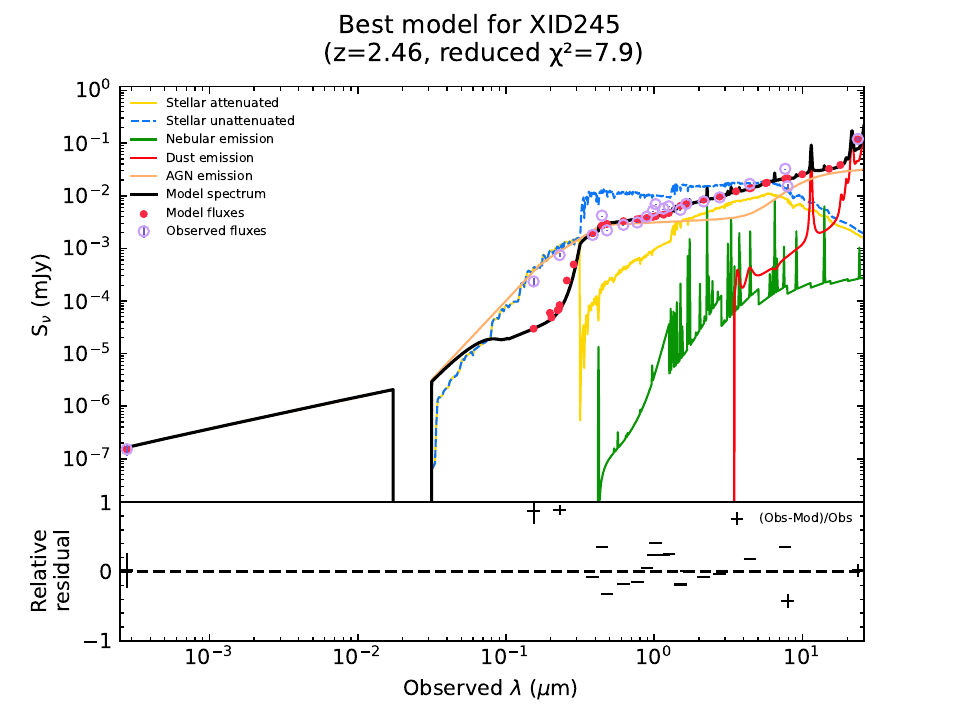}\\ 
\includegraphics[scale=0.30]{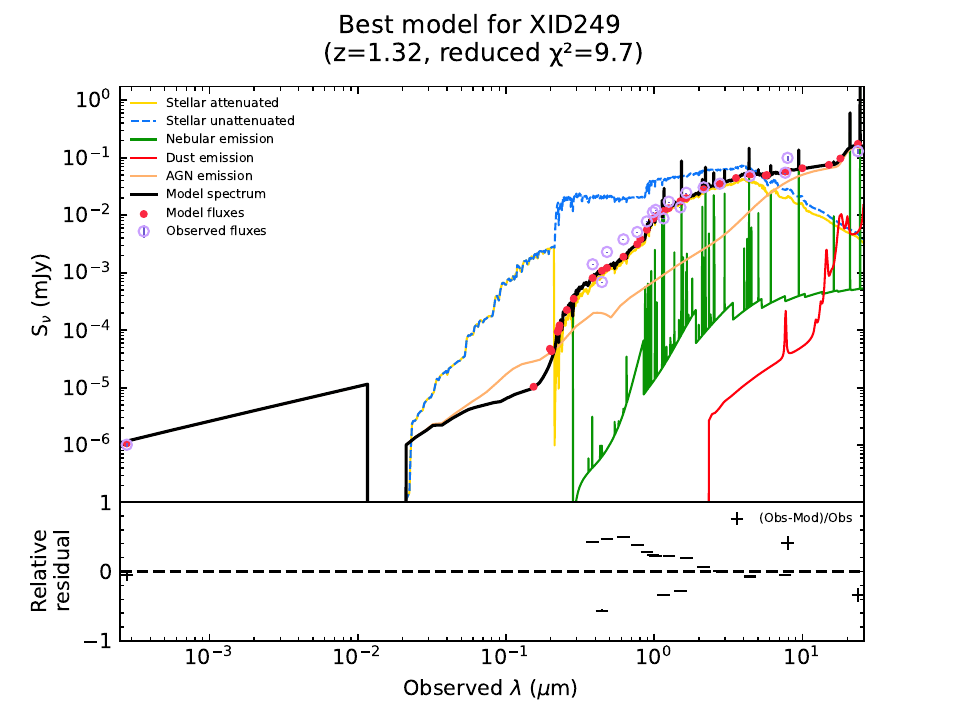}& 
\includegraphics[scale=0.30]{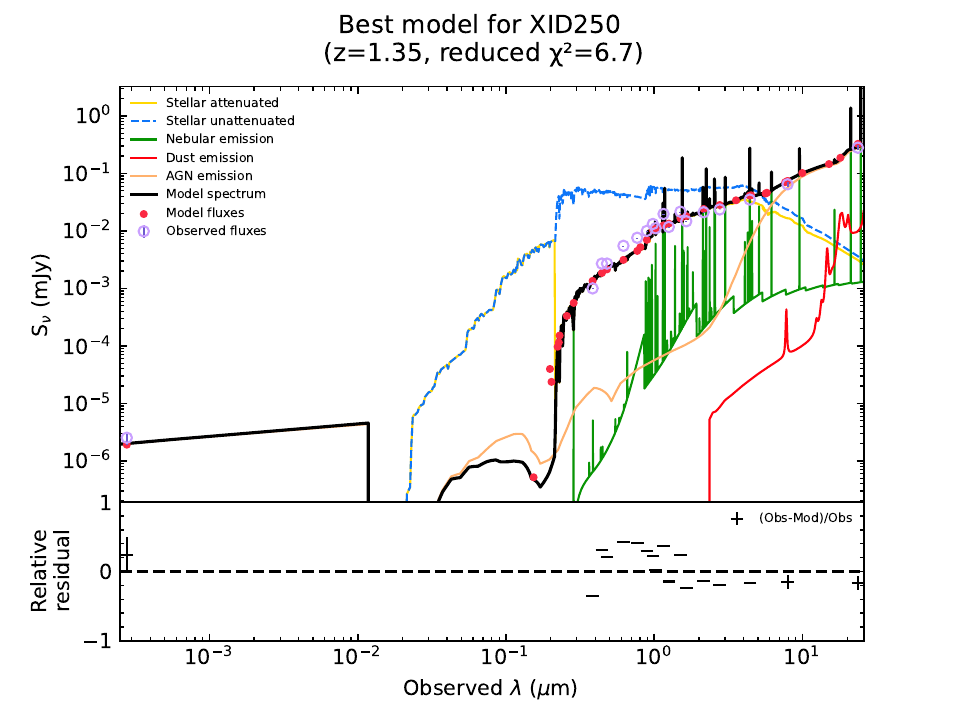}& 
\end{array}$
\end{center}
\caption{The best-fitting SEDs of S1, the AGN with JWST NIRCam coverage.}
\end{figure*}

\renewcommand{\thefigure}{\Alph{section}\arabic{figure}}
\begin{figure*}
\begin{center}$
\begin{array}{lll}

\includegraphics[scale=0.30]{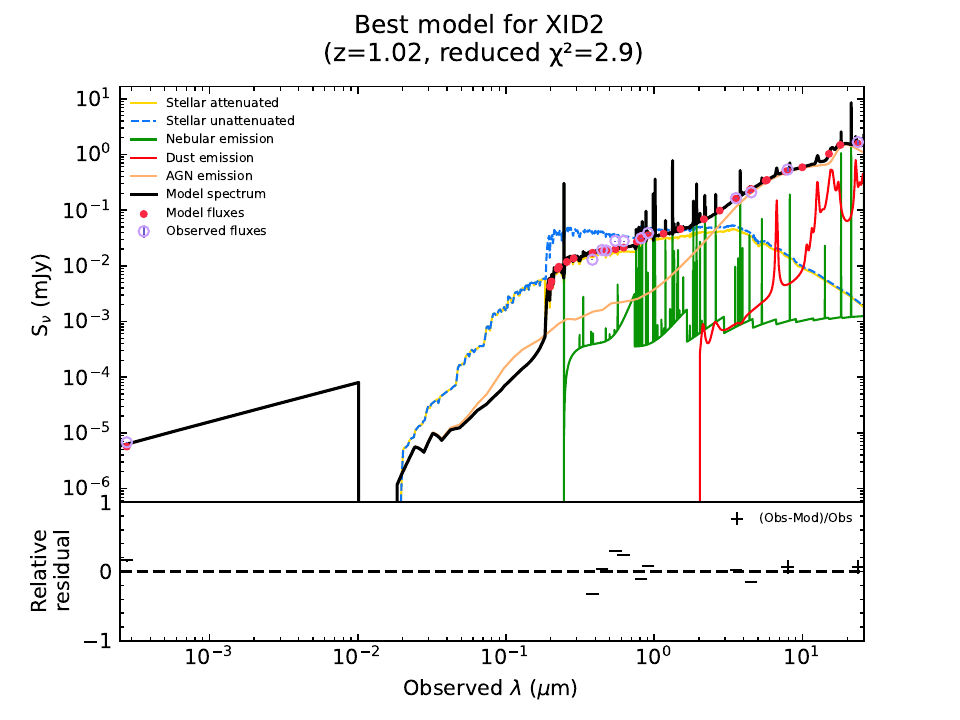}&
\includegraphics[scale=0.30]{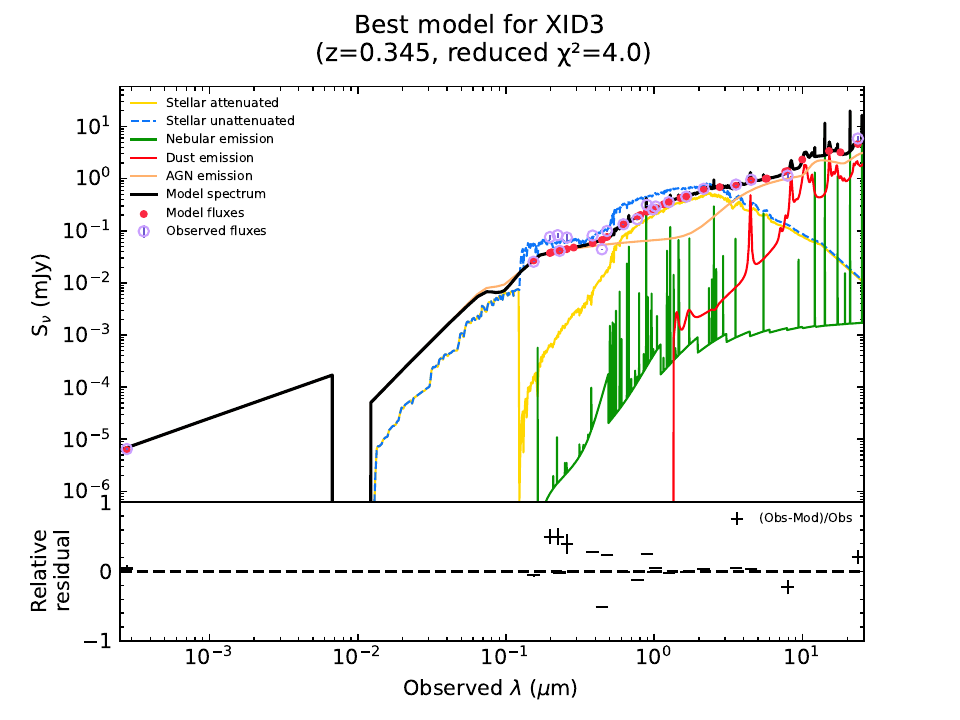}&
\includegraphics[scale=0.30]{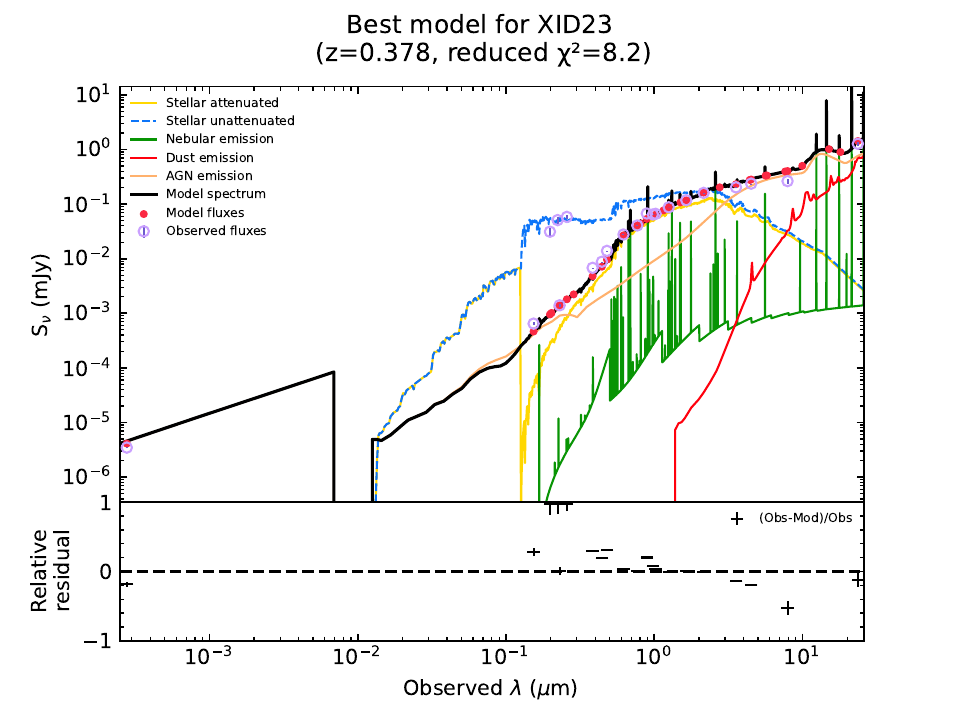}\\
\includegraphics[scale=0.30]{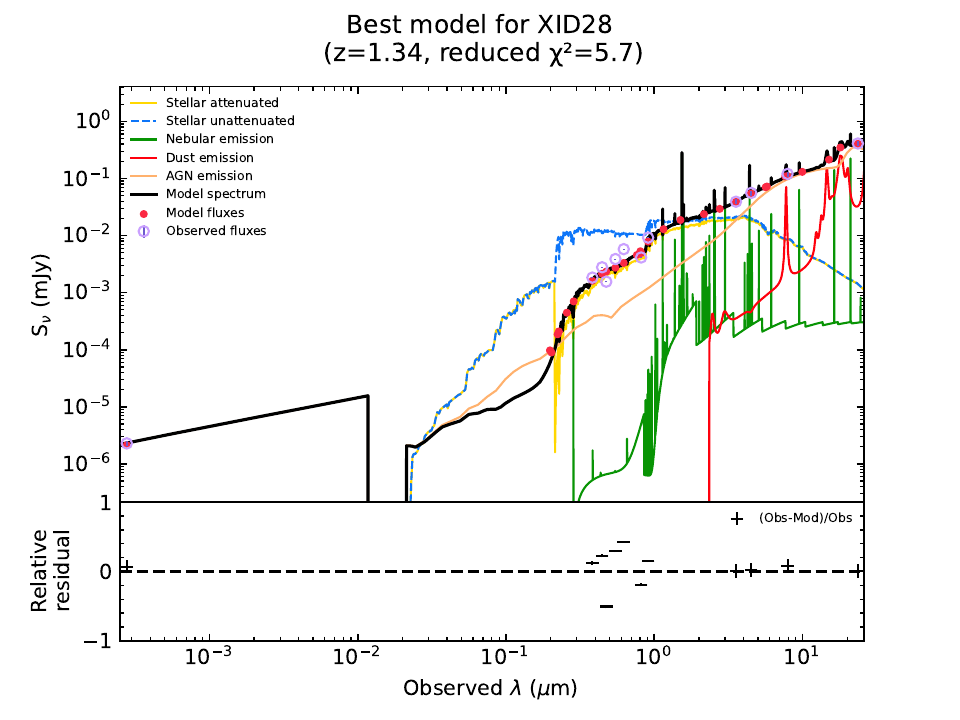}&
\includegraphics[scale=0.30]{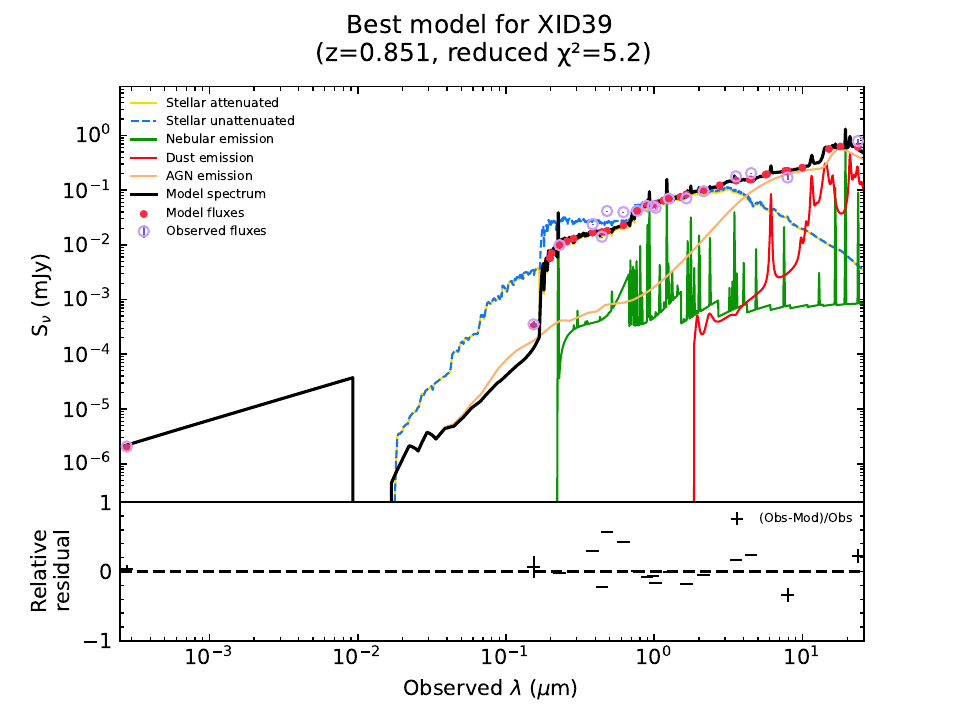}&
\includegraphics[scale=0.30]{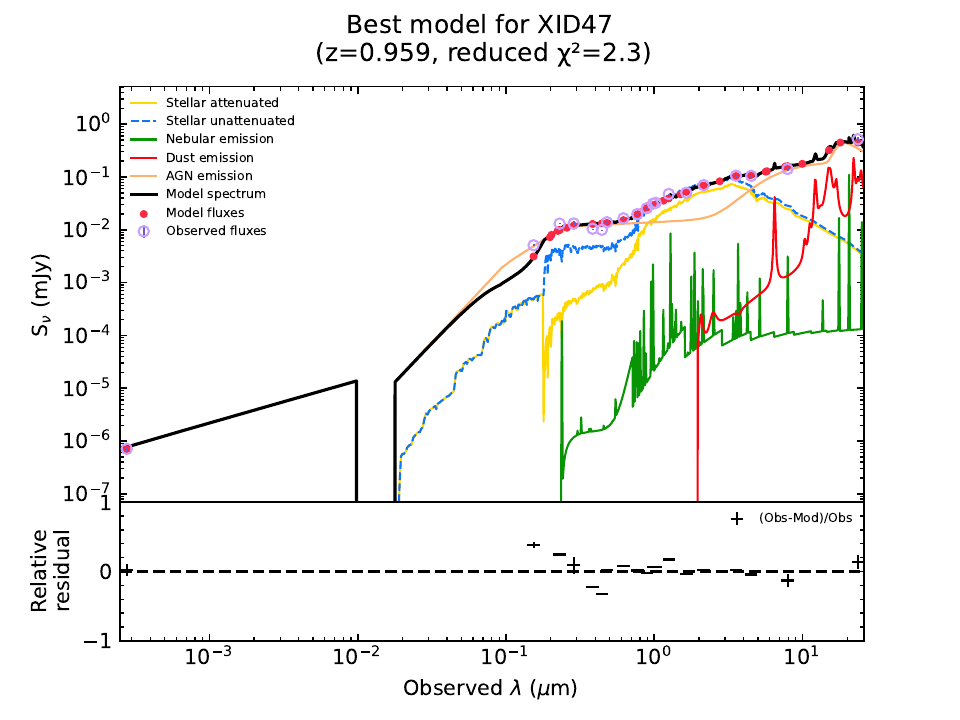}\\
\includegraphics[scale=0.30]{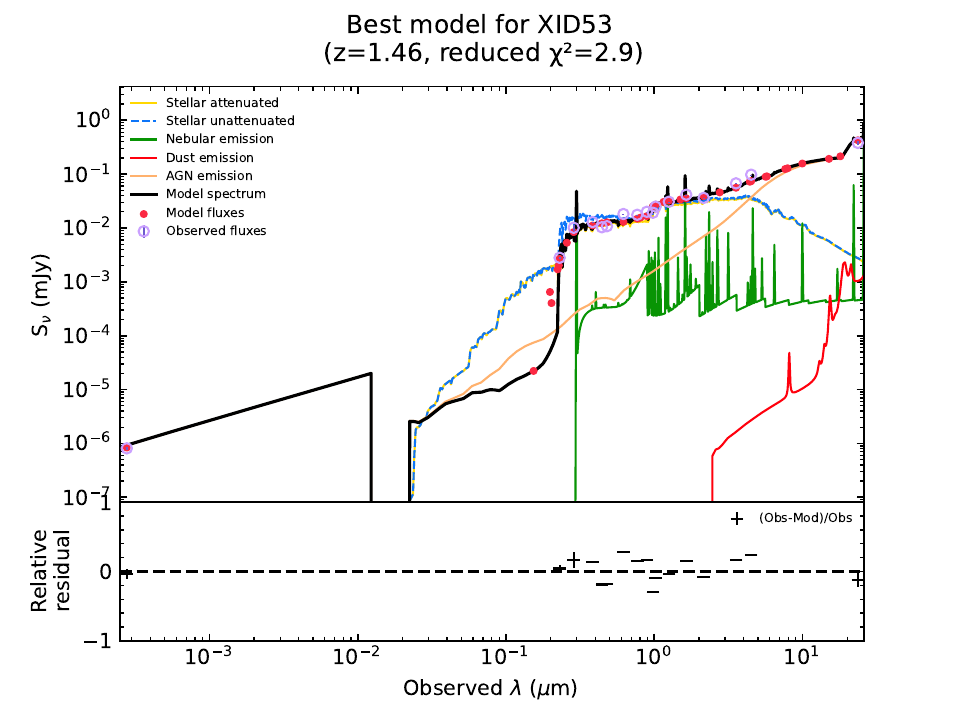}& 
\includegraphics[scale=0.30]{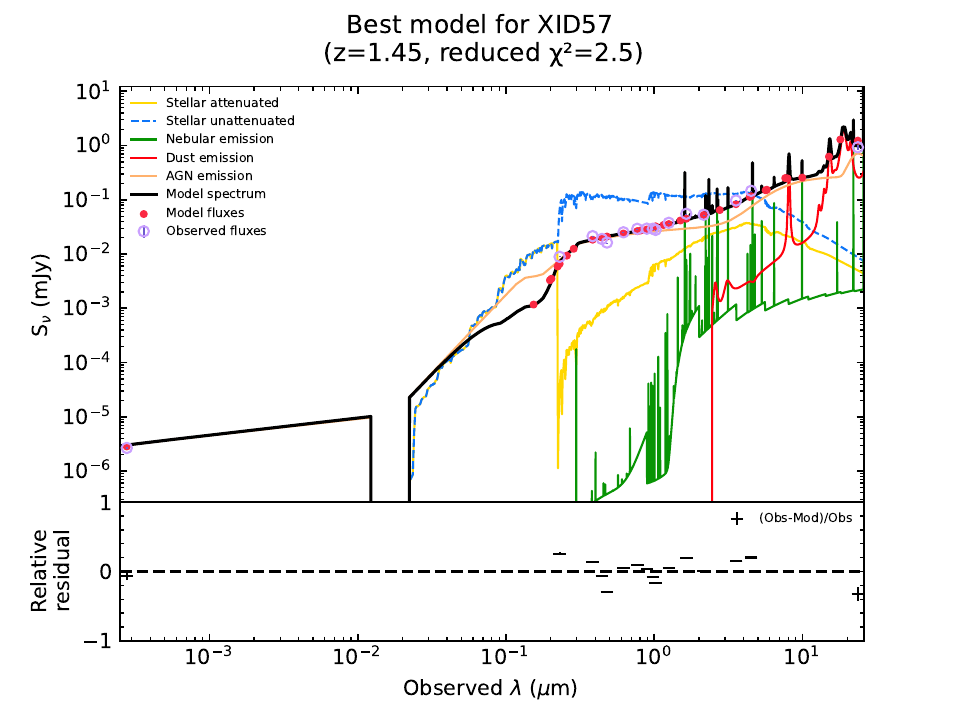}& 
\includegraphics[scale=0.30]{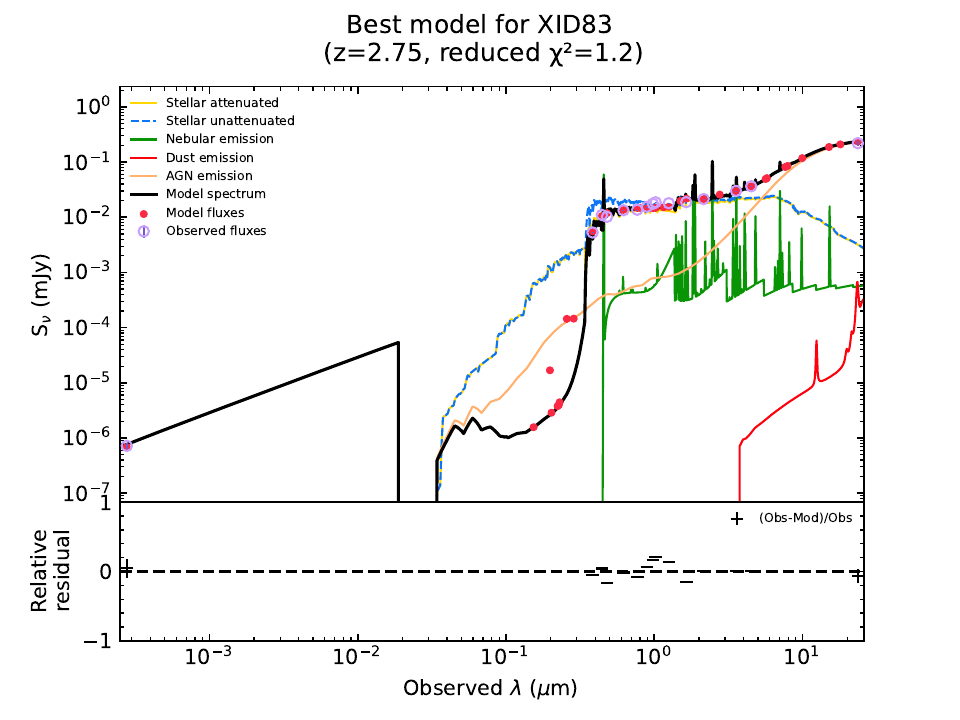}\\ 
\includegraphics[scale=0.30]{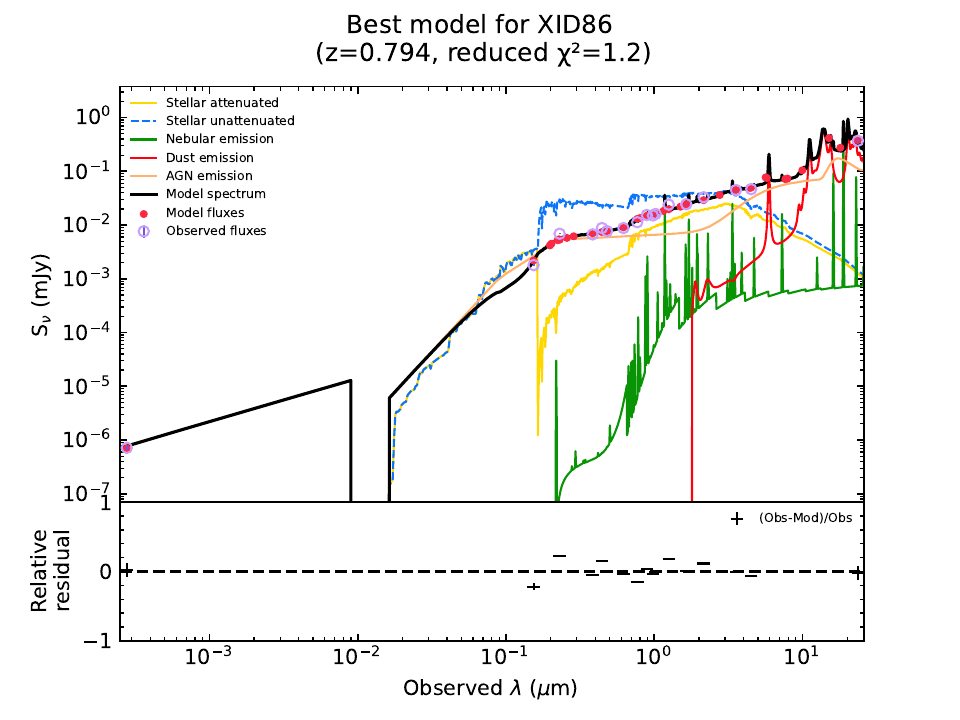}& 
\includegraphics[scale=0.30]{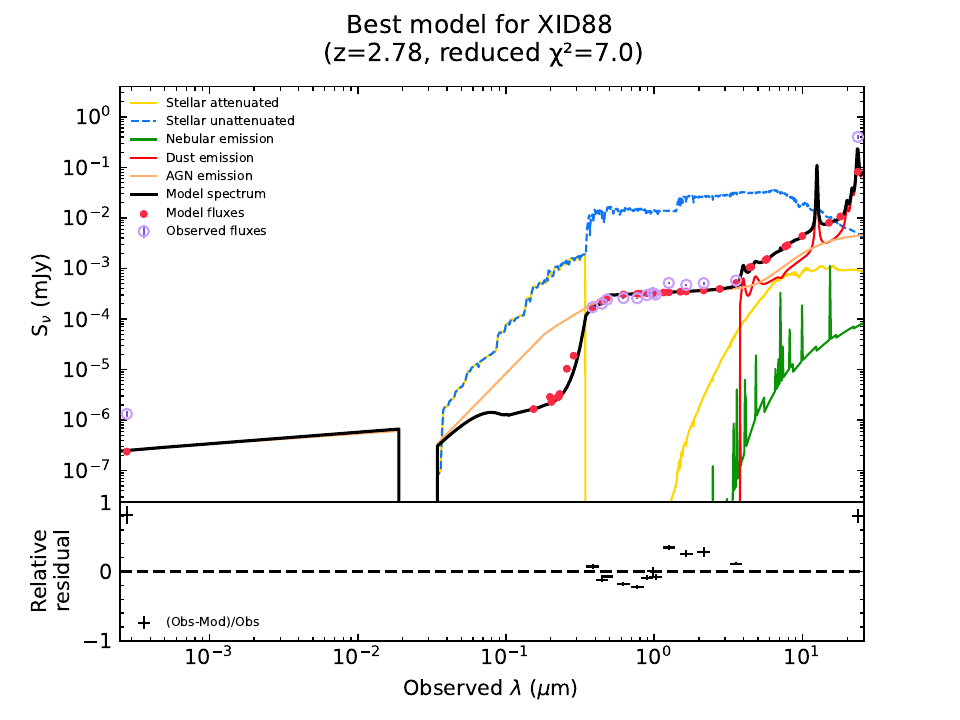}& 
\includegraphics[scale=0.30]{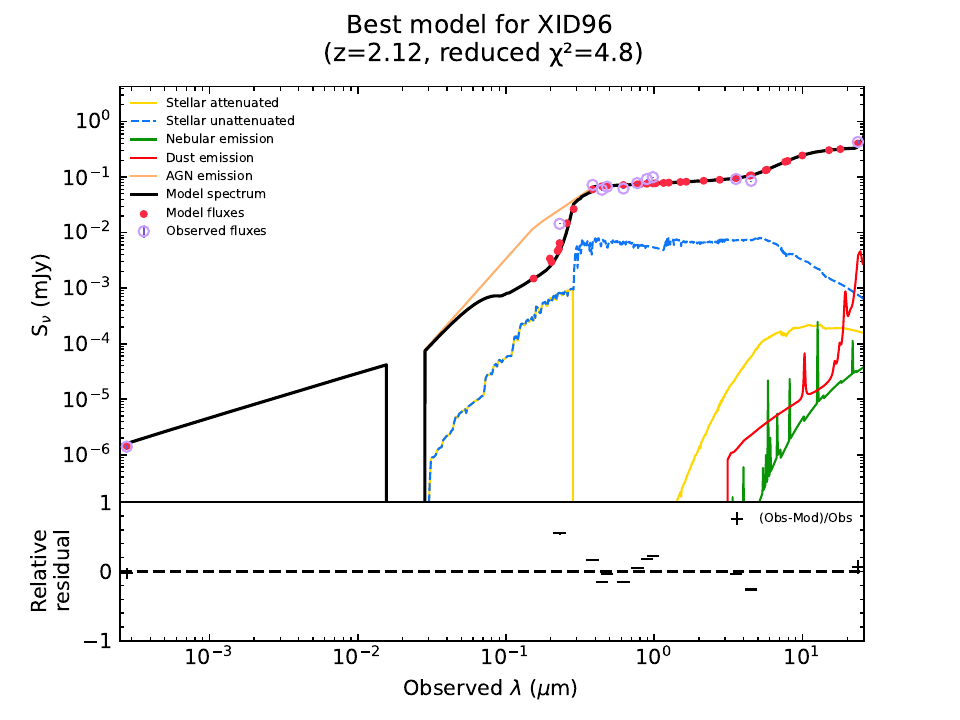}\\ 
\includegraphics[scale=0.30]{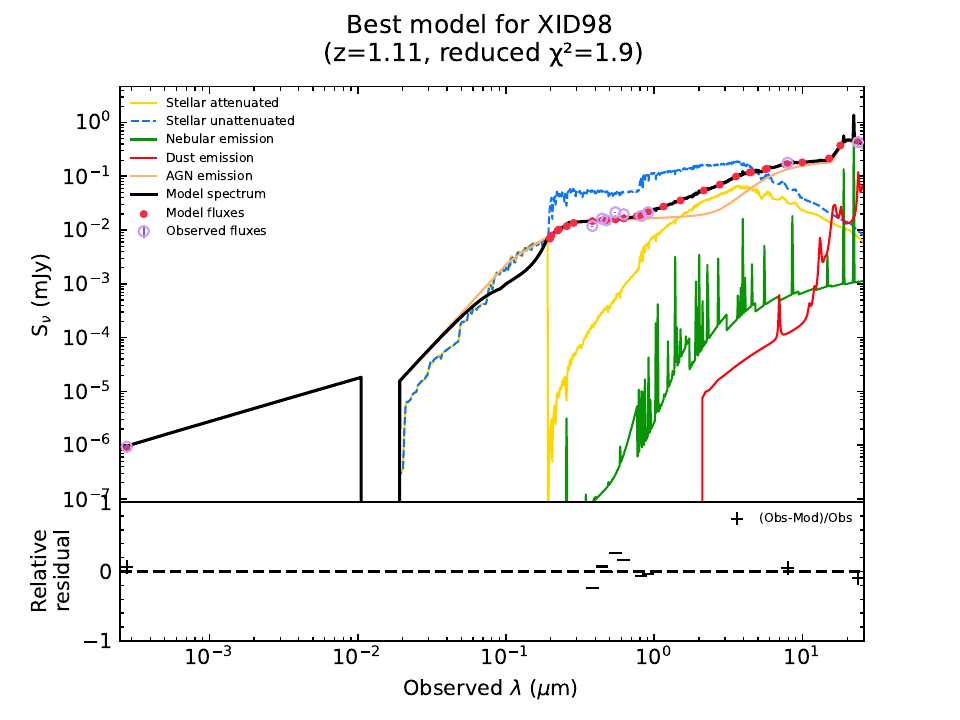}& 
\includegraphics[scale=0.30]{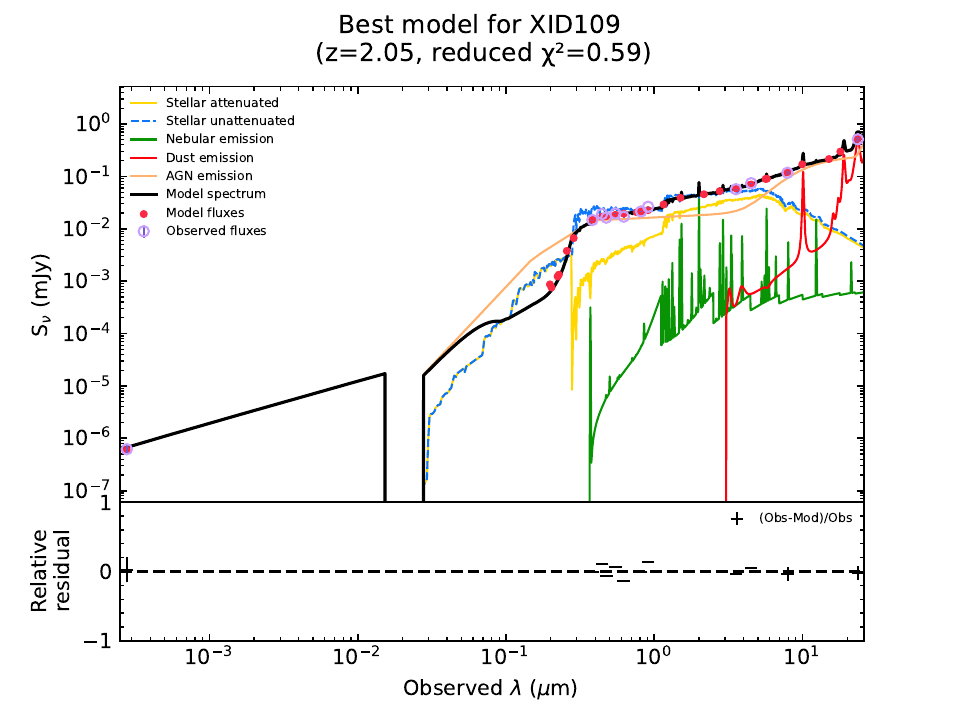}& 
\includegraphics[scale=0.30]{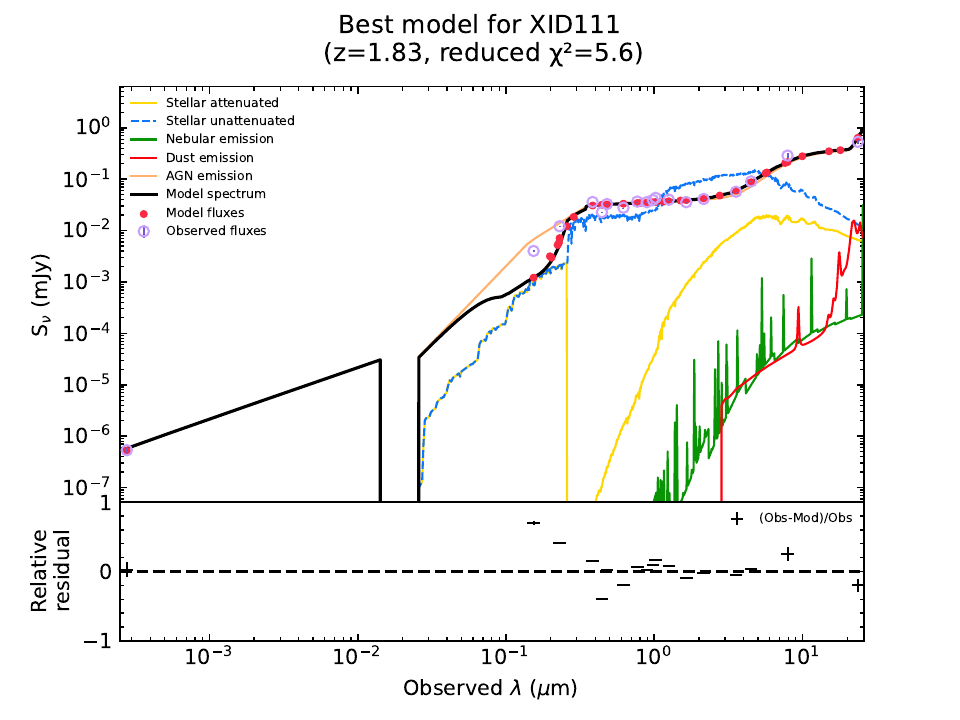}\\ 
\includegraphics[scale=0.30]{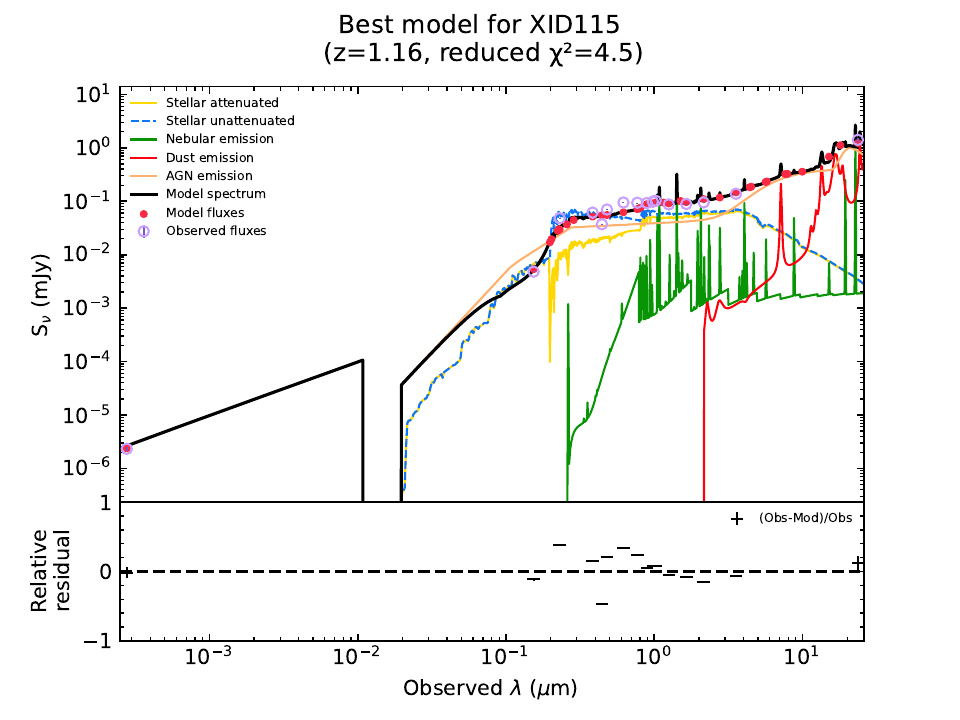}& 
\includegraphics[scale=0.30]{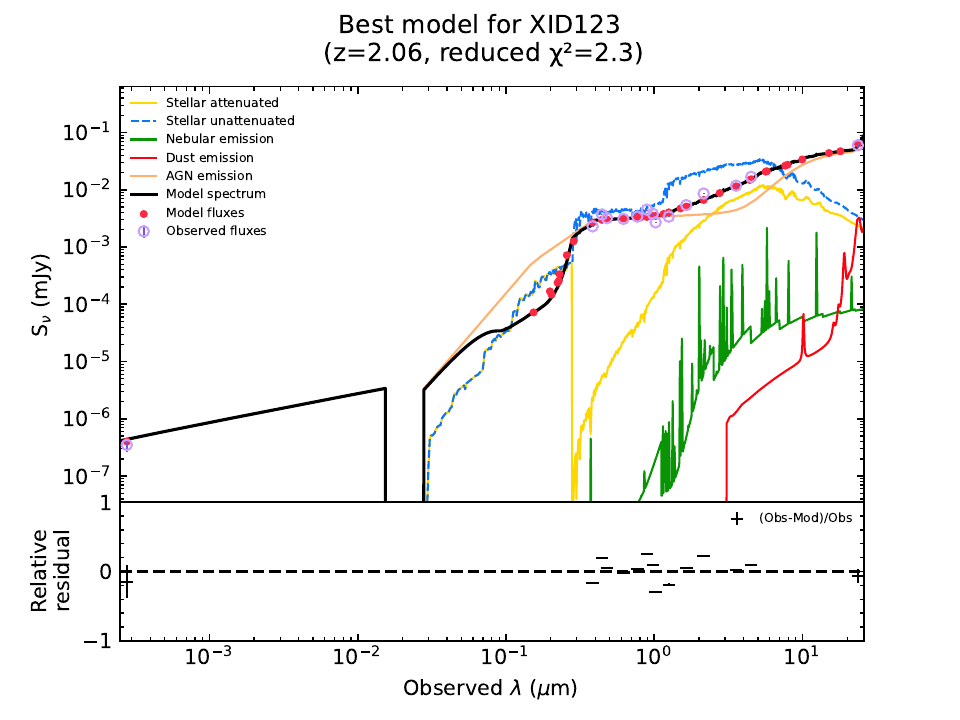}& 
\includegraphics[scale=0.30]{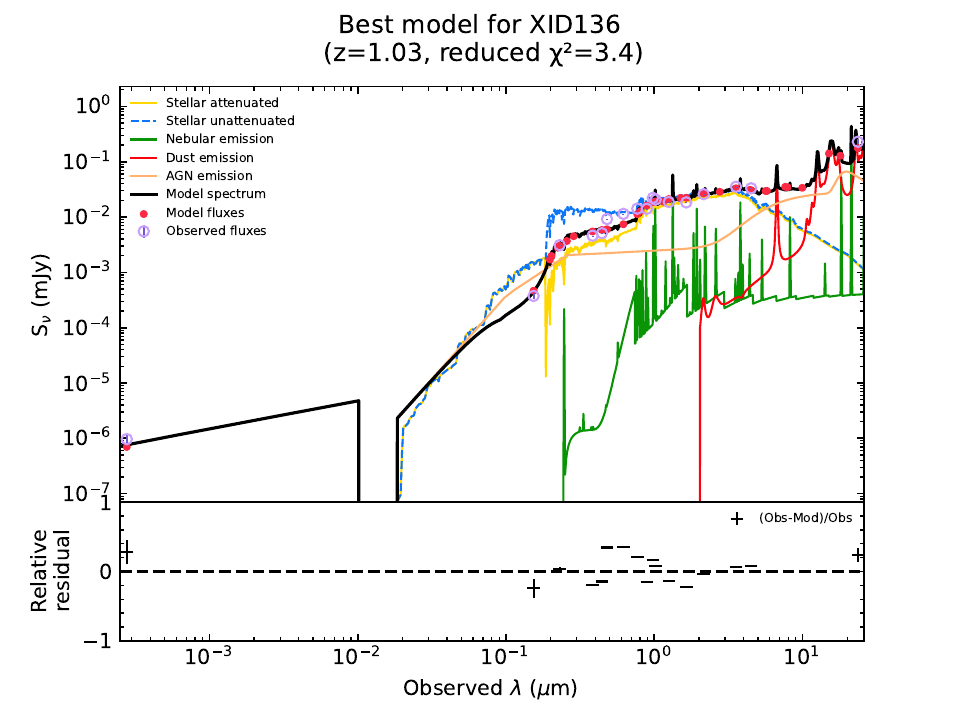}\\ 
\end{array}$
\end{center}
\caption{The best-fitting SEDs of S2, the AGN without JWST NIRCam coverage. The dust emission is shown by the red line. The orange line shows the AGN model. The blue curve shows the stellar light. The observed data is shown by the violet points. The red circles are the model fluxes in the given bands.}
\label{fig:figB4}
\end{figure*}

\renewcommand{\thefigure}{\Alph{section}\arabic{figure}(Cont.)}
\addtocounter{figure}{-1}
\begin{figure*}
\begin{center}$
\begin{array}{lll}
\includegraphics[scale=0.30]{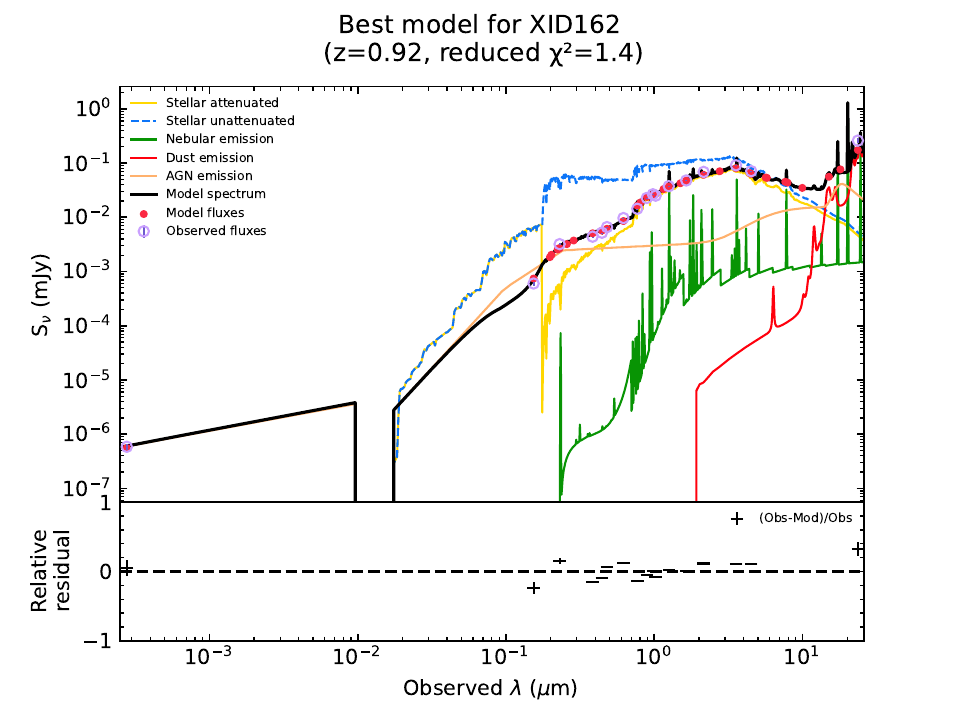}& 
\includegraphics[scale=0.30]{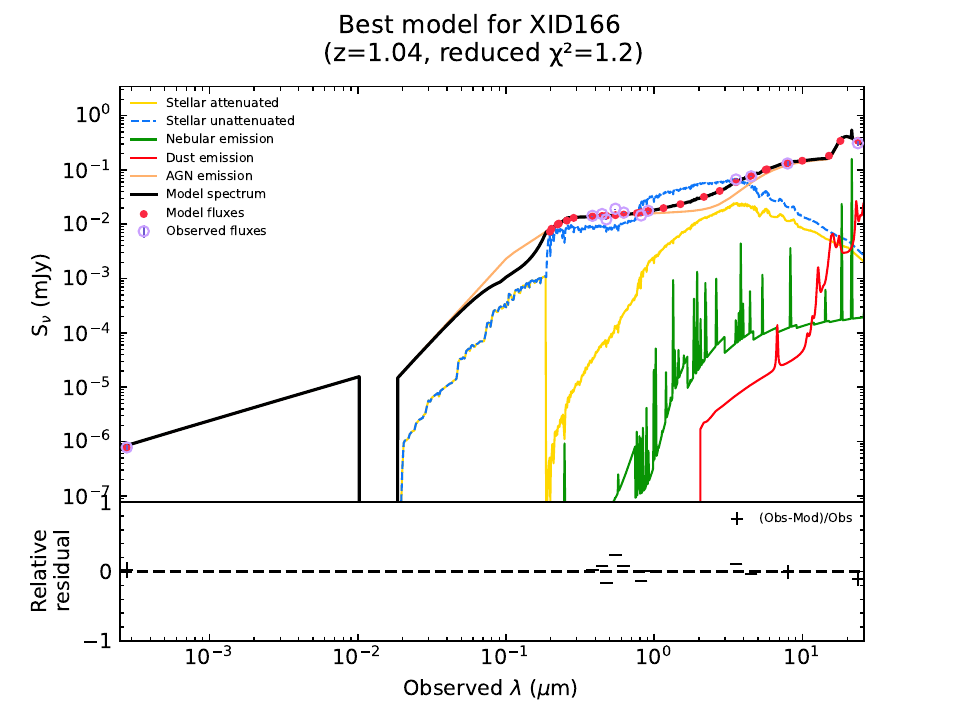}& 
\includegraphics[scale=0.30]{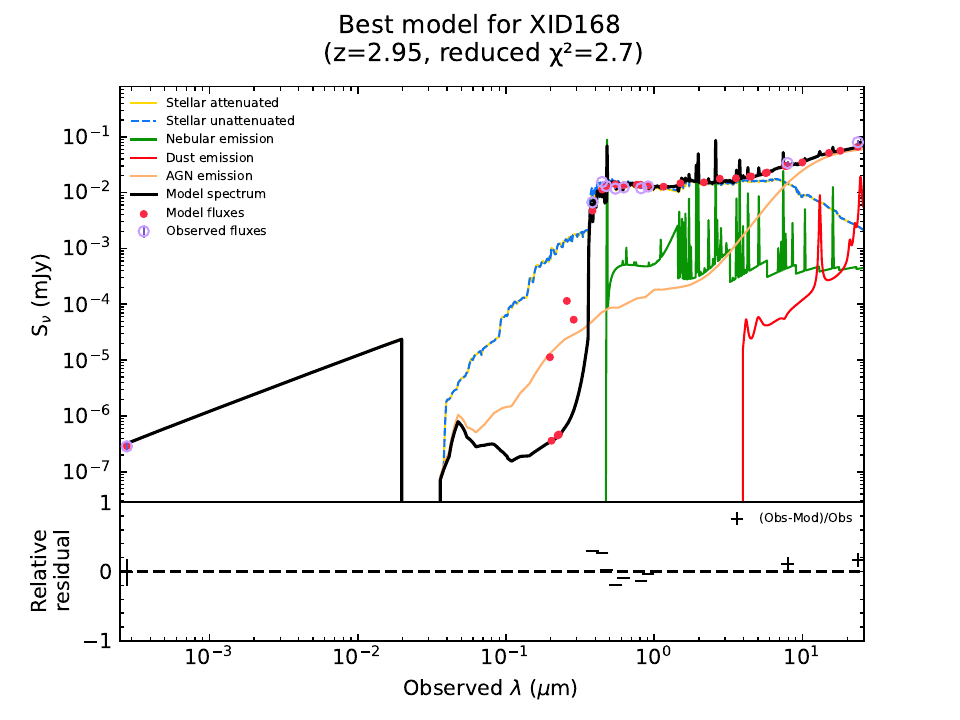}\\ 
\includegraphics[scale=0.30]{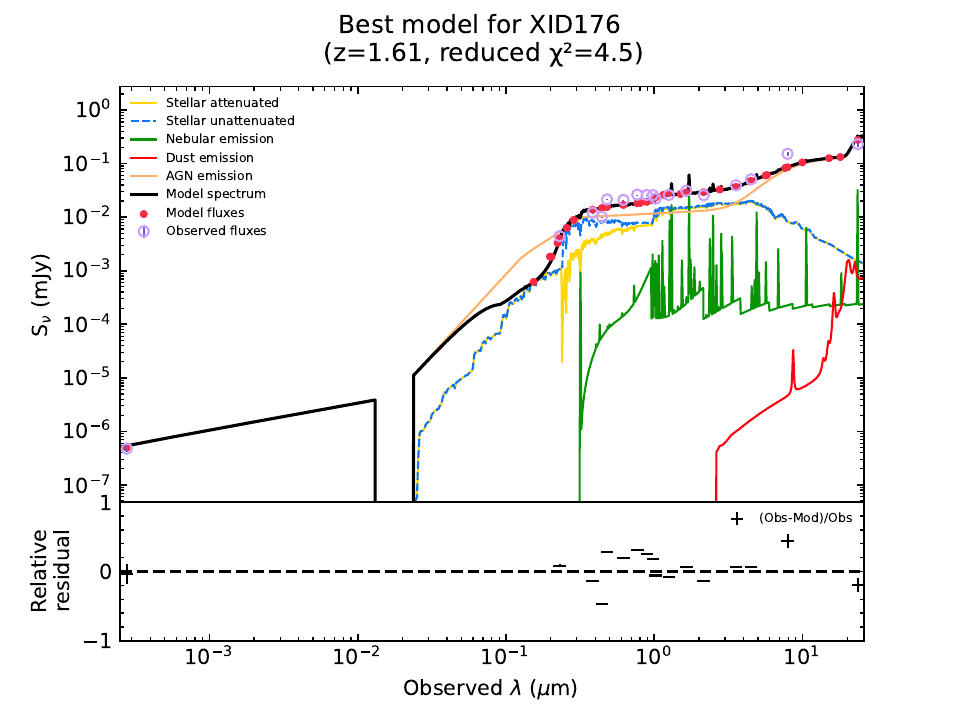}& 
\includegraphics[scale=0.30]{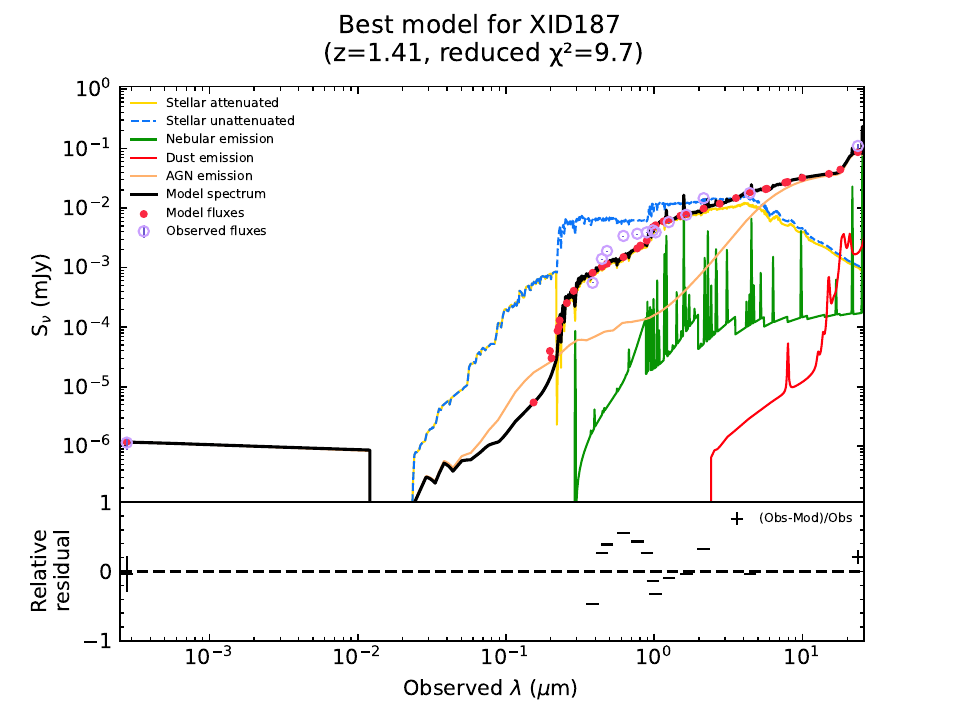}& 
\includegraphics[scale=0.30]{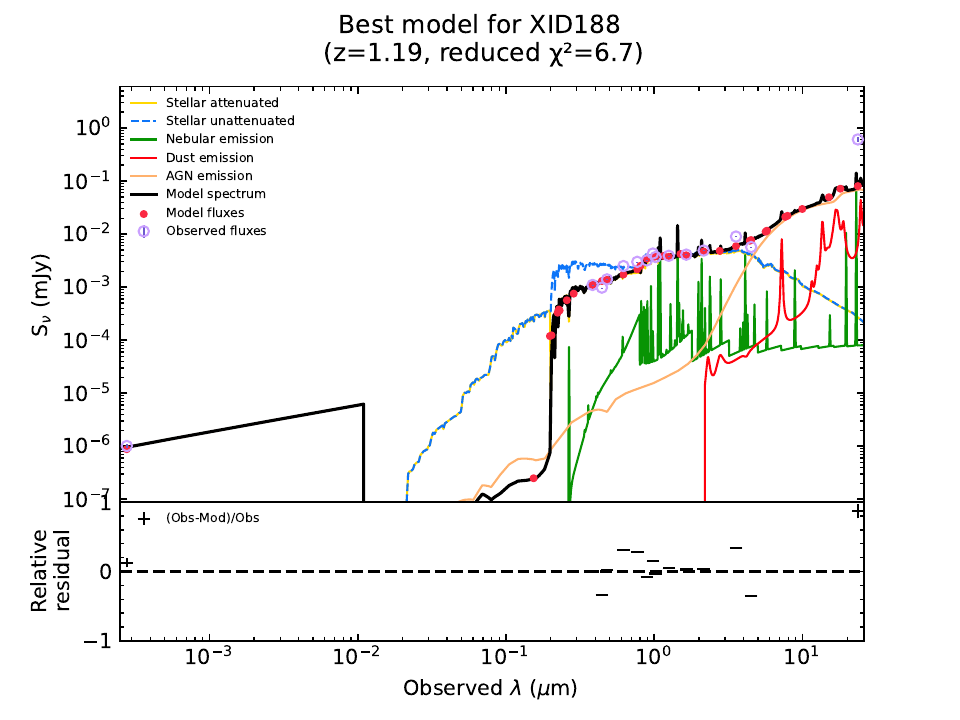}\\ 
\includegraphics[scale=0.30]{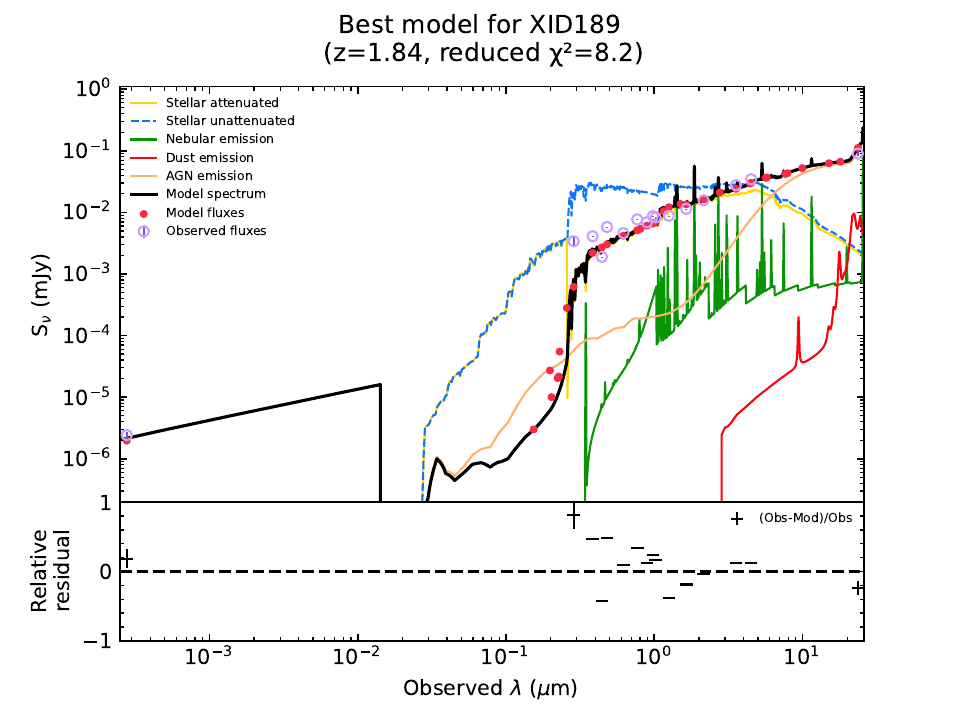}& 
\includegraphics[scale=0.30]{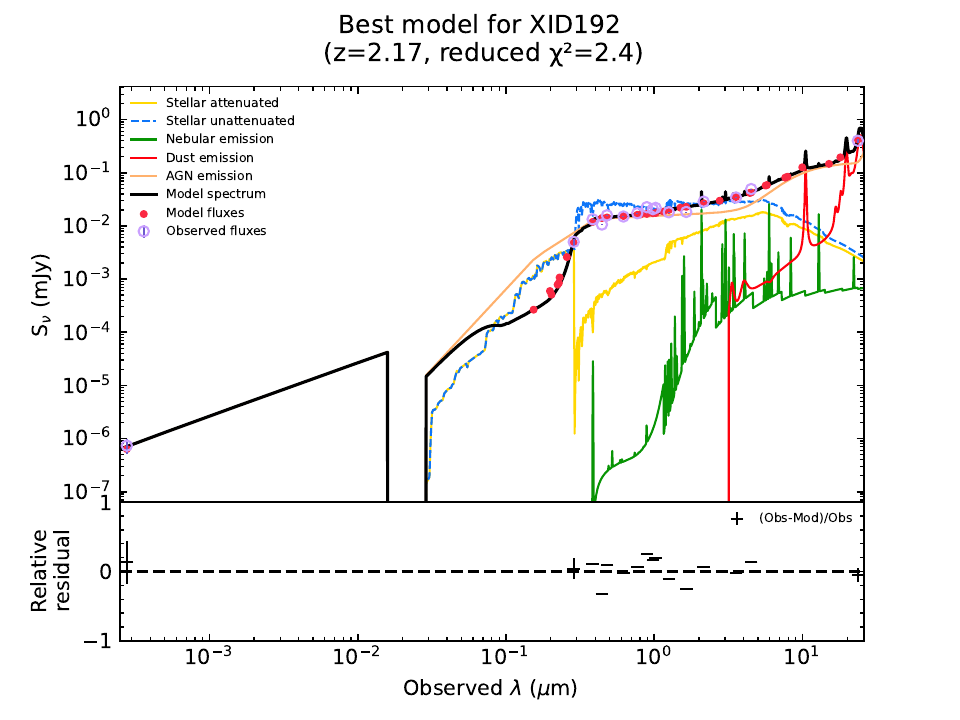}& 
\includegraphics[scale=0.30]{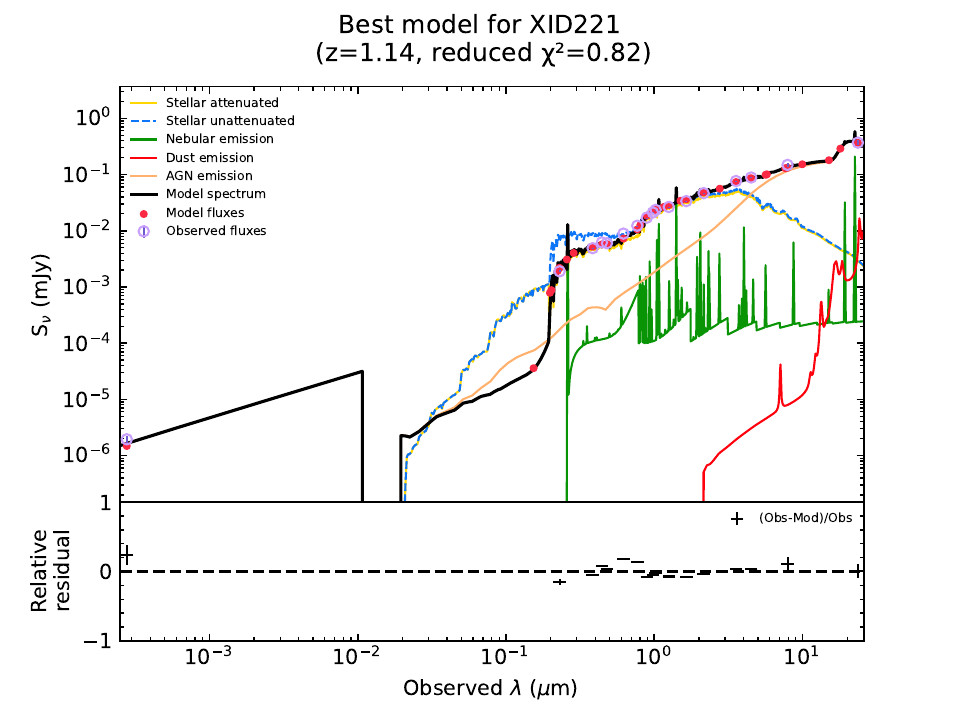}\\ 
\includegraphics[scale=0.30]{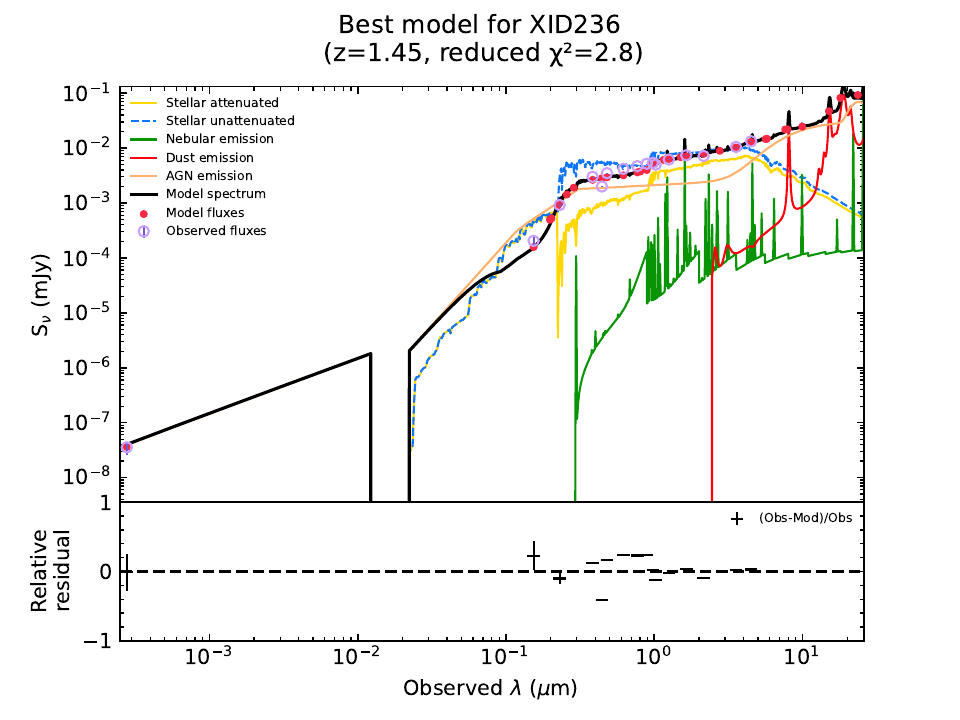}& 
\includegraphics[scale=0.30]{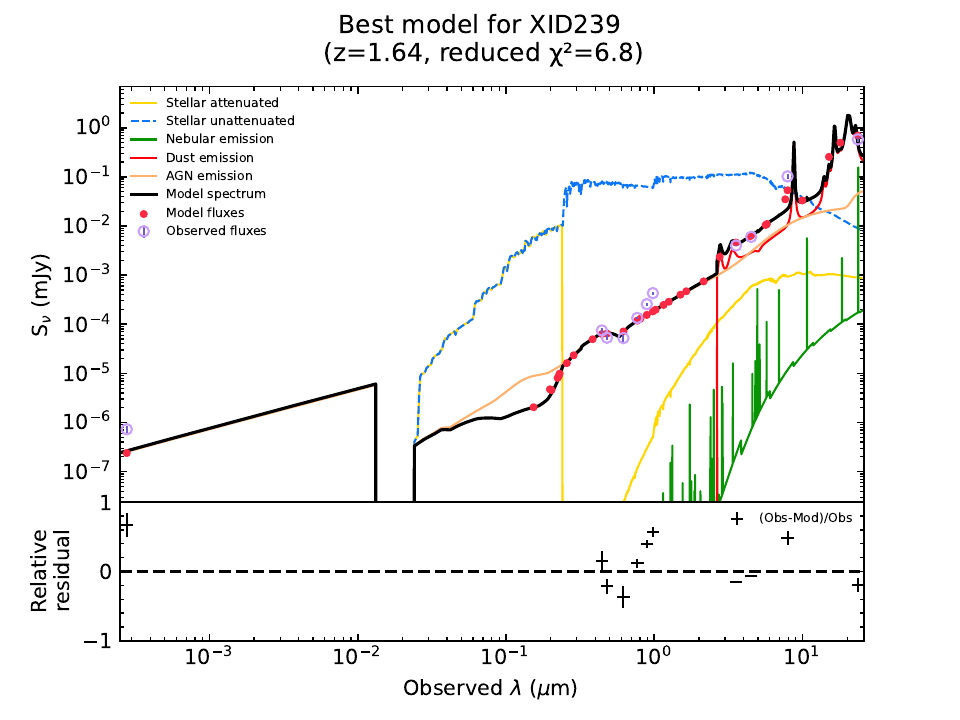}& 
\includegraphics[scale=0.30]{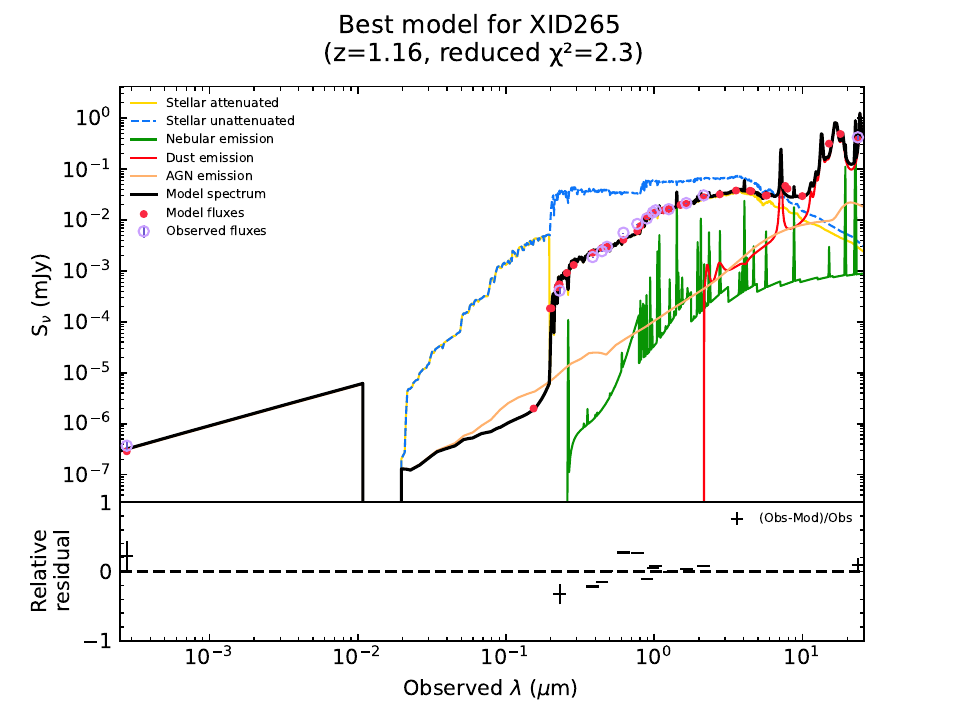}\\ 
\includegraphics[scale=0.30]{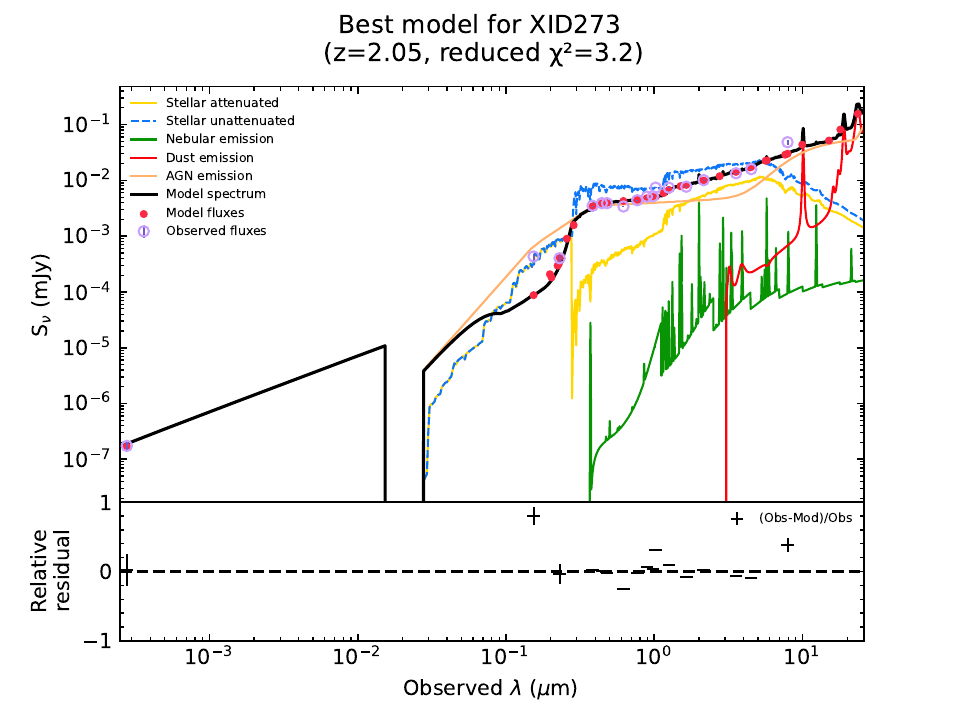}& 
\includegraphics[scale=0.30]{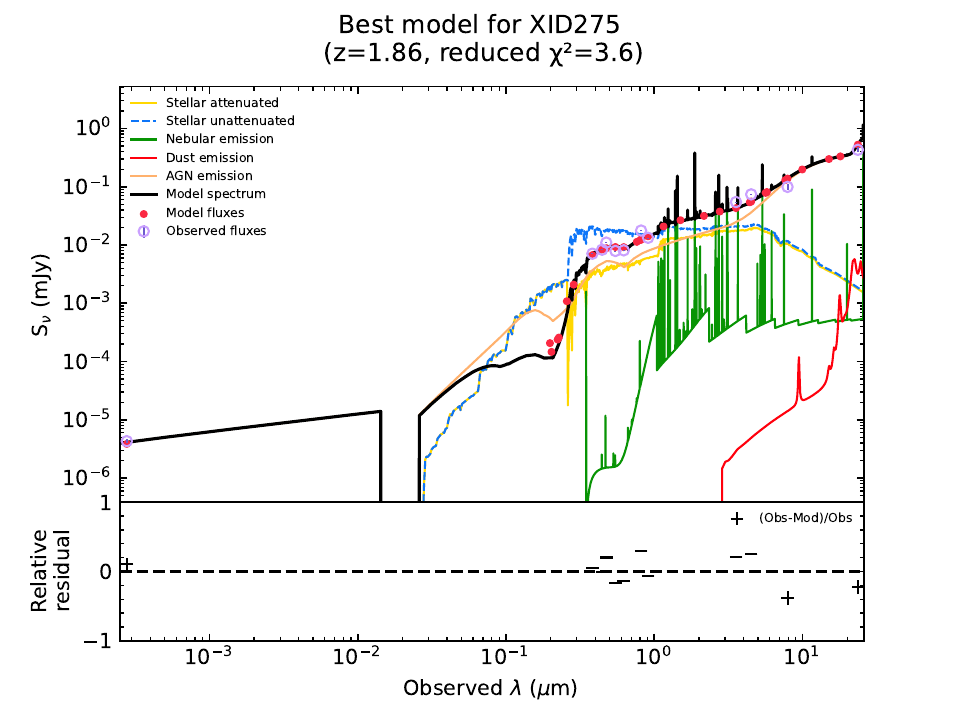}& 
\includegraphics[scale=0.30]{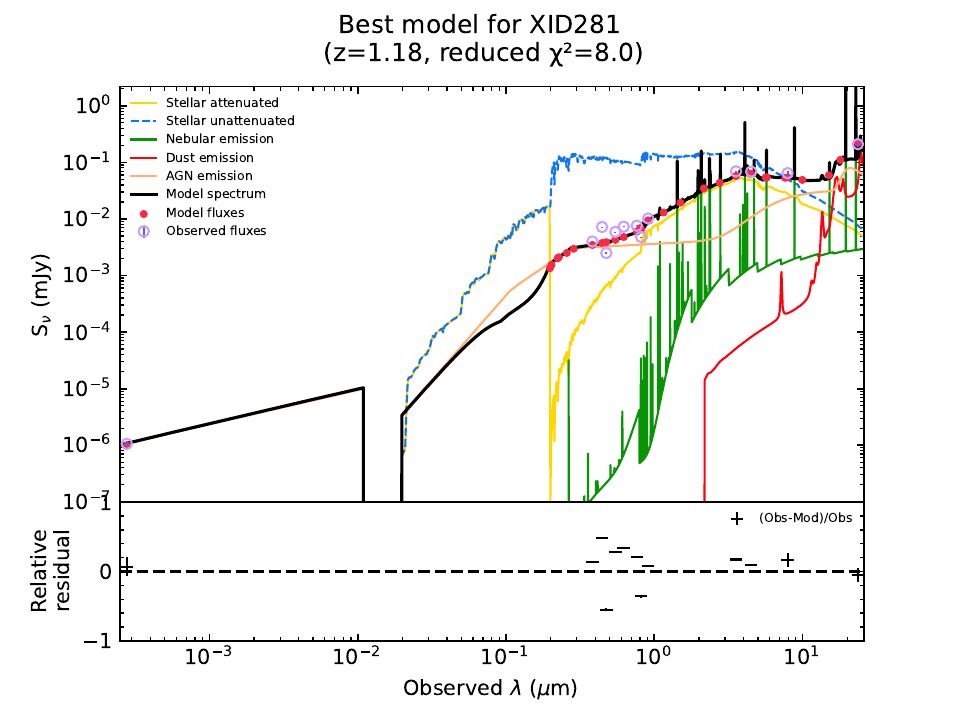}\\ 
\end{array}$
\end{center}
\caption{The best-fitting SEDs of S2, the AGN without JWST NIRCam coverage.}
\end{figure*}


\bibliographystyle{aasjournal}
\bibliography{refs}{}

\end{document}